\documentclass[aps,pra]{revtex4-1}

\usepackage{tabularx} 
\usepackage{amsmath} 
\usepackage{amssymb}
\usepackage{ccaption}
\usepackage{natbib}
\usepackage{overpic}
\usepackage[dvipsnames]{xcolor}
\usepackage{tikz}
\usepackage{amsfonts}
\usepackage{standalone}
\usepackage{color}
\usepackage{siunitx}
\usepackage{array}
\usepackage{multirow}
\usepackage{gensymb}
\usepackage{booktabs}
\usetikzlibrary{fadings}
\usepackage{graphicx} 
\usepackage[margin=1in,letterpaper]{geometry} 
\usepackage[final]{hyperref} 
\hypersetup{
	colorlinks=true,       
	linkcolor=blue,        
	citecolor=blue,        
	filecolor=magenta,     
	urlcolor=blue         
}

\newcommand{\newstuff}[1]{{\leavevmode\color{blue}{#1}}}

\def\DDt#1{\frac{D#1}{Dt}}
\def\DDtL#1{\frac{D^{L}#1}{Dt}}
\def\DDtcal#1{\frac{\mathcal{D}#1}{\mathcal{D}t}}
\def\ddt#1{\frac{\mathrm{d}#1}{\mathrm{d}t}}

\newcommand{\ddp}[2]{\frac{\partial #1}{\partial #2}}

\def\dd#1_#2{\frac{\mathrm{d}#1}{\mathrm{d}#2}}

\newcommand{\eqn}{equation}
\newcommand{\Eqn}{Equation}

\renewcommand{\div}{\nabla \cdot}
\newcommand{\grad}{\nabla}
\newcommand{\curl}{\nabla \times}
\newcommand{\lap}{\nabla^2}
\newcommand{\lapnew}{\tilde{\nabla}^2}

\newcommand{\euler}[1]{\overline{#1}}
\newcommand{\lagrange}[1]{\overline{#1}^L}

\newcommand{\material}[1]{{#1}_{0}}

\newcommand{\tslow}{t_{\mathrm{s}}}
\newcommand{\tfast}{t_{\mathrm{f}}}
\newcommand{\xp}{\boldsymbol{x}_p}

\newcommand{\Vp}{\boldsymbol{V}_p}
\newcommand{\Rey}{\mathrm{Re}}

\newcommand{\fluidv}{\boldsymbol{u}}
\newcommand{\inertv}{\boldsymbol{v}}
\newcommand{\vdiff}{\boldsymbol{q}}
\newcommand{\vfax}{\boldsymbol{q}_\mathrm{F}}
\newcommand{\accforcenew}{\boldsymbol{a}}
\newcommand{\stream}{\boldsymbol{\psi}}

\newcommand{\genv}{\boldsymbol{w}}

\newcommand{\meanvL}{\lagrange{\genv}}
\newcommand{\drift}{\euler{\genv}_{\mathrm{d}}}

\newcommand{\refspace}{\mathcal{X}_0}
\newcommand{\configspace}{\mathcal{X}_t}
\newcommand{\meanconfigspace}{\mathcal{Y}_t}

\newcommand{\x}{\boldsymbol{x}}
\newcommand{\plabel}{\boldsymbol{x}_0}
\newcommand{\xtraj}{\boldsymbol{x}^{\xi}}

\newcommand{\flowmap}[1]{\boldsymbol{X}(#1,t)}

\newcommand{\meanflowmap}[1]{\boldsymbol{Y}(#1,t)}
\newcommand{\meanflowmapinv}[1]{\boldsymbol{Y}^{-1}(#1,t)}
\newcommand{\genvmean}{\boldsymbol{W}}
\newcommand{\fluct}{\boldsymbol{\xi}}

\newcommand{\fluctmap}[1]{\fluct(#1,t)}

\newcommand{\push}[1]{{#1}^{\xi}}
\newcommand{\secondmapbase}{\push{\boldsymbol{X}}}
\newcommand{\secondmap}[1]{\secondmapbase(#1,t)}

\newcommand{\surfb}{S_b}
\newcommand{\meansurfb}{\euler{S}_b}
\newcommand{\surfbref}{S_{b0}}

\newcommand{\surfmap}[1]{\boldsymbol{X}_b(#1,t)}
\newcommand{\meansurfmap}[1]{\boldsymbol{Y}_b(#1,t)}
\newcommand{\fluctb}{\boldsymbol{\xi}_b}
\newcommand{\fluctbhat}{\boldsymbol{\Xi}_b}
\newcommand{\fluctmapb}[1]{\fluctb(#1,t)}
\newcommand{\fluctmapbhat}[1]{\fluctbhat(#1,t)}

\newcommand{\particle}{particle}
\newcommand{\particles}{particles}

\newcommand{\accforce}{\boldsymbol{\alpha}}

\newcommand{\Lop}{\mathcal{L}}
\newcommand{\Basset}{\Lop_{\mathrm{B}}}
\newcommand{\Saffman}{\Lop_{\mathrm{S}}}

\renewcommand{\mp}{m_p}
\newcommand{\mf}{m_f}

\begin{document}

\title{Mean transport of inertial particles in viscous streaming flows}
\author{Mathieu Le Provost}
\author{Jeff D. Eldredge}
\email{jdeldre@ucla.edu}
\affiliation{Mechanical \& Aerospace Engineering Department, University of California, Los Angeles, Los Angeles, CA, 90095, USA}

\begin{abstract}
Viscous streaming has emerged as an effective method to transport, trap, and cluster inertial particles in a fluid. Previous work has shown that this transport is well described by the Maxey--Riley equation augmented with a term representing Saffman lift. However, in its straightforward application to viscous streaming flows, the equation suffers from severe numerical stiffness due to the wide disparity between the time scales of viscous response, oscillation period, and slow mean transport, posing a severe challenge for drawing physical insight on mean particle trajectories. In this work, we develop equations that directly govern the mean transport of particles in oscillatory viscous flows. The derivation of these equations relies on a combination of three key techniques. In the first, we develop an inertial particle velocity field via a small Stokes number expansion of the particle's deviation from that of the fluid. This expansion clearly reveals the primary importance of Fax\'en correction and Saffman lift in effecting the trapping of particles in streaming cells. Then, we apply Generalized Lagrangian Mean theory to unambiguously decompose the transport into fast and slow scales, and ultimately, develop the Lagrangian mean velocity field to govern mean transport. Finally, we carry out an expansion in small oscillation amplitude to simplify the governing equations and to clarify the hierarchy of first- and second-order influences, and particularly, the crucial role of Stokes drift in the mean transport. We demonstrate the final set of equations on the transport of both fluid and inertial particles in configurations involving one cylinder in weak oscillation and two cylinders undergoing such oscillations in sequential intervals. Notably, the new equations allow numerical time steps that are $O(10^{3})$ larger than the existing approach with little sacrifice in accuracy, allowing more efficient predictions of transport.
\end{abstract}

\maketitle

\section{Introduction}
\label{introduction}

Recent developments in the fields of biomedical diagnosis, pollutant treatment, drug delivery, and microfluidics---to name a few---have motivated the need for efficient and fast methods to transport, cluster or trap
inertial particles (small finite-sized particles) in a fluid environment. The particles transported, such as drugs or biological cells, are fragile, and any direct contact creates undesirable stresses on the particles that may cause irreversible damage. Though no method of transport can avoid applying stress, non-contact methods can provide opportunities to distribute the stresses over the particle more uniformly, reducing the possibility for damage. Techniques using ultrasound \citep{hertz1995standing, haake2005positioning}, lasers \citep{ashkin1970acceleration,molloy2002lights}, magnetic effects \citep{van2014integrated}, dielectrophoresis \citep{durr2003microdevices} or inertial hydrodynamics effects \citep{chmela2002chip,blom2003chip} have emerged as some of the most effective methods to manipulate inertial particles.

Another attractive possibility for non-contact particle transport is based on the notion of viscous streaming. A streaming flow is a weak but large-scale steady response of the fluid to oscillatory forcing, brought about through the Reynolds stresses imparted on the fluid. Numerous studies have shown the promises of viscous streaming to transport and trap inertial particles. Classical works have focused on viscous streaming created by a cylinder oscillating weakly in rectilinear motion  \citep{raney1954acoustical,riley1965oscillating,riley1966sphere,chong2013inertial}. Lutz et al.~\citep{lutz2006hydrodynamic} have been able to trap particles in steady streaming eddies arranged in a clover-shaped pattern around a cylindrical post fixed in a micro-channel through which fluid was forced in oscillatory fashion. Chong et al.~have identified the mechanisms that underlie this trapping \citep{chong2013inertial}, and have shown that an arrangement of multiple cylinders forced in sequence with oscillatory motions can be used to construct desired inertial particle trajectories \citep{chong2016transport}. Abadi et al.~\citep{abadi2018closed} have recently designed a closed-loop controller for the position and velocity of inertial particles inside a two-dimensional square box using steady streaming mechanisms. The control actuation is made through four vibrating  piezoelectric beams inclined at $45^\circ$ at each corner of the square box. Very good performance was reported: inertial particles were successfully forced to carry out a variety of prescribed motions, such as eight-branch star trajectories or the transport of a constellation of inertial particles without changing the distance between them. Parthasarathy et al.~\citep{gazzola19} have recently shown that, in a arrangement of two cylinders in a fluid, in which one is actively moved and the other is passively transported by the resulting flow, the passive cylinder's transport is enhanced by adding oscillations to the active cylinder's motion. The additional oscillations generate a streaming flow.

Rigid walls are not the only means for introducing oscillatory motion into the fluid to enact a streaming flow. For example, recent works have shown the potential of using streaming flows created by bubbles undergoing oscillatory volume and shape changes \citep{phan2014single,rallabandi2015three,costalonga2015low,bolanos2017streaming}. Two reasons motivate the use of bubbles for actuation of viscous streaming: Larger amplitude motions can be created compared to rigid bodies, resulting in quadratic increase of the streaming speed \citep{rallabandi2015three}. Also, the bubble--fluid interface allows non-zero tangential velocity, leading to less deceleration of inertial particles in the vicinity of bubbles' surface compared to rigid surfaces.

All of these works have shown the potential for particle manipulation using viscous streaming. In order to devise means of strategically exploiting this mechanism for transport, it is important to have a mathematical model for predicting the particle trajectories effected by a given geometry and forcing modality. Chong et al.~\citep{chong2013inertial} showed that the dynamics of isolated inertial particles in viscous streaming flows are well captured by the Maxey--Riley (MR) equation \citep{maxey1983equation} with the addition of a Saffman lift force \citep{saffman1965lift}. (We refer readers to Michaelides \citep{michaelides1997transient} for a review and history of various transport equations for inertial particles.) In particular, this approach accounts for the local velocity of the fluid in the particle transport, but the particle's influence on the fluid motion can be reasonably neglected.

This prediction of particle transport is challenged by the underlying mechanisms responsible for the transport. A viscous streaming flow has two well-separated time scales: a fast oscillatory scale $\tfast$ and a slow one associated with the steady streaming $\tslow$ \citep{chong2013inertial,thameem2017fast}. Consider a weakly oscillating cylinder with angular frequency $\Omega$, amplitude of oscillation $A$ and radius $R$ such that, $\epsilon=A/R \ll 1$, as is typical in streaming applications. The fast time scale is set by the period of oscillation, $\tfast = T \sim 1/\Omega$. Inertial particles are transported at the characteristic (drift-corrected) speed of the streaming flow, $V_s=\epsilon \Omega A$, in a streaming cell of characteristic size $\delta_{DC}$. Chong et al.~\citep{chong2013inertial} have shown that this streaming cell size remains $\delta_{DC}=O(R)$ over a wide range of Reynolds number $Re = \Omega R^2/\nu$. Hence, the characteristic convective time of an inertial particle around a streaming cell is of order $\tslow \sim \delta_{DC}/V_s \sim R/(\epsilon \Omega A) = 1/(\epsilon^2 \Omega)$. Therefore the slow time scale $\tslow$ is a factor $1/\epsilon^2$ larger than the fast scale, i.e., $\tslow=\tfast/\epsilon^2$ with $\epsilon \ll 1$. An inertial particle of small radius $a \ll R$ has a third, even faster time scale: the time of viscous response when its velocity deviates from that of the fluid. The ratio of this scale to that of the oscillations is measured by the Stokes number, denoted in this paper by $\tau$, which is proportional to $\Rey a^{2}/R^{2}$.

Indeed, in a viscous streaming flow, a particle is continuously wiggling around its mean trajectory at the fast time scale and translating at the slow time scale. In order to discern long-time behaviors of inertial particles in viscous streaming flows, simulations must be carried out over several transport time scales, i.e., several multiples of $\tslow$, and consequently, several thousands of oscillation cycles. Since the oscillatory motion of the fluid---occurring on the fast time scale---has non-negligible influence on the particle transport, the simulations of both the particle transport and the governing equations in the fluid must ostensibly be well-resolved temporally at the scale $\tfast$. Even with the one-way coupling described above, a full simulation of a single particle trajectory is computationally expensive when carried out in this manner.

Clearly, the primary desire is to predict the slow-timescale (i.e, `mean') trajectory of the particle and to seek only the averaged influence of the fast-timescale on this trajectory. In this work, we seek to provide a framework for accelerated prediction of inertial particle trajectories in this fashion. We are not the first to pursue such a strategy. It should be noted that Thameem et al.~\citep{thameem2017fast} and Agarwal et al.~\citep{agarwal2018inertial} have proposed a time-scale separation of the Maxey--Riley equation to derive equations resolved at the slow time scale. However, their approach is restricted to a purely radial velocity field, periodic in time at leading order (in powers of $\epsilon$) and steady at second order.

The approach we propose in this work is also focused on oscillatory fluid velocity fields, though it allows for slow transient changes of such fields, and places no restrictions on the field's spatial structure. We will use the framework of the Generalized Lagrangian Mean (GLM) theory of Andrews and McIntyre~\citep{andrews1978exact} to form an expression for the Lagrangian mean velocity field, $\lagrange{\genv}$, associated with an underlying oscillatory velocity field $\genv$. This velocity field, defined as the time-average velocity of the particle passing through any field point, is a cornerstone of our method: once we have it, we can directly compute the mean particle trajectories, using numerical time steps equal to several oscillation periods. The field $\lagrange{\genv}$ explicitly filters the fast fluctuating component from the mean motion of particles; we will show that the fast component's effect is confined to the Stokes drift. To the best of our knowledge, no previous work has used these tools to efficiently solve the Maxey--Riley equation in a setting of disparate time scales.

We will form Lagrangian mean velocity fields for fluid particles as well as inertial particles. The underlying velocity field for inertial particles will be obtained by an asymptotic expansion of the Maxey--Riley equation in small Stokes number for the deviation of the particle's velocity from that of the fluid. In so doing, we will extend an approach used previously by Maxey~\citep{maxey1987gravitational} and Ferry and Balachandar~\citep{ferry2001fast}; here, we will add the important Fax\'en correction to these earlier treatments. This approach allows us to obtain the inertial particle's velocity from that of the fluid at little extra cost. The problem will be further simplified by applying a separate asymptotic expansion in small oscillation amplitude $\epsilon$. Truncating this expansion at second order, we will arrive at a compact and self-consistent form of equations for the fluid velocity field and the subsequent mean transport of fluid and inertial particles. 

The remainder of the paper is organized as follows. The description of the Maxey--Riley equation and their expansion in small Stokes number will be presented in Section \ref{transport}. The equations for the Lagrangian mean velocity field, their simplification in small-amplitude viscous streaming flow, and the algorithm for the fast Lagrangian-averaged transport of particles will be discussed in Section \ref{sGLM}. Applications of our algorithm to the cases of one and two weakly oscillating cylinders will be presented in Section \ref{results}. Concluding remarks will follow in Section \ref{conclusion}. For the sake of completeness but to ensure clarity of the main aspects of the paper, some details on the asymptotic expansion in small Stokes number and a summary of relevant aspects of GLM theory are relegated to the Appendix.

\section{Basic transport for inertial particles}
\label{transport}

In this work, we are interested in computing the trajectory an inertial particle immersed in an incompressible flow. As discussed in Section~\ref{introduction}, we assume that the coupling between the fluid motion and the particle trajectory is one way: the particle's motion is determined by the local fluid velocity and generates a disturbance field that is unmodified by the proximity to oscillating bodies. Thus, the fluid's time-varying velocity field, $\fluidv(\x,t)$ can be assumed known---for example, through analytical or computational means---without regard for the particle's presence, and our focus is only on obtaining the particle trajectory in this field. The goal of this section is to obtain the general transport equations for inertial particles in a form conducive for the next section, in which we distill this transport into the mean motion.

The fluid's density and kinematic viscosity are denoted by $\rho_f$ and $\nu$, respectively.  The fluid velocity field, $\fluidv$, is governed by the incompressible Navier--Stokes equations,
\begin{equation}
\label{ns}
    \ddp {\fluidv}{t} + \fluidv\cdot\grad\fluidv = -\grad p + \frac{1}{\Rey}\lap \fluidv, \qquad \div\fluidv = 0,
\end{equation}
in which all quantities (including pressure, $p$) have been non-dimensionalized by the uniform fluid density $\rho_f$ and the characteristic length and time scales of the flow. These scales are established by the driving mechanism: The fluid is bounded on the interior by impenetrable surfaces that are either stationary or oscillating with angular frequency $\Omega$---generically, we will refer to these surfaces as `oscillators'. Thus, the characteristic time scale is taken as $1/\Omega$ and, in the case of a cylindrically-shaped oscillator, the characteristic length taken as the cylinder's radius, $R$. The flow Reynolds number, $\Rey$, is thus defined as
 \begin{equation}
    \Rey = \frac{\Omega R^2}{\nu}.
\end{equation}
In viscous streaming applications, we anticipate $\Rey = O(10)$. We assume that the fluid is initially quiescent and that the flow is generated in an infinite domain in which the fluid remains at rest at infinity,
\begin{equation}
\label{icbcinf}
    \fluidv(\x,0) = 0,\qquad \fluidv \rightarrow 0,\quad |\x| \rightarrow \infty,
\end{equation}
though this condition at infinity can easily be replaced with, e.g., a steady uniform flow or stationary enclosing walls. The form of the boundary conditions on the oscillators will be discussed later in the paper. For now, we simply note that the displacement amplitude of the oscillations, $A$, will be assumed small compared with the size of the oscillator. The ratio of these scales is denoted by $\epsilon$, so we are assuming that
\begin{equation}
    \epsilon \equiv A/R \ll 1.
\end{equation}
 
Inertial particles are assumed to be rigid spheres with density $\rho_p$ and radius much smaller than the oscillating object, e.g., $a \ll R$. The particle's mass is denoted by $\mp = 4\pi\rho_p a^3/3$, and the displaced fluid mass by $\mf = 4\pi\rho_f a^3/3 = \mp \rho_f/\rho_p$. We will denote the particle trajectory by $\xp(t)$ and associated velocity by $\Vp(t)$:
\begin{equation}
    \ddt {\xp}  = \Vp(t).
    \label{ptransport}
\end{equation}
It will be assumed that the particle starts each trajectory at the same velocity as the surrounding fluid,
\begin{equation}
\label{initvel}
     \Vp(t_0)=\fluidv(\xp(t_0),t_0).
 \end{equation}
It is also useful to define the particle `slip' velocity, $\Vp(t) - \fluidv(\xp(t),t)$, the particle's velocity relative to the surrounding fluid, which is initially zero by virtue of the initial condition (\ref{initvel}).

\subsection{The Maxey--Riley equation with Saffman lift}

We will assume throughout this work that the Reynolds number redefined on the particle radius is small,
\begin{equation}
\label{Re_a}
    \Rey (a/R)^2 \ll 1.
\end{equation}
In this work, for the transport of an inertial particle, we use a form of the Maxey--Riley (MR) equation \citep{maxey1983equation} that includes the Saffman lift \citep{saffman1965lift}, as was done by Chong et al.~\citep{chong2013inertial} or Ferry and Balachandar~~\citep{ferry2001fast}. If we neglect gravity, the trajectory of an inertial particle is governed by



\begin{equation}
\label{maxeyriley}
\begin{split}
        \mp \ddt \Vp =  {} & 6 \pi \rho_f \nu a \left(\fluidv(\xp(t),t)+\frac{1}{6}a^2\lap \fluidv(\xp(t),t)-\Vp(t)\right)\\
        &+\mf \left.\DDt \fluidv \right|_{\xp(t)}\\
        & -\frac{1}{2}\mf \left[\ddt \Vp -\left.\DDt\fluidv \right|_{\xp(t)}
-\ddt {} \left(\frac{1}{10}a^2\lap\fluidv(\xp(t),t)\right)\right]\\
& +2\sqrt{3} \pi \nu^{1/2}a^2\rho_f \Basset\left[\fluidv(\xp(t),t)+\frac{1}{6}a^2\lap \fluidv(\xp(t),t)-\Vp(t) \right]\\
& +2\sqrt{3} \pi \nu^{1/2}  a^2 \rho_f \Saffman\left[\fluidv(\xp(t),t)-\Vp(t)\right].
\end{split}
\end{equation}
Two different time derivatives act on field quantities in \eqn~(\ref{maxeyriley}): by $\mathrm{d}/{\mathrm{d}t}$ and $D/Dt$ we denote, respectively, the time derivative operators following the particle and the fluid:
\begin{align}
    \ddt {} &=  \frac{\partial}{\partial t} +\Vp(t)\cdot\nabla \label{ddt}\\
    \DDt {} &=  \frac{\partial}{\partial t}+\fluidv\cdot\nabla. \label{DDt}
\end{align}

The set of terms on each line of the right-hand side of \eqn~(\ref{maxeyriley}) represent, respectively, the Stokes drag, the fluid acceleration force, the added mass effects, and finally, the Basset history force and the Saffman lift, with linear operators respectively defined as
\begin{align}
    \Basset[\boldsymbol{f}](t) &=  \sqrt{\frac{3}{\pi}}\int_{-\infty}^t\frac{\mathrm{d}\boldsymbol{f}/\mathrm{d}\tau}{\sqrt{t-\tau}}d\tau,\\
    \Saffman[\boldsymbol{f}](t) &= \frac{3\sqrt{3} J_\infty}{2\pi^2 \sqrt{|\boldsymbol{\omega}(\xp(t),t)|}}\boldsymbol{f}(t)\times \boldsymbol{\omega}(\xp(t),t),
\end{align}
where $\boldsymbol{\omega}=\curl \fluidv$ denotes the associated vorticity of the fluid flow at the location of the particle. 
For the coefficient $J_\infty$, we use $J_\infty = 2.255$: the limit of the lift coefficient function $J(\eta)$ as the ratio $\eta=\Rey_G^{1/2}/\Rey_p$ goes to infinity. (For details on this function $J$, see \citep{mei1992approximate,wang1997role}.) 

The Basset history force is a memory term due to the unsteady diffusion of vorticity from the particle during its traveling history. Several studies \citep{mordant2000velocity,olivieri2014effect} have shown that it can be of significant importance. We retain the term for now for the sake of generality and comparison with previous works, and our scaling analysis below will not reveal it to be clearly smaller than other terms. However, in the context of particle transport in viscous streaming, Chong et al.~\citep{chong2013inertial} have shown empirically that this term is of negligible importance in the current parameter regime and can safely be ignored. We will do the same later in the paper.

\Eqn~(\ref{maxeyriley}) also contains the Fax\'en corrections (the Laplacian of the fluid velocity field), which were shown by Chong et al.~\citep{chong2013inertial} to be crucial in regions of high shear to cause the particle's trajectory to deviate from that of the fluid. That study also demonstrated the important role of the Saffman lift in ultimately trapping the inertial particle at the center of a viscous streaming cell, observed in previous experiments by, for example, Lutz et al.~\cite{lutz2006hydrodynamic}.

Several comments are in order regarding our inclusion of the Saffman lift. This term represents an inertial influence of the fluid when the particle moves relative to the fluid in a region of shear, generating a force on the particle perpendicular to the motion. As such, it is non-linearly dependent on the fluid velocity field. The parameter regime of viscous streaming described in this paper justifies the inclusion of such lift, as we will discuss below.  However, it does not strictly meet all of the conditions under which Saffman derived the expression for lift \cite{saffman1965lift}. That derivation relies on the particle lying in a region of shear that is nearly uniform well beyond a region of length $L_S = (\nu/G)^{1/2}$ (the so-called `Saffman length'), where $G$ is the norm of the local velocity gradient; such shear uniformity enables Saffman's rigorous singular perturbation treatment \cite{saffman1965lift}.

In streaming flows, the particle encounters significant shear within the Stokes boundary layer generated around the oscillating body, a region of thickness $\delta_s = (\nu/\Omega)^{1/2}$. The ratio of the Saffman length to the Stokes boundary thickness should be small to justify the singular perturbation treatment. Here, that ratio is $(\Omega/|\boldsymbol{\omega}|)^{1/2}$, where $|\boldsymbol{\omega}|$ is representative of the instantaneous vorticity in the Stokes layer. In the limit of vanishing oscillation amplitude $\epsilon$, this ratio increasingly fails to abide by the required separation of scales. But this limit is of little practical relevance, as the flow itself vanishes in this limit. In the scenarios described later in this paper, the ratio is of order 1: still not quite sufficient for the strict separation of scales. As Saffman's own analysis of a particle in steady Poiseuille flow showed \cite{saffman1965lift}, this separation of scales is difficult to meet even in many simpler flows. Thus, we interpret the mathematical form here as a representative model of the phenomenon, albeit not fully justified mathematically. We believe that the Saffman lift has served as a useful model in this capacity for many other studies.

Aside from our relaxation of proper scale separation, the other conditions of Saffman's derivation \cite{saffman1965lift}---placed on the various Reynolds numbers---are satisfied in the parameter regime considered in this paper. The shear Reynolds number, $\Rey_G = G a^2/\nu$, describes the squared ratio of the particle size to the Saffman length, and should be much smaller than unity. This Reynolds number is approximately $\Rey_G \sim \Rey (a/R)^2 |\boldsymbol{\omega}|/\Omega$, and thus satisfies its condition by virtue of (\ref{Re_a}). The `slip' Reynolds number, $\Rey_p=a|\Vp - \fluidv| /\nu$, is more difficult to ascertain a priori, but based on the analysis that follows in this paper (demonstrated in equation (\ref{v1})), the slip velocity is dominated by the Fax\'en correction and approximately $a^2|\lap \fluidv|$. In the Stokes layer, where this correction is most active, the Laplacian of the fluid velocity scales like $|\boldsymbol{\omega}|/\delta_s$, and thus $\Rey_p \sim \Rey^{3/2} (a/R)^3 |\boldsymbol{\omega}|/\Omega$, also much less than unity. Using these scalings, the requirement that $\Rey_s/\Rey_G^{1/2} \ll 1$ is also met.

For arbitrary shear flows, different generalizations of the Saffman lift can be found in the literature: Tio et al.~\cite{tio1993dynamics} used expressions involving coordinate-independent fluid shear rate and the norm of the particle slip velocity. However, the form used here, due to Ferry and Balachandar~\citep{ferry2001fast}, is written in a manner that is linearly dependent on the particle velocity. These two formulations are not equivalent for arbitrary shear flows but reduce to the same formula with Saffman's assumptions. In unreported tests, only minor differences in the transport of inertial particles were observed between these two formulas. We retain the second formulation whose linearity in the slip velocity will be helpful for deriving the asymptotic expansion of the Maxey--Riley equation.

It should also be emphasized that the particle transport model omits other effects---namely, hydrodynamic interactions with the wall of the oscillator---that are undoubtedly significant in some parts of the trajectory. In particular, both the tangential and normal components of the particle's motion would be slowed relative to the fluid during encounters with the wall, providing an additional mechanism for the particle to be pushed toward the center of the streaming cell. Rather than compute the full hydrodynamics of these encounters, however, we rely instead on capturing similar effects in a manner that does not require a full coupling with the fluid flow field: the Fax\'en correction to generate a relative slip velocity, the Saffman lift to effect transverse motions, and kinematic constraints to prevent penetration.

Overall, these simplifying assumptions, while omitting some of the physics of the particle transport, are made in order to obtain a model that can predict long-range trajectories in extended arrays of oscillators while preserving the basic mechanisms of particle trapping.


\subsection{The inertial particle velocity field}

The velocity $\Vp(t)$ is clearly a quantity associated with a particle-centered (i.e., Lagrangian) perspective. However, our treatment in this paper benefits greatly from changing our view of particle motion into an Eulerian perspective: the velocity and other quantities observed at a fixed position $\x$ are those attributable to the inertial particle currently occupying that position. That is, we define the {\em inertial particle velocity field} $\inertv(\x,t)$ such that
\begin{equation}
\label{fulltransport}
    \ddt \xp = \inertv(\xp(t),t) = \Vp(t),
\end{equation}
By differentiating this expression, it is clear that the time derivative of $\inertv$ following the particle trajectory (\ref{ddt}) is identical to $\mathrm{d}\Vp/\mathrm{d}t$.

This definition $\inertv(\x,t)$, and much of the remainder of this section, draw closely from the work of Ferry and Balachandar~\citep{ferry2001fast}. We briefly review the treatment here, and adapt it to account for the Fax\'en corrections, which were neglected by Ferry and Balachandar~\citep{ferry2001fast} but which we expect are non-negligible in the current context. It is important to note that the integral curves (i.e., pathlines) of this time-varying inertial particle velocity field describe every possible inertial particle trajectory. The flow map of this field, once obtained, provides a comprehensive solution for inertial particle transport, an extremely valuable result.

However, it should also be noted that the definition of $\inertv$ depends on our choice of initial condition for the particle: with a different choice, a different particle would generally occupy a position $\x$ at time $t$. Ferry and Balachandar~\citep{ferry2001fast} reason that the trajectories for two different choices of initial condition converge toward each other over time, losing memory of their different initial velocities.

From hereon, we will presume that independent and dependent variables have been non-dimensionalized by the characteristic time and length scales of the flow. In our viscous streaming context, it is reasonable to take these, respectively, as the fast flow time scale---the inverse of the oscillation frequency, $1/\Omega$---and the radius $R$ of an oscillating cylinder. Thus, for example, $\fluidv$ and $\inertv$ will henceforth denote the fluid and inertial particle velocity fields scaled by $\Omega R$, $\x$ will be the position scaled by $R$, and time $t$ will represent the dimensional time multiplied by $\Omega$.

Using the definitions presented above and some simple manipulation, we can rewrite (\ref{maxeyriley}) in a more compact dimensionless form:
\begin{equation}
\label{maxeyrileyeulerian}
\begin{split}
        \ddt \inertv =  & \frac{1}{\tau} \left(\fluidv+\vfax-\inertv \right)+\beta \DDt \fluidv
          +\frac{\beta}{5} \ddt{\vfax} +\sqrt{\frac{\beta}{\tau}} \left(\Basset\left[\fluidv+\vfax-\inertv \right]+ \Saffman \left[\fluidv-\inertv \right] \right),
\end{split}
\end{equation} 
in which, for convenience, we have defined a Fax\'en correction velocity,
\begin{equation}
    \label{faxen}
    \vfax = \frac{1}{6}(a/R)^2 \lap \fluidv.
\end{equation}
We have also defined two dimensionless parameters: a density ratio parameter, $\beta$, and a particle Stokes number, $\tau$, respectively, as
\begin{equation}
\label{mrparameters}
\beta \equiv \frac{3}{2\rho_p/\rho_f+1}, \quad \tau\equiv \frac{\Omega a^2}{3\beta \nu}.
\end{equation}
This latter parameter represents the ratio of the characteristic response time of the Stokes drag on the particle to the fastest characteristic flow time scale.

\subsection{Asymptotic expansion of the Maxey--Riley equation in small Stokes number}
\label{Sasymptoticexpansion}

The two parameters introduced in \eqn~(\ref{mrparameters}), $\beta$ and $\tau$, each play an important role in dictating the behavior of the inertial particle motion. For example, $\beta = 1$ represents a neutrally buoyant particle, which, in the absence of the Fax\'en corrections, remains on the same trajectory as a fluid particle. The Stokes number, $\tau$, is proportional to $\Rey (a/R)^2$, which we have already assumed to be small in (\ref{Re_a}), and thus make the same assumption for the Stokes number: $\tau \ll 1$. In the absence of Fax\'en corrections, we would expect this small Stokes number to quickly penalize deviations of the inertial particle's velocity from that of the surrounding fluid (and to render the governing equation (\ref{maxeyrileyeulerian}) numerically stiff). Maxey~\citep{maxey1987gravitational} used this argument to develop an asymptotic expansion in $\tau$ for the inertial particle velocity's departure from that of the fluid. Later, Ferry and Balachandar~\citep{ferry2001fast} extended this expansion to include all of the terms that we have included in the MR equation (\ref{maxeyrileyeulerian}) except for the Fax\'en correction.

In regions of shear, the Fax\'en correction cannot be neglected {\em a priori}. For example, in the Stokes boundary layer formed by an oscillating cylinder, whose thickness scales like $1/\sqrt{\Rey}$, one expects $\vfax \sim \Rey (a/R)^2 \fluidv$, and thus, $\vfax \sim \tau \fluidv$. For the sake of keeping our analysis in this section general, we will not yet explicitly invoke this scaling of $\vfax$. It is important simply to note that the Fax\'en correction is at least comparable to the other dominant terms in the analysis. We will include the Fax\'en correction only as $\vfax$, unadorned with scaling; once its scaling in $\tau$ is determined, its placement in the asymptotic expansion can be adjusted accordingly. We will do so for the case of an oscillating cylinder.

Note that this inclusion changes the apparent target velocity at vanishing $\tau$ from $\fluidv$ to $\fluidv+\vfax$, and also changes the lowest power of $\tau$ in the expansion from $\tau$ to $\tau^{1/2}$, as we shall see. Adapting the approach of Maxey~\citep{maxey1987gravitational} and Ferry and Balachandar~\citep{ferry2001fast}, we write the inertial particle velocity field in terms of the fluid velocity field as
\begin{equation}
\label{vform}
    \inertv=\fluidv + \vfax + \tau^{1/2}\vdiff.
\end{equation}
The deviation of the particle velocity from the target is now borne by the third term on the right-hand side of (\ref{vform}). In Section~\ref{appendix_inertialasymptotics} we present the derivation of the resulting equation for $\vdiff$. In the course of that derivation, we use the expected scaling of $\vfax$ in the Stokes layer surrounding the oscillating cylinder, and furthermore, neglect the Basset memory term as consistent with the analysis of Chong et al.~\citep{chong2013inertial}. The resulting expression is
\begin{equation}
\begin{split}
\label{inertialv}
    \inertv = \fluidv & + \tau  \accforcenew - \tau^{3/2} \beta^{1/2} \Saffman \left[ \accforcenew\right] + O(\tau^2),
\end{split}
\end{equation}
in which the fluid acceleration force, $\accforcenew$, has been defined as
\begin{equation}
\label{accforcenew}
    \accforcenew \equiv (\beta-1) \DDt \fluidv  + \frac{1}{2} \frac{\beta}{\Rey} \lap \fluidv.
\end{equation}

\Eqn~(\ref{inertialv}) forms one of the cornerstones of our proposed method for accelerated simulation of inertial particle transport, since it provides a velocity field that describes this transport everywhere, entirely in terms of the local fluid velocity and its derivatives. However, before we proceed to the distillation of this equation into fast and slow time-scales, we make a few observations. First, it is important to note that the equation reduces to that of Ferry and Balachandar~\citep{ferry2001fast} and Haller et al.~\citep{haller2008inertial} when the Fax\'en correction velocity is omitted. In that situation, the acceleration force (\ref{accforcenew}) reduces to
\begin{equation}
    \accforcenew = (\beta-1)\DDt \fluidv.
\end{equation}

We can observe from this reduced form that, without the Fax\'en corrections, the inertial particle's motion can only depart from that of the fluid if the particle is not neutrally buoyant (i.e., if $\beta \neq 1$). But the retention of these Fax\'en correction terms into the expanded field emphasizes the observations made by Chong et al.~\citep{chong2013inertial}: When the particle is neutrally buoyant or nearly so, deviation of the particle's motion from that of the fluid is solely brought about by the Fax\'en correction velocity, and the particle's subsequent dynamics are dominated by the Saffman lift. In \eqn~(\ref{inertialv}), these observations are confirmed to be the two dominant disturbances from the fluid velocity. For lighter or heavier particles, the dynamics are effected by a mixture of this influence with that from fluid acceleration.

\section{Development of the equations for mean particle transport}
\label{sGLM}

In the previous section, we obtained a velocity field for inertial particle transport that derives from the velocity field of the fluid. The trajectories of both fluid and inertial particles are described by the general transport equation
\begin{equation}
    \ddt \xp = \genv\left(\xp(t),t\right), \quad \xp(0) = \plabel,
\end{equation}
where the generic velocity field $\genv$ can be interpreted as either the fluid velocity field $\fluidv$ or the inertial particle velocity field $\inertv$. In the current discussion it is not important to make the distinction, and we will use the generic term `\particle' to describe either a fluid particle or an inertial particle.

In the flows associated with $\fluidv$ or $\inertv$, the trajectories contain a mixture of fast (fluctuating) and slow (mean) scales. The main objective of this work is to seek the mean trajectories of such \particles~directly, skipping over the integration of the fast scales to the extent possible. This challenge to derive equations resolved only at the slower (or larger) scales exists in many realms of physics, e.g., turbulence or climate modeling. The classical Reynolds decomposition \citep{reynolds1895dynamical} of a fluid quantity (such as velocity or pressure) into a mean and a fluctuating component provides the basic machinery for developing such equations from an Eulerian perspective, i.e., for fluctuations observed from a fixed spatial location. Because we are interested in this paper in seeking the slow-scale transport of individual \particles, we cannot perform a standard Eulerian average of the Maxey--Riley equation; rather, we need to average it in a Lagrangian sense, i.e., for a fixed particle label. Following the work of Andrews and McIntyre~\citep{andrews1978exact,andrews1978wave}, we will introduce a Reynolds-like decomposition of the motion of a particle into a slow (mean) Lagrangian component and a fast (fluctuating) component. In the context of Lagrangian fluid stability, Bernstein \citep{bernstein1958energy} first argued that the fluctuating component of the motion of a particle can be derived from an Eulerian disturbed displacement field, $\fluctmap{\x}$, evaluated at the mean Lagrangian position of the particle. This led to Generalized Lagrangian Mean (GLM) theory, developed by Andrews and McIntyre~\citep{andrews1978exact,andrews1978wave}, who successfully applied the theory to wave problems in the contexts of stratified and rotating fluid flows. Holm analyzed the GLM theory from a geometric point of view and derived the Lagrangian averaged Navier-Stokes-alpha (LANS-$\alpha$) model for turbulent flows~\citep{holm1999fluctuation,chen1999direct,holm2002averaged}.

We will use GLM theory to provide a framework in which to analyze particle transport into fast and slow components. The basic aspects of the theory's application are outlined in Section~\ref{GLMtheory}. Like the Reynolds-averaged Navier--Stokes equations, GLM theory still leaves the treated equations with a closure problem, analogous to the one encountered in the Reynolds-averaged Navier-Stokes equations. However, rather than seek to replace the influence of the fastest scales with a model, we instead account for their influence by explicitly computing the disturbed displacement field, $\fluct$. Following the work of Holm \citep{holm2002averaged}, we formulate a simplified form of the equation for $\fluct$ for small disturbances. In Section~\ref{asymp}, we will clarify this equation via an asymptotic expansion in the small oscillation amplitude that underpins viscous streaming theory.

\subsection{Development of the Lagrangian mean field equations}
\label{GLMtheory}

Given a time-varying Eulerian field $f(\x,t)$ of arbitrary tensor rank, we can define the following averaging operator 
\begin{equation}
\label{eulermean}
    \euler{f}(\x,t)= \frac{1}{T}\int_{t-T}^{t}f(\x,t')dt'.
\end{equation}
In anticipation of the emergence of two timescales, it is important to note that this operator is intended to average the fast scales and leave the slow scales unaffected. Following Holm~\citep{holm2002averaged}, we could formally define $f$ with distinct dependencies on time in these two separate scales. By expanding the averaging operator in the ratio $\tfast/\tslow\ll 1$, it is straightforward to show that the operator's leading-order behavior preserves the slow behavior of $f$, provided that $\tslow \ll T \ll \tfast$. In the case of strictly periodic fast scales, it is sufficient for $T$ to be the period of oscillation or some integer multiple thereof. 

We will refer to the field defined in this fashion as the {\em Eulerian mean field}. The averaging operator has the following properties:
\begin{itemize}
    \item Linearity: For constant scalars $a$, $b$ and two Eulerian fields $f$ and $g$,  $\euler{af+bg}=a\euler{f}+b\euler{g}$
    \item Idempotence: $\euler{\euler{f}}=\euler{f}$
\end{itemize}
It also commutes with spatial and time derivatives, but importantly (and famously in the Navier--Stokes equations), does not commute with the advection operator: That is,
\begin{equation}
    \euler{\fluidv\cdot\grad f} \neq \euler{\fluidv} \cdot \grad \euler{f}.
\end{equation}




In this paper, we seek the mean paths of fluid or inertial particles, and the mean variation of quantities along those paths. The Eulerian mean, assessed at a fixed location, is not the appropriate measure of average in this context. However, it can be used to `induce' a definition of a {\em Lagrangian mean}, $\lagrange{(\cdot)}$: that is, an average of a field taken along the trajectory of a particle, for fixed Lagrangian label. This definition, and several useful tools associated with it, are provided by the Generalized Lagrangian Mean (GLM) theory of Andrews and McIntyre~\citep{andrews1978exact}.

\begin{figure}[t]
    \centering
    \includegraphics[width=0.5\textwidth]{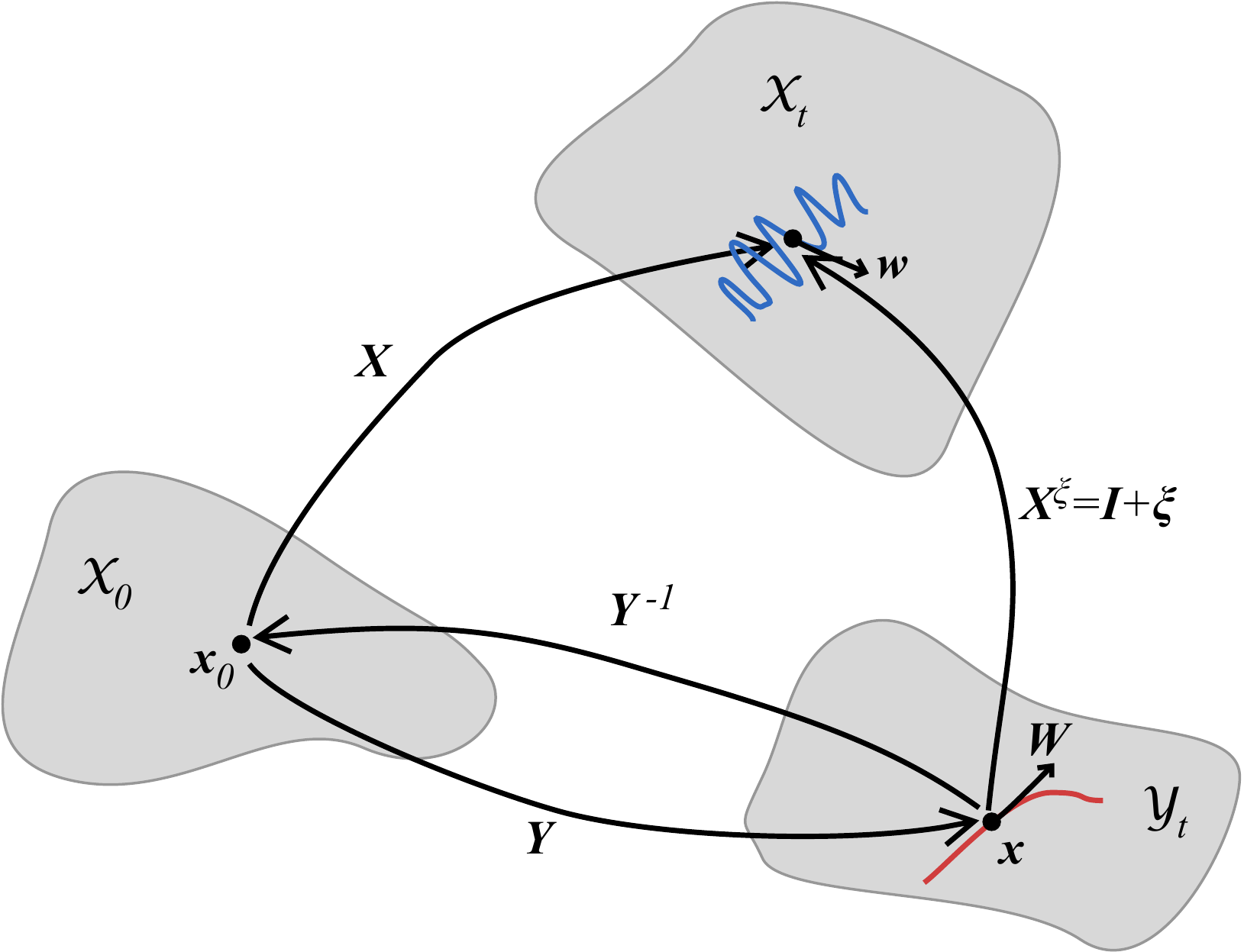}
    \caption{Schematic of flow maps used in this work. $\configspace$ and $\meanconfigspace$ represent slices of the configuration spaces $\configspace \times [0,\infty)$ and $\meanconfigspace \times [0,\infty)$ at some instant $t$. Illustrations of \particle~trajectories are shown as colored curves (though strictly, these trajectories would proceed along the time axis of the respective space).}
    \label{fig:flowmap}
\end{figure}

The basic foundation of GLM theory, illustrated in Figure~\ref{fig:flowmap}, is the separate definitions of an {\em actual configuration space}, $\configspace\times[0,\infty)$, in which \particles~follow the actual (oscillatory) trajectories of the velocity field $\genv$, and a {\em mean configuration space}, $\meanconfigspace\times[0,\infty)$, in which particles follow the mean (smooth) trajectories of this velocity field. From the perspective of any fixed location $\x$ in this mean space, GLM theory describes the difference between the actual and mean positions of particles via an Eulerian {\em disturbed displacement field}, $\fluctmap{\x}$, defined to have zero mean. With this connection between the two spaces, any Eulerian field $\phi$ in the actual space can be described equivalently by an associated Eulerian field $\push{\phi}(\x,t)$ in the mean space, defined in \eqn~(\ref{phirelate}) as
\begin{equation}
\push{\phi}(\x,t) \equiv \phi(\x+\fluctmap{\x},t).
\end{equation}
In other words, this associated field expresses the variation of $\phi$ at fixed $\x$ in the mean space as one follows the disturbed trajectory $\x+\fluctmap{\x}$ in the actual space. The Lagrangian mean of $\phi$ is then sensibly defined by taking the Eulerian mean of $\push{\phi}(\x,t)$, i.e.,
\begin{equation}
\lagrange{\phi}(\x,t) \equiv \euler{\push{\phi}}(\x,t).
\end{equation}
The Lagrangian disturbance $\phi^l$ follows naturally as the difference between the actual value and the Lagrangian mean, i.e.,
\begin{equation}
\phi^l(\x,t) \equiv  \push{\phi}(\x,t) - \lagrange{\phi}(\x,t).
\end{equation}
In particular, the Lagrangian disturbance velocity is defined in (\ref{genvl}). We review other consequences of GLM theory in the appendix.

\subsubsection{The basic equations for mean \particle~transport}

\begin{figure}
  
\tikzset {_63u0u2c9s/.code = {\pgfsetadditionalshadetransform{ \pgftransformshift{\pgfpoint{0 bp } { 0 bp }  }  \pgftransformscale{1 }  }}}
\pgfdeclareradialshading{_ovwuyyfpv}{\pgfpoint{0bp}{0bp}}{rgb(0bp)=(1,1,1);
rgb(0bp)=(1,1,1);
rgb(25bp)=(0,0,0);
rgb(400bp)=(0,0,0)}

  
\tikzset {_jswurlafb/.code = {\pgfsetadditionalshadetransform{ \pgftransformshift{\pgfpoint{0 bp } { 0 bp }  }  \pgftransformscale{1 }  }}}
\pgfdeclareradialshading{_b1n8l59we}{\pgfpoint{0bp}{0bp}}{rgb(0bp)=(1,1,1);
rgb(0bp)=(1,1,1);
rgb(25bp)=(0,0,0);
rgb(400bp)=(0,0,0)}

  
\tikzset {_h9kk8xc9r/.code = {\pgfsetadditionalshadetransform{ \pgftransformshift{\pgfpoint{0 bp } { 0 bp }  }  \pgftransformscale{1 }  }}}
\pgfdeclareradialshading{_yqy05sk51}{\pgfpoint{0bp}{0bp}}{rgb(0bp)=(1,1,1);
rgb(0bp)=(1,1,1);
rgb(25bp)=(0,0,0);
rgb(400bp)=(0,0,0)}
\tikzset{every picture/.style={line width=0.75pt}} 

\begin{tikzpicture}[x=0.75pt,y=0.75pt,yscale=-1,xscale=1]

\draw [color={rgb, 255:red, 57; green, 81; blue, 180 }  ,draw opacity=1 ][line width=2.25]    (100,122) .. controls (140,92) and (191.5,119) .. (242.5,115) ;

\draw [color={rgb, 255:red, 57; green, 81; blue, 180 }  ,draw opacity=1 ][line width=2.25]    (242.5,115) .. controls (293.5,119) and (383.5,47) .. (460.5,74) ;

\draw [color={rgb, 255:red, 56; green, 153; blue, 46 }  ,draw opacity=1 ][line width=2.25]    (100,122) .. controls (140.5,104) and (134.5,73) .. (170.5,65) ;

\path  [shading=_ovwuyyfpv,_63u0u2c9s] (93,122) .. controls (93,118.13) and (96.13,115) .. (100,115) .. controls (103.87,115) and (107,118.13) .. (107,122) .. controls (107,125.87) and (103.87,129) .. (100,129) .. controls (96.13,129) and (93,125.87) .. (93,122) -- cycle ; 
 \draw   (93,122) .. controls (93,118.13) and (96.13,115) .. (100,115) .. controls (103.87,115) and (107,118.13) .. (107,122) .. controls (107,125.87) and (103.87,129) .. (100,129) .. controls (96.13,129) and (93,125.87) .. (93,122) -- cycle ; 

\draw [color={rgb, 255:red, 56; green, 153; blue, 46 }  ,draw opacity=1 ][line width=2.25]    (170.5,65) .. controls (214.5,66) and (241.5,234) .. (281.5,204) ;

\draw [color={rgb, 255:red, 56; green, 150; blue, 46 }  ,draw opacity=1 ][line width=2.25]    (281.5,204) .. controls (321.5,174) and (300.5,93) .. (340.5,63) ;

\draw [color={rgb, 255:red, 53; green, 146; blue, 43 }  ,draw opacity=1 ][line width=2.25]    (340.5,63) .. controls (380.5,33) and (409.5,123) .. (449.5,93) ;

\draw [color={rgb, 255:red, 55; green, 150; blue, 45 }  ,draw opacity=1 ][line width=2.25]    (449.5,93) .. controls (489.5,63) and (556.5,28) .. (549.5,93) ;

\draw [color={rgb, 255:red, 58; green, 83; blue, 184 }  ,draw opacity=1 ][line width=2.25]    (460.5,74) .. controls (511.46,86.74) and (524,111.97) .. (579.07,122.38) ;
\draw [shift={(582.5,123)}, rotate = 189.78] [color={rgb, 255:red, 58; green, 83; blue, 184 }  ,draw opacity=1 ][line width=2.25]    (17.49,-5.26) .. controls (11.12,-2.23) and (5.29,-0.48) .. (0,0) .. controls (5.29,0.48) and (11.12,2.23) .. (17.49,5.26)   ;



\draw [color={rgb, 255:red, 55; green, 150; blue, 45 }  ,draw opacity=1 ][line width=2.25]    (549.5,93) .. controls (536.83,176.85) and (550.77,186.55) .. (592.25,174.01) ;
\draw [shift={(595.5,173)}, rotate = 522.35] [color={rgb, 255:red, 55; green, 150; blue, 45 }  ,draw opacity=1 ][line width=2.25]    (17.49,-5.26) .. controls (11.12,-2.23) and (5.29,-0.48) .. (0,0) .. controls (5.29,0.48) and (11.12,2.23) .. (17.49,5.26)   ;

\draw [color={rgb, 255:red, 0; green, 0; blue, 0 }  ,draw opacity=1 ][line width=1.5]    (251.53,191.94) -- (272.9,221.95) ;
\draw [shift={(274.64,224.39)}, rotate = 234.54] [color={rgb, 255:red, 0; green, 0; blue, 0 }  ,draw opacity=1 ][line width=1.5]    (14.21,-4.28) .. controls (9.04,-1.82) and (4.3,-0.39) .. (0,0) .. controls (4.3,0.39) and (9.04,1.82) .. (14.21,4.28)   ;

\draw [line width=1.5]  [dash pattern={on 1.69pt off 2.76pt}]  (255,121) -- (250.53,192.94) ;

\draw [line width=1.5]    (262,114) -- (304.51,111.2) ;
\draw [shift={(307.5,111)}, rotate = 536.23] [color={rgb, 255:red, 0; green, 0; blue, 0 }  ][line width=1.5]    (14.21,-4.28) .. controls (9.04,-1.82) and (4.3,-0.39) .. (0,0) .. controls (4.3,0.39) and (9.04,1.82) .. (14.21,4.28)   ;

\path  [shading=_yqy05sk51,_h9kk8xc9r] (248,114) .. controls (248,110.13) and (251.13,107) .. (255,107) .. controls (258.87,107) and (262,110.13) .. (262,114) .. controls (262,117.87) and (258.87,121) .. (255,121) .. controls (251.13,121) and (248,117.87) .. (248,114) -- cycle ; 
 \draw   (248,114) .. controls (248,110.13) and (251.13,107) .. (255,107) .. controls (258.87,107) and (262,110.13) .. (262,114) .. controls (262,117.87) and (258.87,121) .. (255,121) .. controls (251.13,121) and (248,117.87) .. (248,114) -- cycle ; 

\draw [color={rgb, 255:red, 53; green, 146; blue, 43 }  ,draw opacity=1 ][line width=2.25]    (439.38,246.5) -- (485.38,246.04) ;
\draw [shift={(489.38,246)}, rotate = 539.4300000000001] [color={rgb, 255:red, 53; green, 146; blue, 43 }  ,draw opacity=1 ][line width=2.25]    (17.49,-5.26) .. controls (11.12,-2.23) and (5.29,-0.48) .. (0,0) .. controls (5.29,0.48) and (11.12,2.23) .. (17.49,5.26)   ;

\draw [color={rgb, 255:red, 58; green, 83; blue, 184 }  ,draw opacity=1 ][line width=2.25]    (440.38,280.5) -- (486.38,280.04) ;
\draw [shift={(490.38,280)}, rotate = 539.4300000000001] [color={rgb, 255:red, 58; green, 83; blue, 184 }  ,draw opacity=1 ][line width=2.25]    (17.49,-5.26) .. controls (11.12,-2.23) and (5.29,-0.48) .. (0,0) .. controls (5.29,0.48) and (11.12,2.23) .. (17.49,5.26)   ;

\draw (92,99) node [scale=1.2]  {$\plabel$};
\draw (94,147) node [scale=1.2]  {$t=0$};
\draw (217,201) node [scale=1.2]  {$\push{\genv}(\lagrange{\x},t)$};
\draw (237,125) node [scale=1.2]  {$\lagrange{\x}$};
\draw (263,152) node [scale=1.2]  {$\fluct$};
\draw (265,81) node [scale=1.2]  {$\lagrange{\genv}(\lagrange{\x},t)$};
\draw (564,244) node   {Actual trajectory};
\draw (563,278) node   {Mean trajectory};


\end{tikzpicture}
\caption{Mean (green) and actual (blue) particle trajectories with initial position $\plabel$. The position $\xtraj$ is the \textit{actual} position whose mean is $\x$. Adapted from B\"{u}hler et al.~\citep{buhler1998non}.}
\label{diagram_GLM}
\end{figure}
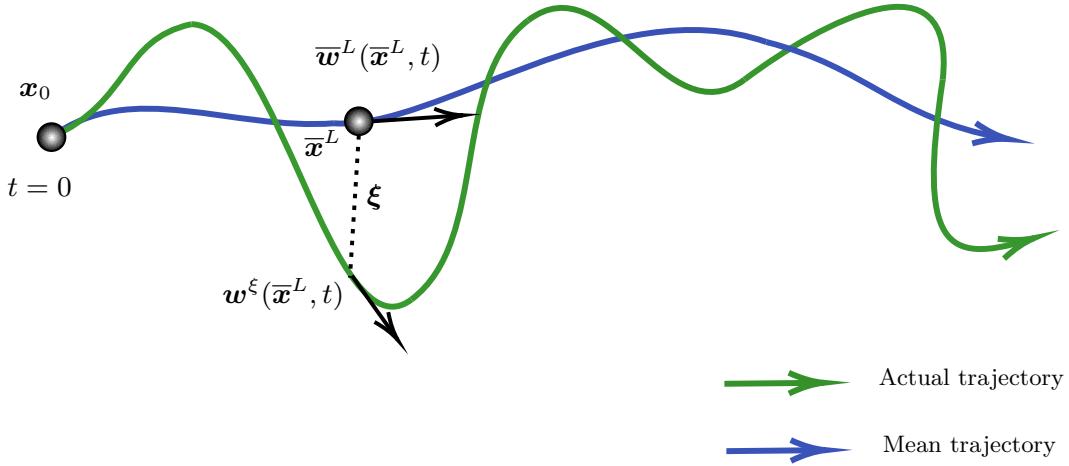

Using the notation for GLM theory defined in Section~\ref{appendix_glm} and illustrated in Figure~\ref{fig:flowmap}, our objective is to seek the mean flow map $\meanflowmap{\cdot}$ for particular values of the \particle~label $\plabel$. The equation generating this trajectory for a specific \particle~$\plabel$ follows directly from equation~(\ref{meanvel}). When we substitute the velocity with the Lagrangian mean velocity using relation~(\ref{genmeanv}), we obtain the kinematic equation for a mean \particle~trajectory:
\begin{equation}
\label{meantraj}
    \ddt {\lagrange{\x}} = \lagrange{\genv}(\lagrange{\x}(t),t), \quad \lagrange{\x}(0) = \plabel,
\end{equation}
where we have used the shorthand notation $\lagrange{\x}(t) \equiv \meanflowmap{\plabel}$ to denote the mean trajectory of a single \particle, $\plabel$, and explicitly included its initial condition. By definition, the Lagrangian mean velocity field requires averaging while following the actual trajectory of the \particle~in $\configspace\times[0,\infty)$, obtained by adding the local disturbed displacement, $\fluctmap{\lagrange{\x}(t)}$ to the trajectory in $\meanconfigspace\times[0,\infty)$. The disturbed displacement field can be generated from the velocity field via its transport equation~(\ref{flucteqn}), which we rewrite here with relevant definitions for the purpose of elucidating the underlying (and thus far, exact) computational problem:
\begin{align}
    \ddp {\fluct}{t} &= -\lagrange{\genv}\cdot\grad\fluct + \push{\genv} - \lagrange{\genv}, \nonumber\\
    \fluct(\x,0) &= 0, \label{flucteqnfull}\\
    \push{\genv}(\x,t) &= \genv(\x+\fluctmap{\x},t), \nonumber \\
    \lagrange{\genv}(\x,t) &= \frac{1}{T} \int_{t-T}^{t} \push{\genv}(\x,t')\,\mathrm{d}t'. \nonumber
\end{align}
We have included here the initial condition on $\fluct$, which was established by requiring that $\meanconfigspace = \configspace = \refspace$ at $t = 0$. Figure~\ref{diagram_GLM} illustrates the relationships between the mean and actual trajectories.

The set of equations (\ref{meantraj}) and (\ref{flucteqnfull}) does not obviously achieve our goal of `skipping over' the fast (oscillatory) time scales of the flow to accelerate the solution for mean trajectories. However, it is important to observe that the coupled equations~(\ref{flucteqnfull}) are Eulerian in the mean configuration space $\meanconfigspace\times[0,\infty)$. Furthermore, when they are supplied with the actual velocity field, $\genv$, they can be solved {\em a priori} to generate the (slowly-varying) Lagrangian mean field $\lagrange{\genv}(\x,t)$, either simultaneously with $\genv$ or in a subsequent procedure. With this Lagrangian mean velocity field in hand, the slow \particle~trajectories are easily obtained {\em with no further regard for the fast (oscillatory) timescale} by integrating \eqn~(\ref{meantraj}).

In Section~\ref{asympmeanv}, we will clarify features of the equations (\ref{flucteqnfull}) that can be used to simplify our task. In particular, we will make the assumption that the disturbed displacement field $\fluct$ is small everywhere (compared with the characteristic length scale of the flow, e.g., $R$) and exploit this assumption to reduce the equations to a more computationally tractable form.

\subsubsection{A note on the disturbed displacement field on moving surfaces}
\label{sec:surfaces}

In the context of this paper, the problems we wish to solve involve flows whose fastest time-scales are generated by impenetrable surfaces undergoing motion consisting entirely of rapid disturbances. The partial differential equation in (\ref{flucteqnfull}) for the disturbed displacement field, $\fluctmap{\x}$, is already closed---it purely describes transport and has no need of boundary conditions. However, it is nonetheless useful to reconcile the surface motion (which we will assume is prescribed) with our definition of the mean configuration space $\meanconfigspace\times[0,\infty)$ and the field $\fluctmap{\x}$; intuitively, we should expect that the surface motion is somewhat simpler in this space. In fact, we will restrict our attention in this work to {\em surfaces that are at rest in the mean configuration space}, and consider the implications of this restriction. After that, we will discuss its rationale.

Let us consider a subset of the \particles~in $\refspace$ to comprise a reference surface, $\surfbref$. This surface represents the interface between the body and the surrounding fluid at $t = 0$. We can also interpret $\surfbref$ as comprising multiple disconnected surfaces in case there are many bodies. The (actual) subsequent motion of any surface \particle~$\plabel \in \surfbref$ is described by a map $\surfmap{\plabel}$, and the collection of all such points is denoted by $\surfb(t)$. For the velocity field $\genv$, whether it represents the fluid or inertial particle motion, we will require that the no-slip and no-penetration conditions are both enforced at the surface. (It should be apparent from the inertial particle velocity field (\ref{inertialv}) that this field does not satisfy these conditions even if the fluid velocity does. We will augment the field with a constraint, to be discussed in Section~\ref{sec:algorithm}.) We thus insist that $\flowmap{\plabel} = \surfmap{\plabel}$ for all $\plabel \in \surfbref$, which ensures that
\begin{equation}
\label{noslip}
    \material{\genv}(\plabel,t) = \ddp {\surfmap{\plabel}}{t}
\end{equation}
for all such \particles. 

We also define the map $\meansurfmap{\plabel}$ and insist that it, too, agrees with the overall map to $\meanconfigspace\times[0,\infty)$ for all \particles~in $\surfbref$. But, in line with our assumption that the surface is at rest in the mean configuration space, we require that this map is the identity: 
\begin{equation}
    \meanflowmap{\plabel} = \meansurfmap{\plabel} \equiv \plabel, \quad \plabel \in \surfbref.
\end{equation}
Applying our Lagrangian form of the Reynolds decomposition (\ref{lagrangereynolds}) to \particles~in $\surfbref$, we can relate these surface maps to the disturbed displacement field:
\begin{equation}
\label{fluctbcref}
    \fluctmap{\plabel} = \fluctmapb{\plabel} \equiv \surfmap{\plabel} - \plabel, \quad \plabel \in \surfbref.
\end{equation}
where we have defined a {\em surface displacement field}, $\fluctmapb{\x}$ for \particles~in the surface. Note that this relationship also ensures that $\fluct(\plabel,0) = 0$, as desired.

To express this relationship in the usual Eulerian form of the field $\fluct$, we will define $\meansurfb$ as the fixed position of the surface in the mean configuration space. By assumption, $\meansurfb = \surfbref$, and clearly, the coordinates of any fixed location on this surface $\x \in \meansurfb$ are equal to the \particle~label there, $\x = \plabel$. Thus, we obtain the following simple expression for the disturbed displacement field at points on the surface:
\begin{equation}
\label{fluctbc}
    \fluctmap{\x} = \fluctmapb{\x} \equiv \surfmap{\x} - \x,\quad \x \in \meansurfb.
\end{equation}
Applying the GLM axiom that the mean of $\fluct$ vanishes identically, it is clear that requiring the surface to remain at rest in the mean configuration space is equivalent to requiring that $\euler{\fluctb} = 0$, or equivalently,
\begin{equation}
\label{meansurf}
    \euler{\surfmap{\x}} = \x.
\end{equation}
That is, the location $\x \in \meansurfb$, as intuitively expected, is identically the mean position of this moving surface point on $\surfb(t)$ in the full configuration space $\configspace\times[0,\infty)$, and $\fluctmap{\x}$ entirely describes its motion in that space. It should also be clear that, being stationary in the mean configuration space implies that the Lagrangian mean velocity, $\lagrange{\genv}(\x,t)$, is exactly zero at all points on the surface:
\begin{equation}
\label{wLbc}
    \lagrange{\genv}(\x,t) = 0,\quad \x \in \meansurfb.
\end{equation}

In the specific case of the fluid velocity field, $\genv\rightarrow\fluidv$, the no-slip boundary condition (\ref{noslip}) can be rewritten in terms of the surface displacement field $\fluctb$ as
\begin{equation}
\label{noslipfluidvel}
    \fluidv(\plabel+\fluctmapb{\plabel},t) = \ddp {\fluctmapb{\plabel}}{t}, \quad \plabel \in \surfbref,
\end{equation}
where $\euler{\fluctb} = 0$ and $\plabel$ coincides with the mean location of the surface point on $\surfb(t)$.

As an example of an admissible surface motion, relevant for the results we will show later in this work, let us consider a rigid body in oscillatory translational motion. Then we can write the full surface map for any surface label $\plabel \in \surfbref$ as
\begin{equation}
    \surfmap{\plabel} = \x_c(t) + \left(\plabel - \x_{c0} \right),
\end{equation}
where $\x_c(t)$ is the time-varying position of the centroid of the body, and $\x_c(0) = \x_{c0}$ is that centroid's initial location. Applying the restriction to this motion, the surface point must remain fixed at $\plabel$ in the mean configuration space, and, to ensure that this truly is the mean space,
\begin{equation}
    \lagrange{\x}_c = \x_{c0}.
\end{equation}
That is, the centroid must start at its mean position. The resulting disturbed displacement field on the mean surface is described by
\begin{equation}
\label{osciltrans}
    \fluctmap{\x} = \fluctmapb{\x} \equiv \surfmap{\x} - \x = \x_c(t) - \x_{c0}, \quad \x \in \meansurfb.
\end{equation}
In other words, when the surface is in rigid translation, the disturbed displacement field is uniform. It is straightforward to conceive of other admissible surface motions that would generate streaming flows, including oscillatory rigid rotation or time-varying deformations about a stationary mean surface.

We should observe that our restriction (\ref{meansurf}) precludes combinations of faster (fluctuating) motions with slower motions that have non-zero mean. We make this restriction to avoid ambiguity involving our definition of the Eulerian mean (\ref{eulermean}) when applied in the mean configuration space at fixed positions on or near the surface. If such a surface were moving in this space, then it would move relative to this fixed averaging position during the averaging interval, obscuring the decomposition of the surface's motion. Of course, by construction, any motion in the mean space is presumed to occur on a much slower time-scale than the averaging interval. In fact, if we rely on two independent measures of time, slow and fast---as, for example, by Holm~\citep{holm1999fluctuation} and others---then the surface can be treated as stationary with respect to the time averaging over the fast scales. However, we have chosen to use only a single measure of time, primarily because the time-scale associated with the slow motion effected by viscous streaming only presents itself {\em a posteriori}. Nevertheless, the viscous streaming flows we focus on in this work do not contain such slow surface motions.

In passing, we note that there is nothing in the analysis of this section that prevents us from applying it to surfaces formed from bubble (liquid--gas) interfaces. However, in that case, one would have to allow for mean transport within the mean surface. That is, although $\meansurfb$ would still be stationary, $\meansurfmap{\cdot}$ would no longer be the identity, but would allow for mean transport on the surface.

\subsection{Asymptotic reduction for small oscillation amplitude}
\label{asymp}

The full governing equations describing the mean transport of fluid and inertial particles have now been specified. These include the equations for the velocity fields themselves---the Navier--Stokes equations (\ref{ns}) for the fluid velocity field $\fluidv$ and \eqn~(\ref{inertialv}) for the inertial particle velocity field $\inertv$ induced by this fluid motion, and the initial condition and boundary conditions on the fluid velocity (\ref{icbcinf}) and (\ref{noslipfluidvel}). They also include the equations for mean transport in these velocity fields, including~(\ref{flucteqnfull}) for the Lagrangian mean $\lagrange{\genv}$ of either of the velocity fields, and equation~(\ref{meantraj}) for the mean transport $\lagrange{\x}(t)$ of any particle, fluid or inertial. Collectively, they represent a map from a given surface displacement field of the oscillator(s), $\fluctb$, to the resulting mean transport of the particles in the fluid.

In this section, we aim to simplify the calculation of this mean transport by exploiting the small amplitude, $\epsilon = A/R \ll 1$, of the oscillations described by $\fluctb$. We will help our cause by exposing the oscillation amplitude with a {\em unit form}, $\fluctbhat$, of the surface displacement field:
\begin{equation}
    \fluctmapb{\x} = \epsilon \fluctmapbhat{\x},
\end{equation}
where \newstuff{$\fluctbhat(\x,t) = O(1)$}. With the driving mechanism proportional to $\epsilon$, we expect that all other quantities, including either velocity field embodied by $\genv$, and its associated disturbed displacement field, $\fluct$, are themselves proportional to $\epsilon$ at leading order (and we do not expect flow instabilities to emerge in this parameter regime that might change this fact). Thus, we will expand all such quantities in powers of $\epsilon$,
\begin{equation}
\label{asympt}
    \genv = \epsilon \genv_1 + \epsilon^2 \genv_2 + O(\epsilon^3), \quad \fluct = \epsilon \fluct_1 + \epsilon^2 \fluct_2 + O(\epsilon^3),
\end{equation}
where $\genv$, as usual, could be either the fluid velocity $\fluidv$ or the inertial particle velocity $\inertv$. We will also use the same expansion for the pressure, $p$ (which, more precisely, represents the pressure disturbance from ambient).

\subsubsection{Reduction of the Navier--Stokes equations}

Let us first introduce the asymptotic expansions for $\fluidv$ and $p$ into the Navier--Stokes equations (\ref{ns}). We can also do the same for the boundary condition (\ref{noslipfluidvel}) and expand the velocity in a Taylor series about $\epsilon = 0$. Equating powers of $\epsilon$, it is easy to show that we get
\begin{equation}
\label{ns1}
    \ddp {\fluidv_1}{t} - \frac{1}{\Rey}\lap \fluidv_1 +\grad p_1 = 0,\qquad \div \fluidv_1 = 0,
    \end{equation}
    and the associated initial condition $\fluidv_1(\x,0) = 0$ and boundary conditions
\begin{equation}
\label{bc1}
    \fluidv_1(\x,t) = \ddp {\fluctmapbhat{\x}}{t}, \, \x \in \meansurfb,\qquad \fluidv_{1} \rightarrow 0, \, |\x| \rightarrow \infty.
    \end{equation}    
At the next level, we get
\begin{equation}
\label{ns2}
    \ddp {\fluidv_2}{t} - \frac{1}{\Rey}\lap \fluidv_2 +\grad p_2 = -\fluidv_1\cdot\grad\fluidv_1,\qquad \div \fluidv_2 = 0,
\end{equation}
with initial condition $\fluidv_2(\x,0) = 0$ and boundary condition
\begin{equation}
\label{bc2}
    \fluidv_2(\x,t) = -\fluctmapbhat{\x}\cdot\grad\fluidv_1(\x,t), \, \x \in \meansurfb, \qquad \fluidv_{2} \rightarrow 0, \, |\x| \rightarrow \infty.
\end{equation}

Equations (\ref{ns1})--(\ref{bc2}) describe the dominant fluid behavior in a viscous streaming problem for a given unit surface motion, $\fluctmapbhat{\x}$. It is important to note that the equations for $\fluidv_1$ and $\fluidv_2$ are both unsteady Stokes equations and linear. The non-linear effects enter the second-order equation, as a forcing term involving only the first-order flow field. The boundary conditions are applied at the mean location of the cylinder surface. Because of this, the second-order boundary condition contains a correction to account for the application of the first-order boundary condition at this mean location rather than its actual location. In fact, as will be revealed in equations~(\ref{expandgenv}) and (\ref{meanv1}) below, the boundary conditions on $\fluidv_1$ and $\fluidv_2$ ensure that fluid particles initially on the surface remain on the surface and that the Lagrangian mean fluid velocity $\lagrange{\fluidv}$ will remain zero on the mean surface to $O(\epsilon^3)$.

There are two significant advantages gained by this asymptotic treatment of the governing equation. First, the geometry of the problem, including that of the oscillators, is fixed, which greatly improves the computational efficiency of the solution procedure; and second, the flow field generated at each asymptotic level has unit order of magnitude, which reduces the effects of numerical error on the solution.

With this expansion of the fluid velocity in small amplitude, the inertial particle velocity field, $\inertv$, provided by \eqn~(\ref{inertialv}), can be asymptotically expanded in the same manner, though with the inclusion of an intermediate term at $\epsilon^{3/2}$ due to the Saffman lift, $\inertv = \epsilon \inertv_{1} + \epsilon^{3/2} \inertv_{3/2} + \epsilon^{2}\inertv_{2}$. It is straightforward to show that the leading contributions are given by
\begin{equation}
\label{v1}
\inertv_{1} = \fluidv_{1} + \tau \accforcenew_{1}, 
\end{equation}
and 
\begin{equation}
\label{v2pre}
\inertv_{3/2} = -\tau^{3/2} \beta^{1/2} \frac{3\sqrt{3} J_\infty}{2\pi^2 \sqrt{|\boldsymbol{\omega}_{1}|}}\accforcenew_1\times \boldsymbol{\omega}_{1}, \qquad \inertv_{2} = \fluidv_{2} + \tau \accforcenew_{2},
\end{equation}

where $\boldsymbol{\omega}_{1} = \curl \fluidv_{1}$ and where the leading-order acceleration forces $\accforcenew_1$ and $\accforcenew_2$ are given by
\begin{equation}
\accforcenew_1 = (\beta-1) \ddp {\fluidv_{1}}{t}  + \frac{1}{2} \frac{\beta}{\Rey} \lap \fluidv_{1}
\end{equation}
and
\begin{equation}
\accforcenew_2 = (\beta-1) \left(\ddp {\fluidv_{2}}{t}  + \fluidv_{1}\cdot\grad\fluidv_{1}\right) + \frac{1}{2} \frac{\beta}{\Rey} \lap \fluidv_{2}.
\end{equation}
In order to unify our later discussions on averaging the trajectories of fluid and inertial particles under the generic velocity $\genv$, we take some liberty in asymptotic notation by lumping $\epsilon^{3/2} \inertv_{3/2} + \epsilon^2 \inertv_{2}$ into a single second-order term, $\epsilon^2 \inertv_{2}$, where
\begin{equation}
\label{v2}
\inertv_{2} = -\epsilon^{-1/2}\tau^{3/2} \beta^{1/2} \frac{3\sqrt{3} J_\infty}{2\pi^2 \sqrt{|\boldsymbol{\omega}_{1}|}}\accforcenew_1\times \boldsymbol{\omega}_{1} + \fluidv_{2} + \tau \accforcenew_{2}.
\end{equation}
Though there is some awkwardness in this notation with a negative power of $\epsilon$, there should be no ambiguity, and the $O(\epsilon^{3/2})$ term is easily recovered in every expression that follows.



With the advantages presented by the mean configuration space, it is worth wondering whether we might formulate and solve governing equations directly for $\push{\fluidv}$ in this space to provide a more direct path to the Lagrangian mean velocity field. Such equations have been derived, for example, by Andrews and McIntyre~\citep{andrews1978exact,andrews1978wave}. These equations introduce new quantities, such as the pseudo-momentum density field, $-\fluidv^{l}\cdot\grad\fluct$, that couple the disturbed displacement field into the equations. We have chosen instead to solve for $\fluidv$ in the actual configuration space and then follow the procedure described in Section~\ref{asympmeanv} to relate this velocity (or the inertial particle velocity) to its corresponding Lagrangian mean field. Ultimately, after asymptotic expansions in $\epsilon$ have been used to reduce the equations, one can show that both procedures reduce to the same result.

\subsubsection{Reduction of the Lagrangian mean velocity field}
\label{asympmeanv}

Now let us apply our asymptotic expansions (\ref{asympt}) to the particle transport equations. We start with the equation for $\push{\genv}$ in (\ref{flucteqnfull}) and carry out a Taylor expansion about $\epsilon = 0$. We get
\begin{equation}
\label{expandgenv}
    \push{\genv}(\x,t) = \epsilon \genv_1(\x,t) + \epsilon^2 \left(\genv_2(\x,t) + \fluct_1\cdot\grad\genv_1(\x,t) \right) + O(\epsilon^3).
\end{equation}
It should be noted that this expanded form of $\push{\genv}$ relates the Eulerian forms of the actual velocity field in our two spaces: between the value at fixed $\x$ in the mean configuration space and its value at the mean location to which $\x$ is mapped in the actual configuration space. We can then substitute this expanded form of $\push{\genv}$ into the definition of $\lagrange{\genv}$ in (\ref{flucteqnfull}) and easily get an expanded form of this mean velocity:
\begin{equation}
\label{asymptmeanv}
    \lagrange{\genv} = \epsilon \euler{\genv}_1 + \epsilon^2 \left(\euler{\genv}_2 + \euler{\fluct_1\cdot\grad\genv_1} \right) + O(\epsilon^3).
\end{equation}
This shows that, at leading order, the Lagrangian mean velocity at some location $\x$ in $\meanconfigspace$ is equal to the Eulerian mean of the leading-order velocity at the same location in $\configspace$. At the next order, however, an additional term appears: the Eulerian mean velocity is modified by the {\em Stokes drift velocity}, $\drift \equiv \euler{\fluct_1\cdot\grad\genv_1}$. When the fluid velocity field is purely oscillatory (i.e., without transient behavior), it can be easily verified that {\em the leading velocity has zero Eulerian mean, $\euler{\genv}_1 = 0$, for both types of particles}. Thus, the Stokes drift has an essential role in determining the mean trajectories of particles. For inertial particles, where $\genv$ is taken to be $\inertv$, the Saffman lift in (\ref{inertialv}) generates non-zero Eulerian mean at smaller order ($\epsilon^{3/2}$) than for a fluid, but the Stokes drift still cannot be neglected in such a case.

Now let us complete the asymptotic analysis by substituting the expansion of $\fluct$ and both expanded forms of the velocities into the equation for $\fluct$ in (\ref{flucteqnfull}) and equating like powers of $\epsilon$. At the leading two asymptotic levels, we get
\begin{align}
    \ddp {\fluct_1}{t} &= \genv_1 - \euler{\genv}_1, \\
    \ddp {\fluct_2}{t} &= \genv_2 - \euler{\genv}_2 -\euler{\genv}_1\cdot \grad \fluct_1 + \fluct_1\cdot\grad\genv_1 - \euler{\fluct_1\cdot\grad\genv_1}.
\end{align}
First, let us note that at each asymptotic level these equations preserve the zero mean of $\fluct$, thereby ensuring that we remain within the constraints of GLM theory. Second, we observe that we can obtain a completely self-consistent algorithm for generating the Lagrangian mean velocity field by retaining only the equation for $\fluct_1$. In other words, for a given pair of asymptotic velocity fields $\genv_{1}$ and $\genv_{2}$, we can solve
\begin{align}
      \ddp {\fluct_1}{t} &= \genv_1 - \euler{\genv}_1, \label{flucteqnsimple} \\
          \drift &\equiv \euler{\fluct_1\cdot\grad\genv_1},  \label{drift} \\
    \lagrange{\genv}(\x,t) &= \epsilon \euler{\genv}_1(\x,t) + \epsilon^2 \left(\euler{\genv}_2(\x,t) + \drift(\x,t) \right), \label{meanv1}
\end{align}
to generate a Lagrangian mean field, $\meanvL$, valid to $O(\epsilon^{5/2})$. This reduced form only requires the leading {\em Eulerian} velocity fluctuation, $\genv_1 - \euler{\genv}_1$ to obtain the required disturbed displacement field, and thence, the Stokes drift's important contribution, $\drift$, to the Lagrangian mean velocity field for \particle~transport.

It is worth making a few other notes on the Lagrangian velocity before we close this section. To support the first two notes, let us first develop an alternative form of the Stokes drift (\ref{drift}). If we substitute (\ref{flucteqnsimple}) for $\genv_{1}$ and remember that $\euler{\fluct}_{1} = 0$, then this Stokes drift can be written as $\euler{\fluct_{1}\cdot\grad \partial\fluct_{1,t}}$, where $\fluct_{1,t}$ denotes $\partial\fluct_{1}/\partial t$. Using our definition of the Eulerian mean in (\ref{eulermean}), we can integrate this form of the Stokes drift by parts:
\begin{equation}
\drift = \frac{1}{T} \int_{t-T}^{t} \fluct_{1}\cdot\grad\fluct_{1,t'}\,\mathrm{d}t' = \frac{1}{T} \left[ \fluct_{1}\cdot\grad\fluct_{1}\right]_{t-T}^{t} - \frac{1}{T} \int_{t-T}^{t} \fluct_{1,t'}\cdot\grad\fluct_{1}\,\mathrm{d}t'
\end{equation}
with $T$ the averaging interval. The first term on the right-hand side of this equation is identically zero when the field $\fluct_{1}$ is periodic and $T$ is an integer multiple of the period. For transient problems, in which $T$ is taken to be much larger than the oscillatory time-scale, the term does not strictly vanish. However, it should be noted that, had we formally defined separate fast and slow measures of time, then $T$ would be much longer than this fast time scale (while the slow time is effectively held fixed). The term would vanish in this limit. For this reason, we argue that it can be neglected in general cases without consequence. Thus, the Stokes drift can also be written as $\drift = -\euler{\fluct_{1,t}\cdot\grad \fluct_{1}}$, or, alternatively, as a combination of the two forms, with the velocity substituted back in,
\begin{equation}
\label{driftsymm}
\drift  = \frac{1}{2} \left(\euler{\fluct_{1}\cdot\grad \genv_{1} - \genv_{1}\cdot\grad \fluct_{1}}\right).
\end{equation}
This latter form of the Stokes drift has several merits, as we discuss below.

\paragraph{Mean trajectories in an incompressible velocity field.} In the case when the velocity field $\genv$ is incompressible, $\div\genv = 0$, and the Lagrangian mean of this field is steady, $\partial \meanvL/\partial t = 0$, then we can obtain mean particle trajectories directly from the contours of a {\em Lagrangian streamfunction field} associated with $\lagrange{\genv}$ \citep{longuet1953mass}, as we will show here. When $\genv$ is divergence-free, then $\genv_{1}$ and $\genv_{2}$ and their means are, as well. Each of these means can thus be derived from a streamfunction field, which we will denote by $\euler{\stream}_{1}$ and $\euler{\stream}_{2}$, respectively (where, e.g., $\euler{\genv}_{1} = \curl \euler{\stream}_{1}$). It is less obvious that the Stokes drift term, $\drift = \euler{\fluct_1\cdot\grad\genv_1}$ can be derived from a streamfunction, as well.  However, when the velocity field $\genv_{1}$ is divergence-free, then by (\ref{flucteqnsimple}), so are $\fluct_{1}$ and its time derivative, and the right-hand side of (\ref{driftsymm}) can thus be written as $\euler{\div (\fluct_{1}\genv_{1})/2 - \div(\genv_{1}\fluct_{1})/2}$. After applying a standard vector identity on this latter form, the Lagrangian mean of an incompressible periodic velocity field can be written as
\begin{equation}
\lagrange{\genv} = \curl \lagrange{\stream},
\end{equation}
where the Lagrangian streamfunction, $\lagrange{\stream}$, is defined as
\begin{equation}
\label{streamL}
\lagrange{\stream} = \epsilon \euler{\stream}_{1} + \epsilon^{2} \left( \euler{\stream}_{2} + \euler{\stream}_{\mathrm{d}}\right),
\end{equation}
and we have defined a Stokes drift streamfunction, $\euler{\stream}_{\mathrm{d}}$, as
\begin{equation}
\euler{\stream}_{\mathrm{d}} = \frac{1}{2}\euler{ \genv_{1} \times \fluct_{1}}.
\end{equation}

\paragraph{Numerical computation of the Lagrangian mean velocity field.} In our present viscous streaming context, the assumptions necessary to derive the trajectories from a Lagrangian streamfunction are only fulfilled by the fluid velocity field. The inertial particle velocity field is not divergence-free due to the Saffman lift, and we must compute trajectories by numerically integrating the mean transport equation~(\ref{meantraj}). For most problems, this numerical integration must make use of a $\meanvL$ field numerically computed from grid velocity data. This grid-based approximation inevitably introduces error, and the specific form of this error can be highly deleterious. The error is most clearly revealed in the cases in which we have the Lagrangian streamfunction available to verify our result---that is, cases in which $\meanvL$ is divergence-free. Most contours of the Lagrangian streamfunction form closed loops, whereas numerically computed trajectories generally fail to close unless we adopt a numerical approximation that has certain key properties. In particular, if, for two vector fields $\boldsymbol{a}$ and $\boldsymbol{b}$, the underlying approximation satisfies the product rule $\boldsymbol{a}\cdot\grad\boldsymbol{b} = \div (\boldsymbol{a}\boldsymbol{b}) - (\div\boldsymbol{a})\boldsymbol{b}$ in a {\em discrete} sense, and if the {\em discrete} divergence vanishes when the continuous divergence does, then the form of the Stokes drift given by \eqn~(\ref{driftsymm}) greatly mitigates the mismatch between the numerically-computed particle trajectories and the Lagrangian streamlines. Even when the mean particle trajectories cannot be otherwise obtained from a Lagrangian streamfunction, form (\ref{driftsymm}) retains many of its benefits for reducing the accumulated error in the trajectories.

\paragraph{Reconciliation with surface motion.} Finally, it is useful to reconcile equations (\ref{flucteqnsimple})--(\ref{meanv1}) with the expected behavior of these quantities on the surface $\meansurfb$ at this order of approximation. Most obviously, the first-order disturbed displacement field $\fluct_1$ on this surface is described by the fluctuating velocity of the moving surface $\surfb(t)$ evaluated at its mean location in the actual space. Furthermore, by (\ref{wLbc}), we restrict the Lagrangian mean velocity to be exactly zero on $\meansurfb$. Equation (\ref{meanv1}) shows that this restriction requires that $\euler{\genv}_1 = 0$ at the mean location of the surface in actual space; at second order, it is further required that
\begin{equation}
    \euler{\genv}_2 = - \drift,
\end{equation}
which reflects that the Eulerian mean velocity at this mean location is not exactly zero, but must vary with the displacement of the surface from this location.


\subsection{Algorithm for Lagrangian averaged transport of particles}
\label{sec:algorithm}

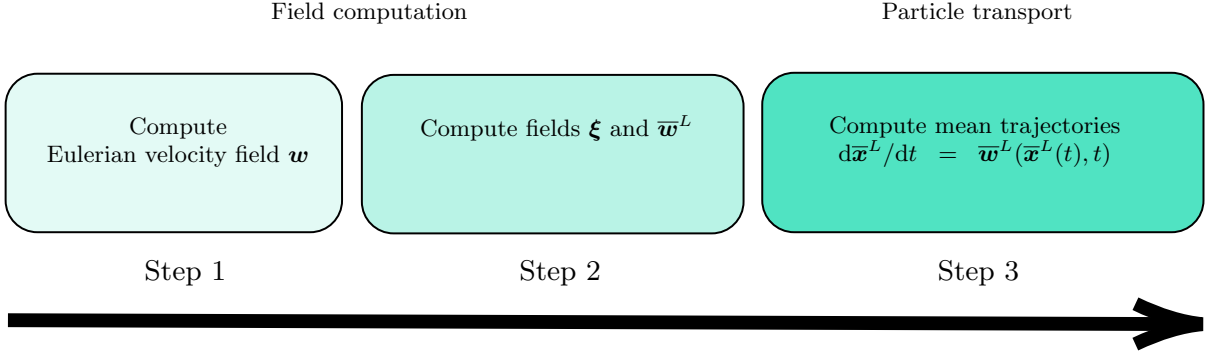
\begin{figure}[t]
\centering
\tikzset{every picture/.style={line width=0.75pt}} 

\begin{tikzpicture}[x=0.75pt,y=0.75pt,yscale=-1,xscale=1]

\draw  [fill={rgb, 255:red, 80; green, 227; blue, 194 }  ,fill opacity=0.16 ] (19.33,106.59) .. controls (19.33,97.75) and (26.5,90.59) .. (35.33,90.59) -- (173,90.59) .. controls (181.84,90.59) and (189,97.75) .. (189,106.59) -- (189,154.59) .. controls (189,163.42) and (181.84,170.59) .. (173,170.59) -- (35.33,170.59) .. controls (26.5,170.59) and (19.33,163.42) .. (19.33,154.59) -- cycle ;
\draw  [fill={rgb, 255:red, 80; green, 227; blue, 194 }  ,fill opacity=0.4 ] (199,107.18) .. controls (199,98.35) and (206.16,91.18) .. (215,91.18) -- (375.5,91.18) .. controls (384.34,91.18) and (391.5,98.35) .. (391.5,107.18) -- (391.5,155.18) .. controls (391.5,164.02) and (384.34,171.18) .. (375.5,171.18) -- (215,171.18) .. controls (206.16,171.18) and (199,164.02) .. (199,155.18) -- cycle ;
\draw  [fill={rgb, 255:red, 80; green, 227; blue, 194 }  ,fill opacity=1 ] (400.92,106.21) .. controls (400.92,97.32) and (408.12,90.12) .. (417.01,90.12) -- (607.38,90.12) .. controls (616.27,90.12) and (623.47,97.32) .. (623.47,106.21) -- (623.47,154.49) .. controls (623.47,163.38) and (616.27,170.59) .. (607.38,170.59) -- (417.01,170.59) .. controls (408.12,170.59) and (400.92,163.38) .. (400.92,154.49) -- cycle ;
\draw [line width=5.25]    (20.5,215) -- (615,216.97) ;
\draw [shift={(623,217)}, rotate = 180.19] [color={rgb, 255:red, 0; green, 0; blue, 0 }  ][line width=5.25]    (33.88,-10.2) .. controls (21.55,-4.33) and (10.25,-0.93) .. (0,0) .. controls (10.25,0.93) and (21.55,4.33) .. (33.88,10.2)   ;

\draw (202.8,59.87) node   {Field computation};
\draw (106.29,125) node[text width=120pt,align=center]  {Compute \\ Eulerian velocity field $\genv$};
\draw (297.25,118) node[text width=150pt,align=center]   {Compute fields $\fluct$ and $\meanvL$};
\draw (508.19,125) node[text width=180pt,align=center]   {Compute mean trajectories \\
$\mathrm{d} \lagrange{\x}/\mathrm{d}t = \lagrange{\genv}(\lagrange{\x}(t),t)$
};
\draw (509.41,59.75) node   {Particle transport};
\draw (110,191) node [scale=1.2]  {Step 1};
\draw (299,191) node [scale=1.2]  {Step 2};
\draw (510,191) node [scale=1.2]  {Step 3};

\end{tikzpicture}
\caption{Overall algorithm for the fast Lagrangian averaged transport of particles.}
\label{fig:algorithm_diagram}
\end{figure}

With the governing equations now developed and reduced for $O(\epsilon)$ oscillations, we can now summarize the proposed algorithm for computing the fast Lagrangian averaged transport of fluid or inertial particles in viscous streaming flows. The algorithm, illustrated in Figure~\ref{fig:algorithm_diagram}, involves three steps. The first step consists of computing the underlying velocity field, $\genv$, for a given surface motion $\epsilon \fluctbhat(\x,t)$. If our interest is in fluid particle trajectories, $\genv$ represents the fluid velocity field $\fluidv$. This velocity is assembled from $\fluidv = \epsilon \fluidv_{1} + \epsilon^{2}\fluidv_{2}$, using the solutions of the sequence of unsteady Stokes equations (\ref{ns1})--(\ref{bc2}). In this work, these equations are solved numerically with the Immersed Boundary Projection Method \citep{taira2007immersed,liska2017fast}. For inertial particles, $\genv$ corresponds to the particle velocity field $\inertv$, induced by the fluid velocity $\fluidv$ via the expansion in small Stokes number (\ref{inertialv}). This velocity field's own asymptotic expansion in $\epsilon$, where needed, is provided through $O(\epsilon^{2})$ by equations~(\ref{v1}) and (\ref{v2}).

In the second step of the algorithm, the leading-order disturbed displacement field, $\fluct_{1}$, is computed by integrating equation~(\ref{flucteqnsimple}), rewritten here for reference:
\begin{equation}
\ddp {\fluct_{1}}{t} = \genv_1 - \euler{\genv}_1.
\end{equation}
In the case of fluid particles, we simply integrate this equation simultaneously with the Navier--Stokes equations in the first step. For inertial particles, we integrate with a third-order Runge-Kutta method, using cubic splines to interpolate the underlying time-discretized velocity field. From this field, the Lagrangian mean velocity field $\meanvL$ is computed from \eqn~(\ref{meanv1}) with the Stokes drift calculated with \eqn~(\ref{driftsymm}), i.e.,
\begin{equation}
\meanvL = \epsilon\euler{\genv}_1 + \epsilon^2 \left(\euler{\genv}_2+ \frac{1}{2} \left(\euler{\fluct_{1}\cdot\grad \genv_{1} - \genv_{1}\cdot\grad \fluct_{1}}\right) \right).
\end{equation}

Finally, in the third step, the Lagrangian mean trajectory of each particle is computed by integrating the mean transport \eqn~(\ref{meantraj}) from some initial location $\plabel$, with a fifth-order Adams--Bashforth method. We generally use a time step that is 10 times larger than the period of oscillation. As we observed in Section~\ref{sec:surfaces}, the inertial particle velocity field (\ref{inertialv}) does not inherently satisfy the no-slip or no-flow-through conditions, even if the fluid velocity field does. This is also true of the solution from the full Maxey--Riley equation, and Chong et al.~\citep{chong2016transport} handled the issue by adding a penalty force inspired from lubrication theory to prevent penetration of inertial particles through the surfaces of oscillators. Here, we use an alternative approach wherein we augment the transport equation with a constraint that the particle remain in the region external to the oscillators. The constraint is posed as
\begin{equation}
\label{nopen}
H_{\delta}(d(\x)) = 0,
\end{equation}
where $d(\x)$ is a signed distance field with respect to the oscillator boundaries---positive in the interior of the oscillator and negative in the exterior---and $H_{\delta}$ is a smooth Heaviside function used previously by Li et al.~\citep{li2003overview}, defined as
\begin{equation}
     H_\delta(z) =   \left\{
\begin{array}{ll}
      0 & z< -\delta, \\
      \frac{1}{2}\left(1+ z/\delta + \pi^{-1} \sin ( \pi z/\delta)\right) & |z|\leq \delta, \\
      1 & z> \delta,
\end{array} 
\right.
 \end{equation}
where $\delta$ is a smoothing distance set equal to the particle radius $a$ for all our simulations. The constrained system of ordinary differential equations is then solved with the manifold projection method described by Hairer et al.~\citep{hairer2006geometric}.

For both types of particles, all Eulerian means in the algorithm are computed with the time average defined in \eqn~(\ref{eulermean}).  The averaging interval $T$ is taken to be 10 periods of oscillation. This ensures that, during transient phases, the interval is long compared with the fast time scale but short compared with the trajectory. When the flow reaches a stationary periodic state, the averaging interval can be reduced without consequence to a single period of oscillation.


\section{Results}
\label{results}

In this section, we present the results from the application of the particle transport algorithm to two representative viscous streaming flows. First, in Section~\ref{validation_asymptotic}, we verify that our asymptotic expansion (\ref{inertialv}) of the inertial particle velocity field in small Stokes number produces time-resolved trajectories that are accurate when compared with the solution of these trajectories from the full Maxey--Riley equation (\ref{maxeyriley}). Then, in Sections~\ref{validation_xi_fluid} and \ref{validation_xi_inertial} we investigate the performance of the mean transport algorithm when applied to fluid and inertial particles, respectively, compared with the fully time-resolved integration of these trajectories, and, in the case of fluid particles, with the Lagrangian streamlines.

The two viscous streaming flows we consider in this work each consist of a flow generated by a rigid cylinder of radius $R$ in weak oscillatory translation with angular frequency $\Omega$. An isolated cylinder in such motion creates four streaming cells arranged along 45 degree rays~\citep{holtsmark1954boundary}; in arrays of multiple cylinders, the cells are still present, though somewhat deformed by the presence of other cylinders. Inertial particles tend to become trapped in these streaming cells, as evident from previous work \citep{lutz2006hydrodynamic,chong2013inertial,chong2016transport}. Our focus in this paper is primarily to confirm the various aspects of the proposed transport algorithm. The first case consists of a single cylinder, while the second case consists of two cylinders that oscillate in sequence: one cylinder oscillates while the other remains stationary, then, after a certain interval, they exchange their roles. In this second case, we are particularly interested in the second cylinder's ability to draw an inertial particle originally trapped near the first cylinder towards one of its own streaming cells.

For both cases, the surface displacement $\fluctb$ of any oscillator is described entirely by the motion of the cylinder's centroid, as expressed in \eqn~(\ref{osciltrans}). In our investigations of this section, the centroid motion is purely sinusoidal, $\x_{c}(t) = \x_{c0} + \epsilon \sin t \,\boldsymbol{e}_{x}$, where $\epsilon = A/R$. (We continue to non-dimensionalize all quantities in this section with $\Omega$ and $R$, as discussed just before \eqn~(\ref{maxeyrileyeulerian}).) In terms of the unit form of this surface displacement, $\fluctb = \epsilon \fluctbhat$, we can express the motion as
\begin{equation}
\fluctmapbhat{\x} = \sin t\,\boldsymbol{e}_{x}.
\end{equation}
Throughout, the Reynolds number, $\Rey$, is held fixed at 40, and the amplitude, $\epsilon$, for any oscillator is $0.1$. The Stokes number of the inertial particles is set at $\tau = 0.1$ and the particle density ratio at $\rho_{p}/\rho_{f} = 0.95$, which correspond to $\beta = 1.034$ and a particle of radius $a/R = 0.088$. 

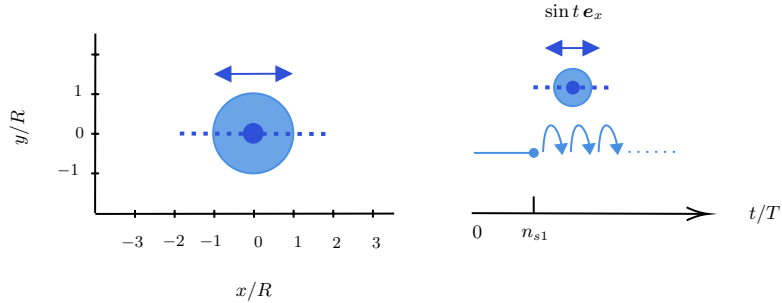
\begin{figure}[tb]
    \centering
    \tikzset{every picture/.style={line width=0.75pt}} 

\begin{tikzpicture}[x=0.75pt,y=0.75pt,yscale=-1,xscale=1]

\draw    (455.2,138.6) -- (572.5,138.99) ;
\draw [shift={(574.5,139)}, rotate = 180.19] [color={rgb, 255:red, 0; green, 0; blue, 0 }  ][line width=0.75]    (10.93,-3.29) .. controls (6.95,-1.4) and (3.31,-0.3) .. (0,0) .. controls (3.31,0.3) and (6.95,1.4) .. (10.93,3.29)   ;

\draw [color={rgb, 255:red, 74; green, 144; blue, 226 }  ,draw opacity=1 ]   (456.3,108.2) -- (486.5,108.2) ;

\draw  [color={rgb, 255:red, 74; green, 144; blue, 226 }  ,draw opacity=1 ][fill={rgb, 255:red, 74; green, 144; blue, 226 }  ,fill opacity=1 ] (484.5,108.2) .. controls (484.5,107.1) and (485.4,106.2) .. (486.5,106.2) .. controls (487.6,106.2) and (488.5,107.1) .. (488.5,108.2) .. controls (488.5,109.3) and (487.6,110.2) .. (486.5,110.2) .. controls (485.4,110.2) and (484.5,109.3) .. (484.5,108.2) -- cycle ;
\draw [color={rgb, 255:red, 74; green, 144; blue, 226 }  ,draw opacity=1 ]   (491.18,107.81) .. controls (492.18,83.21) and (500.98,97.61) .. (500.78,108.41) ;

\draw  [color={rgb, 255:red, 74; green, 144; blue, 226 }  ,draw opacity=1 ][fill={rgb, 255:red, 74; green, 144; blue, 226 }  ,fill opacity=1 ] (500.78,108.41) -- (497.59,104.37) -- (502.95,103.74) -- cycle ;
\draw [color={rgb, 255:red, 74; green, 144; blue, 226 }  ,draw opacity=1 ]   (519.18,107.81) .. controls (520.18,83.21) and (528.98,97.61) .. (528.78,108.41) ;

\draw  [color={rgb, 255:red, 74; green, 144; blue, 226 }  ,draw opacity=1 ][fill={rgb, 255:red, 74; green, 144; blue, 226 }  ,fill opacity=1 ] (528.78,108.41) -- (525.59,104.37) -- (530.95,103.74) -- cycle ;
\draw [color={rgb, 255:red, 74; green, 144; blue, 226 }  ,draw opacity=1 ]   (505.18,107.81) .. controls (506.18,83.21) and (514.98,97.61) .. (514.78,108.41) ;

\draw  [color={rgb, 255:red, 74; green, 144; blue, 226 }  ,draw opacity=1 ][fill={rgb, 255:red, 74; green, 144; blue, 226 }  ,fill opacity=1 ] (514.78,108.41) -- (511.59,104.37) -- (516.95,103.74) -- cycle ;
\draw [color={rgb, 255:red, 74; green, 144; blue, 226 }  ,draw opacity=1 ] [dash pattern={on 0.84pt off 2.51pt}]  (534.25,107.68) -- (560.5,107.68) ;

\draw    (486.46,130.16) -- (486.46,138.56) ;

\draw    (265.38,139) -- (416.08,138.83) (285.37,136.98) -- (285.38,140.98)(305.37,136.96) -- (305.38,140.96)(325.37,136.93) -- (325.38,140.93)(345.37,136.91) -- (345.38,140.91)(365.37,136.89) -- (365.38,140.89)(385.37,136.87) -- (385.38,140.87)(405.37,136.85) -- (405.38,140.85) ;

\draw  [color={rgb, 255:red, 74; green, 144; blue, 226 }  ,draw opacity=1 ][fill={rgb, 255:red, 74; green, 144; blue, 226 }  ,fill opacity=0.81 ] (496.73,75.18) .. controls (496.73,69.97) and (500.95,65.76) .. (506.15,65.76) .. controls (511.35,65.76) and (515.57,69.97) .. (515.57,75.18) .. controls (515.57,80.38) and (511.35,84.59) .. (506.15,84.59) .. controls (500.95,84.59) and (496.73,80.38) .. (496.73,75.18) -- cycle ;
\draw  [color={rgb, 255:red, 47; green, 81; blue, 218 }  ,draw opacity=1 ][fill={rgb, 255:red, 47; green, 81; blue, 218 }  ,fill opacity=1 ] (503.25,75.25) .. controls (503.25,73.59) and (504.59,72.25) .. (506.25,72.25) .. controls (507.91,72.25) and (509.25,73.59) .. (509.25,75.25) .. controls (509.25,76.91) and (507.91,78.25) .. (506.25,78.25) .. controls (504.59,78.25) and (503.25,76.91) .. (503.25,75.25) -- cycle ;
\draw [color={rgb, 255:red, 47; green, 81; blue, 218 }  ,draw opacity=1 ]   (493.97,55.28) -- (517.97,55.28) ;
\draw [shift={(519.97,55.28)}, rotate = 180] [fill={rgb, 255:red, 47; green, 81; blue, 218 }  ,fill opacity=1 ][line width=0.75]  [draw opacity=0] (8.93,-4.29) -- (0,0) -- (8.93,4.29) -- cycle    ;
\draw [shift={(491.98,55.28)}, rotate = 0] [fill={rgb, 255:red, 47; green, 81; blue, 218 }  ,fill opacity=1 ][line width=0.75]  [draw opacity=0] (8.93,-4.29) -- (0,0) -- (8.93,4.29) -- cycle    ;
\draw [color={rgb, 255:red, 47; green, 81; blue, 218 }  ,draw opacity=1 ][line width=1.5]  [dash pattern={on 1.69pt off 2.76pt}]  (486.3,75.1) -- (526,75.25) ;

\draw    (265.25,138.55) -- (265.38,48) (263.28,118.55) -- (267.28,118.55)(263.31,98.55) -- (267.31,98.55)(263.33,78.55) -- (267.33,78.55)(263.36,58.55) -- (267.36,58.55) ;

\draw  [color={rgb, 255:red, 74; green, 144; blue, 226 }  ,draw opacity=1 ][fill={rgb, 255:red, 74; green, 144; blue, 226 }  ,fill opacity=0.81 ] (324.8,98.33) .. controls (324.8,87.17) and (333.84,78.13) .. (345,78.13) .. controls (356.16,78.13) and (365.2,87.17) .. (365.2,98.33) .. controls (365.2,109.48) and (356.16,118.53) .. (345,118.53) .. controls (333.84,118.53) and (324.8,109.48) .. (324.8,98.33) -- cycle ;
\draw [color={rgb, 255:red, 47; green, 81; blue, 218 }  ,draw opacity=1 ]   (327.1,68.39) -- (363.1,68.21) ;
\draw [shift={(365.1,68.2)}, rotate = 539.71] [fill={rgb, 255:red, 47; green, 81; blue, 218 }  ,fill opacity=1 ][line width=0.75]  [draw opacity=0] (8.93,-4.29) -- (0,0) -- (8.93,4.29) -- cycle    ;
\draw [shift={(325.1,68.4)}, rotate = 359.71] [fill={rgb, 255:red, 47; green, 81; blue, 218 }  ,fill opacity=1 ][line width=0.75]  [draw opacity=0] (8.93,-4.29) -- (0,0) -- (8.93,4.29) -- cycle    ;
\draw [color={rgb, 255:red, 47; green, 81; blue, 218 }  ,draw opacity=1 ][line width=1.5]  [dash pattern={on 1.69pt off 2.76pt}]  (307.97,98.43) -- (384.48,98.67) ;

\draw  [color={rgb, 255:red, 47; green, 81; blue, 218 }  ,draw opacity=1 ][fill={rgb, 255:red, 47; green, 81; blue, 218 }  ,fill opacity=1 ] (340.26,98.33) .. controls (340.26,95.71) and (342.38,93.59) .. (345,93.59) .. controls (347.62,93.59) and (349.74,95.71) .. (349.74,98.33) .. controls (349.74,100.94) and (347.62,103.06) .. (345,103.06) .. controls (342.38,103.06) and (340.26,100.94) .. (340.26,98.33) -- cycle ;

\draw (458,148) node [scale=0.8]  {$0$};
\draw (487.33,149.33) node [scale=0.8]  {$n_{s1}$};
\draw (603,139) node [scale=0.8]  {$t/T$};
\draw (345.33,178) node [scale=0.8]  {$x/R$};
\draw (227,95.33) node [scale=0.8,rotate=-270]  {$y/R$};
\draw (323.67,152.67) node [scale=0.7]  {$-1$};
\draw (347.58,153.17) node [scale=0.7]  {$0$};
\draw (257.6,97.8) node [scale=0.7]  {$0$};
\draw (256.67,77) node [scale=0.7]  {$1$};
\draw (251.67,117) node [scale=0.7]  {$-1$};
\draw (367.58,152.83) node [scale=0.7]  {$1$};
\draw (387.08,153.33) node [scale=0.7]  {$2$};
\draw (407.08,153.33) node [scale=0.7]  {$3$};
\draw (305,153) node [scale=0.7]  {$-2$};
\draw (284.08,153.33) node [scale=0.7]  {$-3$};
\draw (507,37) node [scale=0.8]  {$\sin t \, \boldsymbol{e}_{x}$};

\end{tikzpicture}
    \caption{Diagram of the oscillating cylinder (right), and a time sequence illustrating the repeated oscillation cycles.}
    \label{fig:1cylinder}
\end{figure}

\begin{figure}[tb]
    \centering
    \input{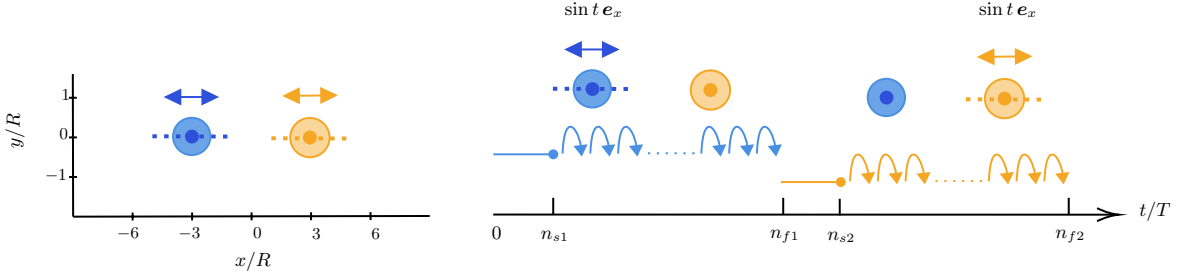}
    \caption{Configuration and oscillation sequence with two cylinders adapted from the work of Chong et al.~\citep{chong2016transport}.}
    \label{fig:2cylinder}
\end{figure}

A schematic of the unit surface motion is depicted in Figures \ref{fig:1cylinder} and \ref{fig:2cylinder}. The blue dot represents the position of the centroid of the cylinder, and the blue dashed line depicts the unit surface displacement applied at points on the fixed cylinder surface. The right diagram in each figure depicts the manner in which we generate flow fields over long time horizons. After each change in the oscillator motion, the flow does not become statistically stationary until viscous diffusion has had sufficient time to act. Once this transient phase has ended and the flow's mean has become stationary, the solution over the last oscillation cycle is re-used as many times as necessary to generate the flow field's history.  In the conditions specified above, we find that the flow becomes statistically stationary in the region within 6 radii of the oscillator after around $n_{s1}=20$ periods of oscillation. We have two such transient phases in the case of two oscillators. As Figure~\ref{fig:2cylinder} shows, these oscillators are arranged 6 radii apart along the same axis on which they oscillate, and the periodic flow solution developed by each oscillator is repeatedly recycled as needed in the respective intervening periods. For particle trapping purposes in this conditions, we find that these recycling intervals require $n_{f1} - n_{s1} = 25000$ periods and $n_{f2} - n_{s2} = 40000$ periods, respectively.

Throughout this investigation we will rely on fluid velocity fields obtained by numerical solution of the unsteady Stokes equations with a procedure based on the immersed boundary projection method with lattice Green's function \citep{taira2007immersed,liska2017fast}. The validation of this procedure, including its convergence to the analytical solution in the case of a single cylinder in oscillatory translation, has been confirmed but is omitted from this work for brevity.  We note that the simulations reported here are carried out on a Cartesian grid with spacing $\Delta x/R = 0.02$ and a time-step size $\Omega\Delta t/(2\pi) = 0.004$, or 250 time steps per period, to satisfy the viscous stability constraint. The computational domain in both cases, $[-6R,6R] \times [-6R,6R]$, is relatively more compact than required by other numerical methods due to the use of the lattice Green's function and associated viscous integrating factor \citep{liska2017fast}.

\subsection{Small Stokes expansion of inertial particle velocity field}
\label{validation_asymptotic}

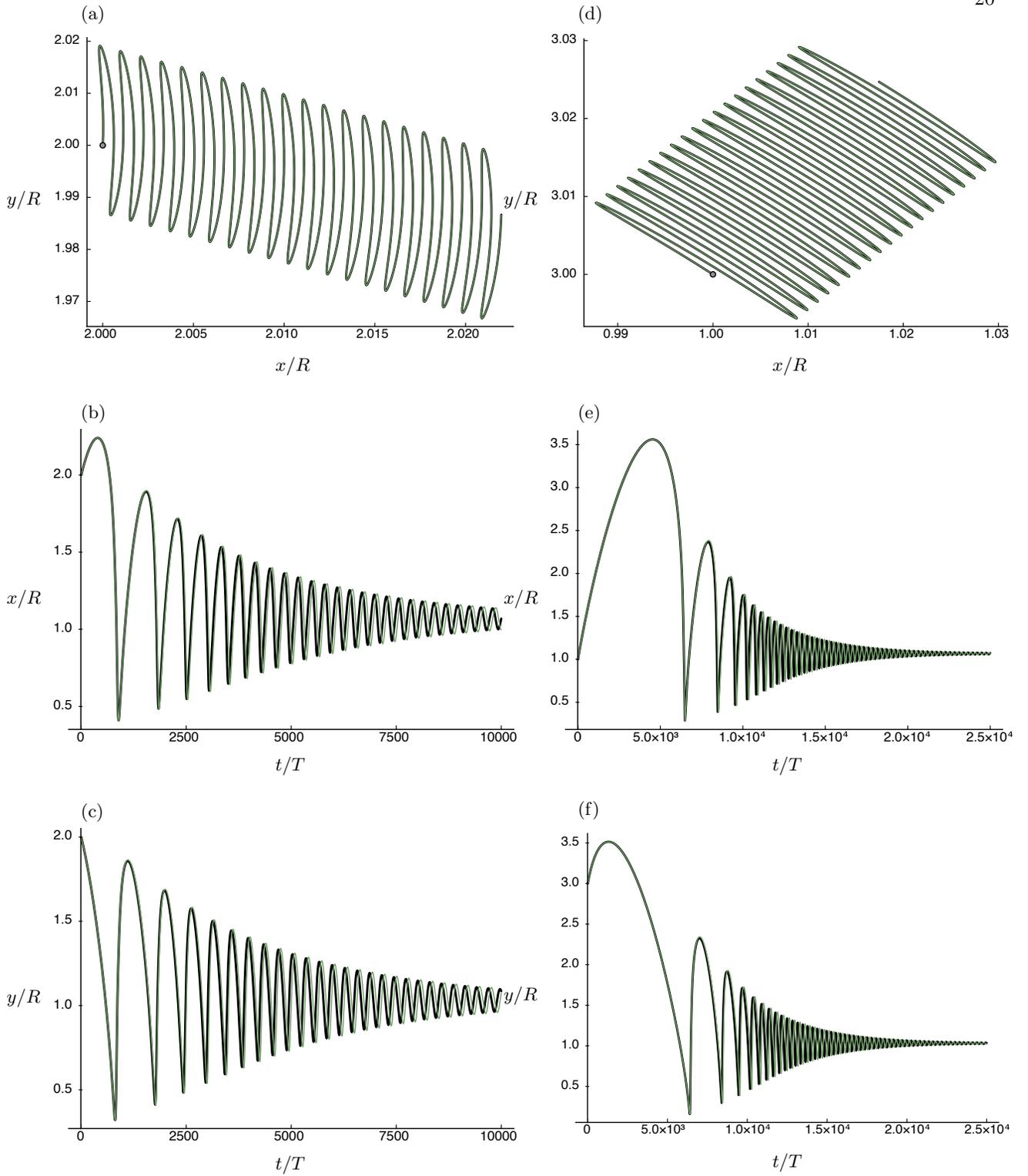
\begin{figure}[tp]
    \centering
    \begin{minipage}[h!]{0.45\textwidth}
    	\centering
        \begin{overpic}[width = 8cm]{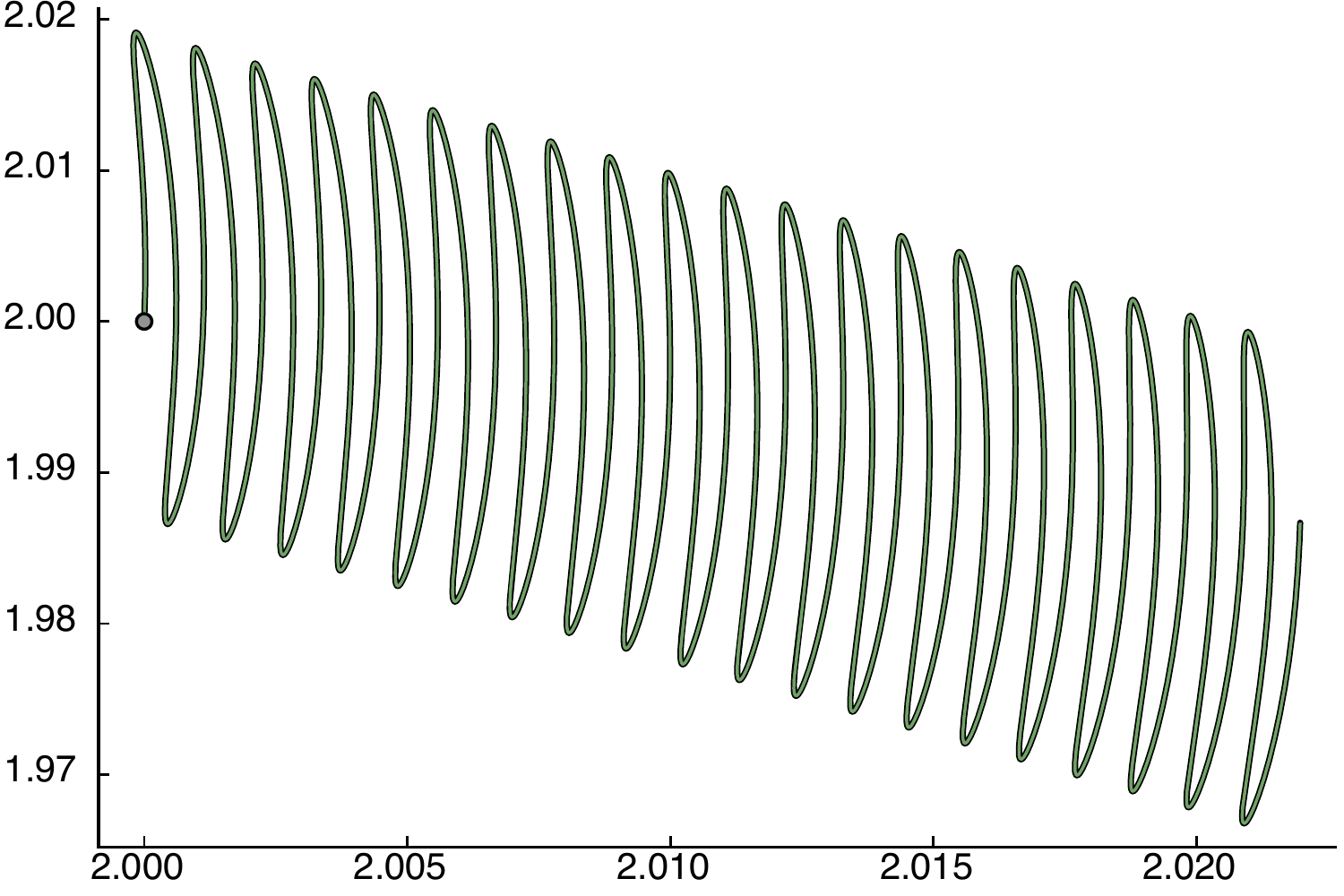}
        \put(48,-6){$x/R$}
        \put(-10,30){$y/R$}
        \put(6,70){(a)}
        \end{overpic}

    \end{minipage}%
    ~
    \hspace{0.05\textwidth}
    \begin{minipage}[h!]{0.45\textwidth}
        \centering
            \begin{overpic}[width = 8cm]{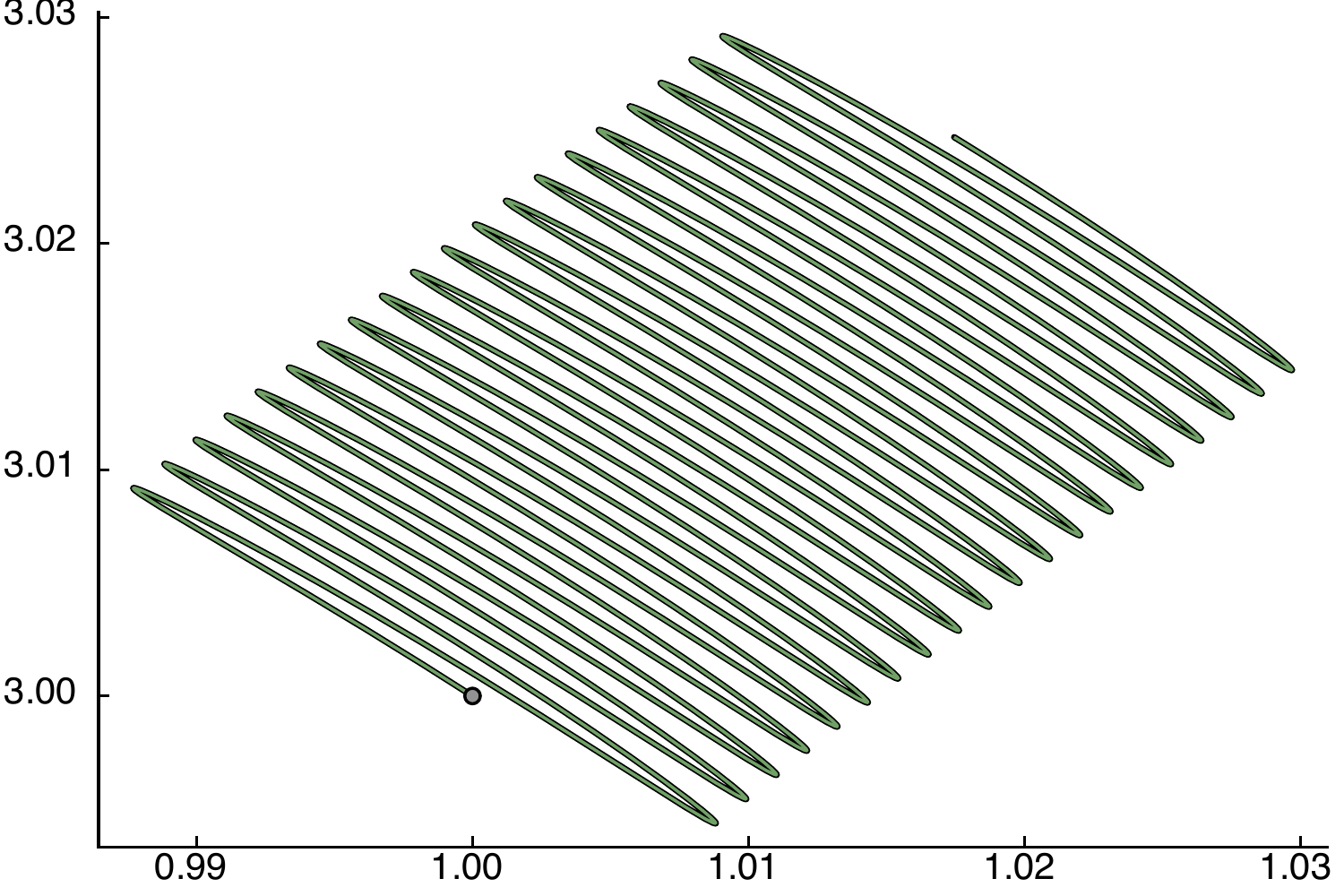}
            \put(48,-6){$x/R$}
            \put(-10,30){$y/R$}
            \put(6,70){(d)}
            \end{overpic}
    \end{minipage}%

    \par\vspace{15mm}
    
    \begin{minipage}[h!]{0.45\textwidth}
    \centering
    \begin{overpic}[width = 8cm]{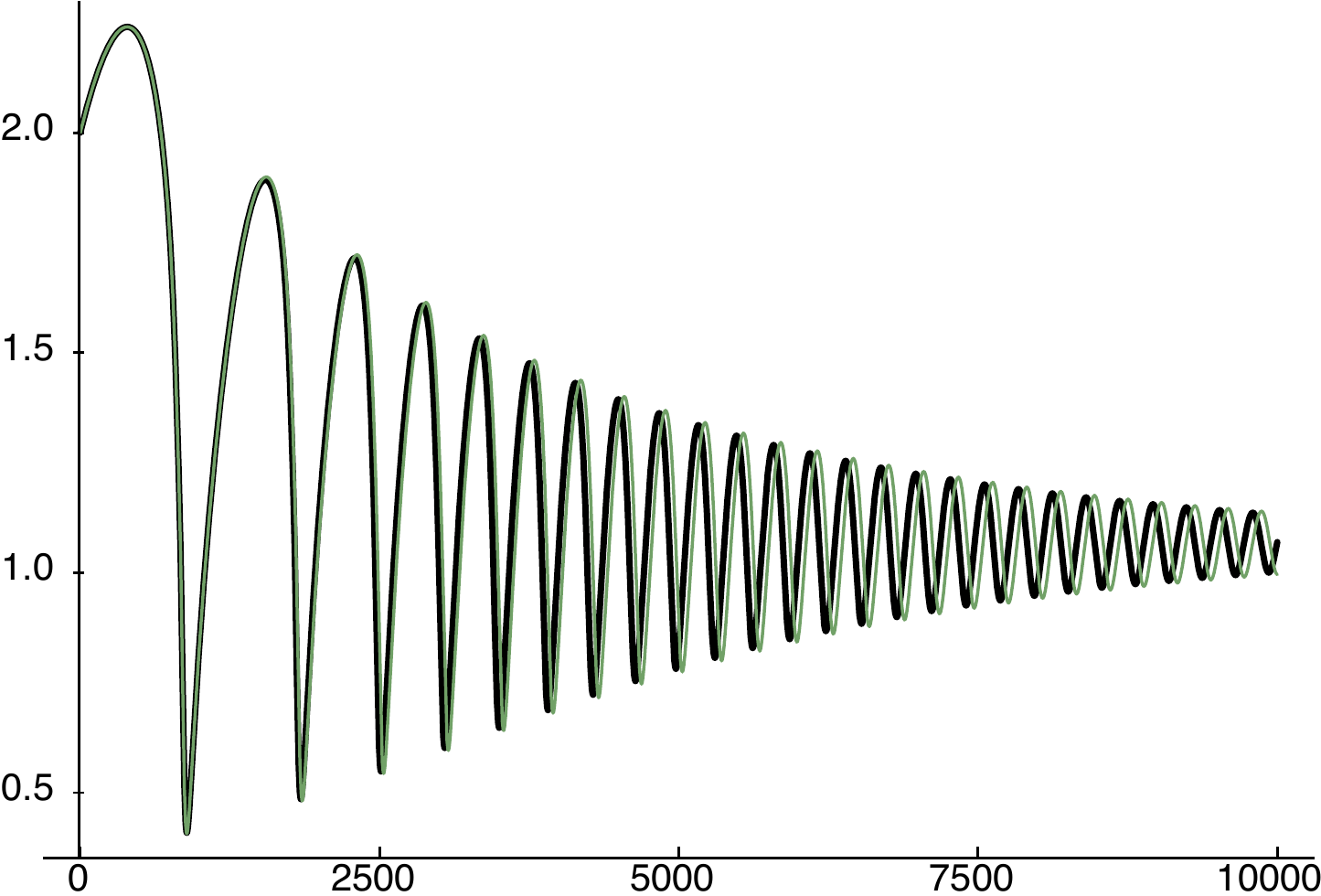}
    \put(48,-6){$t/T$}
    \put(-10,30){$x/R$}
    \put(6,70){(b)}
    \end{overpic}
    \end{minipage}%
    ~
    \hspace{0.05\textwidth}
    \begin{minipage}[h!]{0.45\textwidth}
    \centering
    \begin{overpic}[width = 8cm]{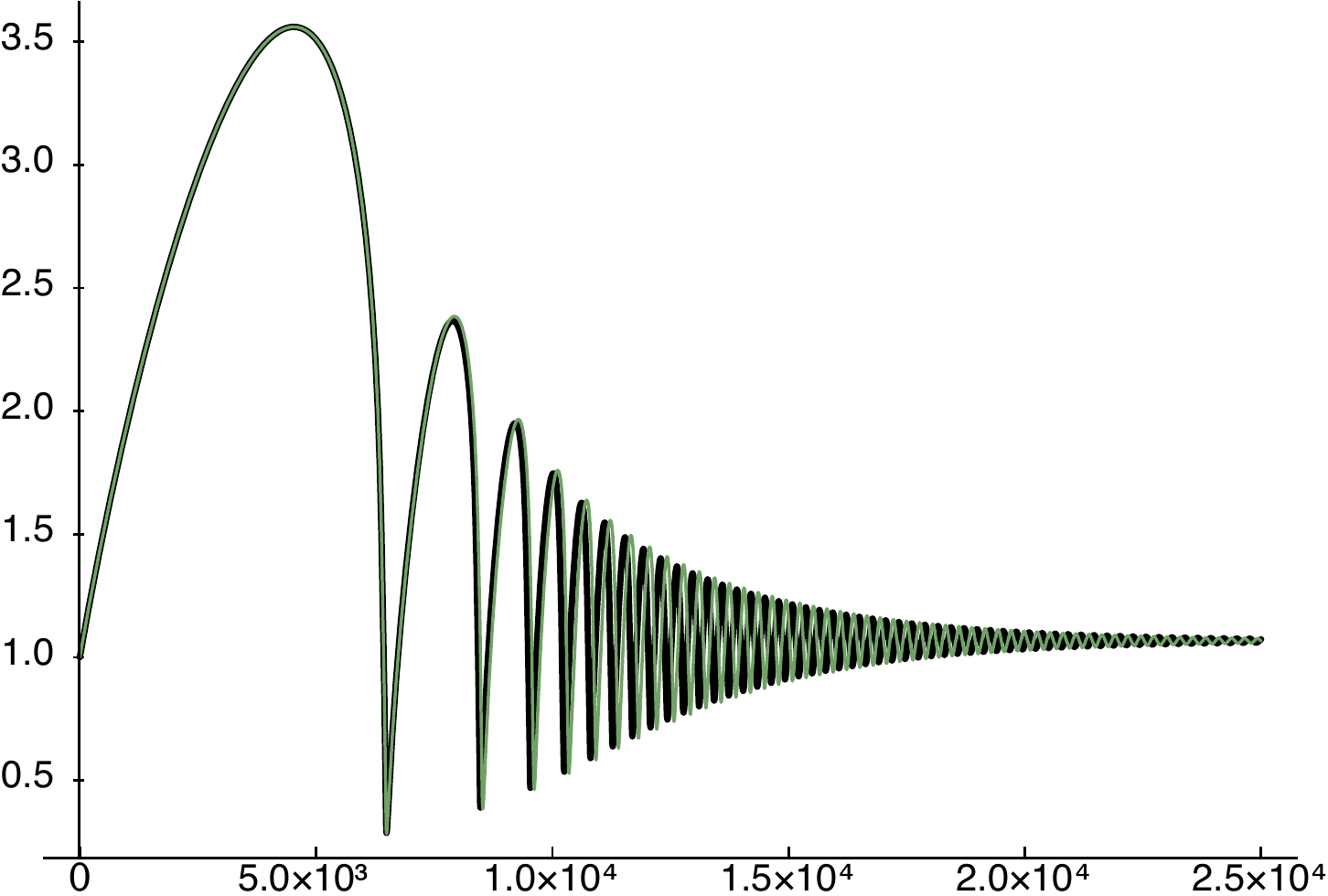}
    \put(48,-6){$t/T$}
    \put(-10,30){$x/R$}
    \put(6,70){(e)}
    \end{overpic}
    \end{minipage}
    
        \par\vspace{15mm}
    \begin{minipage}[h!]{0.45\textwidth}
    \centering
    \begin{overpic}[width = 8cm]{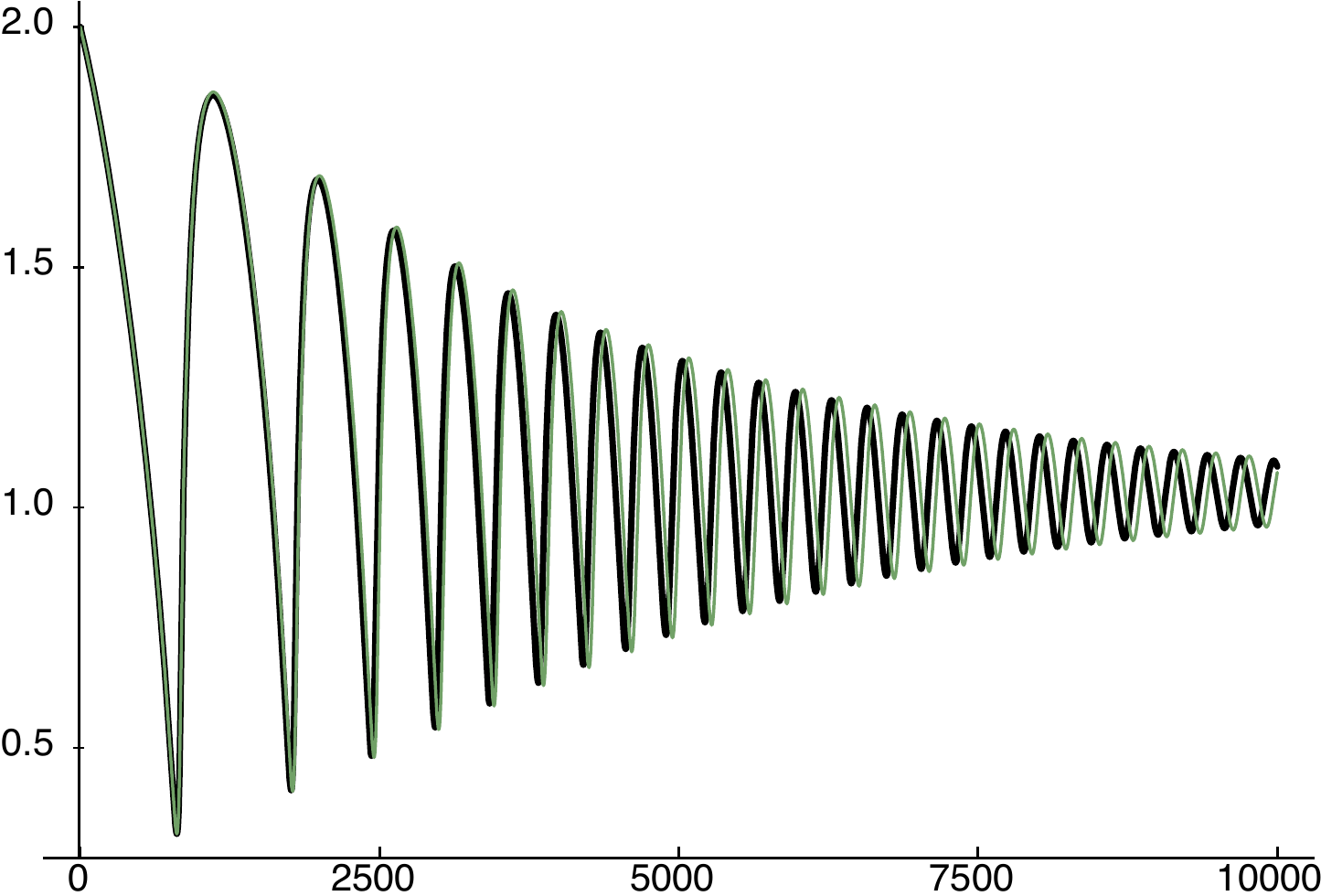}
    \put(48,-6){$t/T$}
    \put(-10,30){$y/R$}
    \put(6,70){(c)}
    \end{overpic}
    \end{minipage}%
    ~
    \hspace{0.05\textwidth}
    \begin{minipage}[h!]{0.45\textwidth}
    \centering
    \begin{overpic}[width = 8cm]{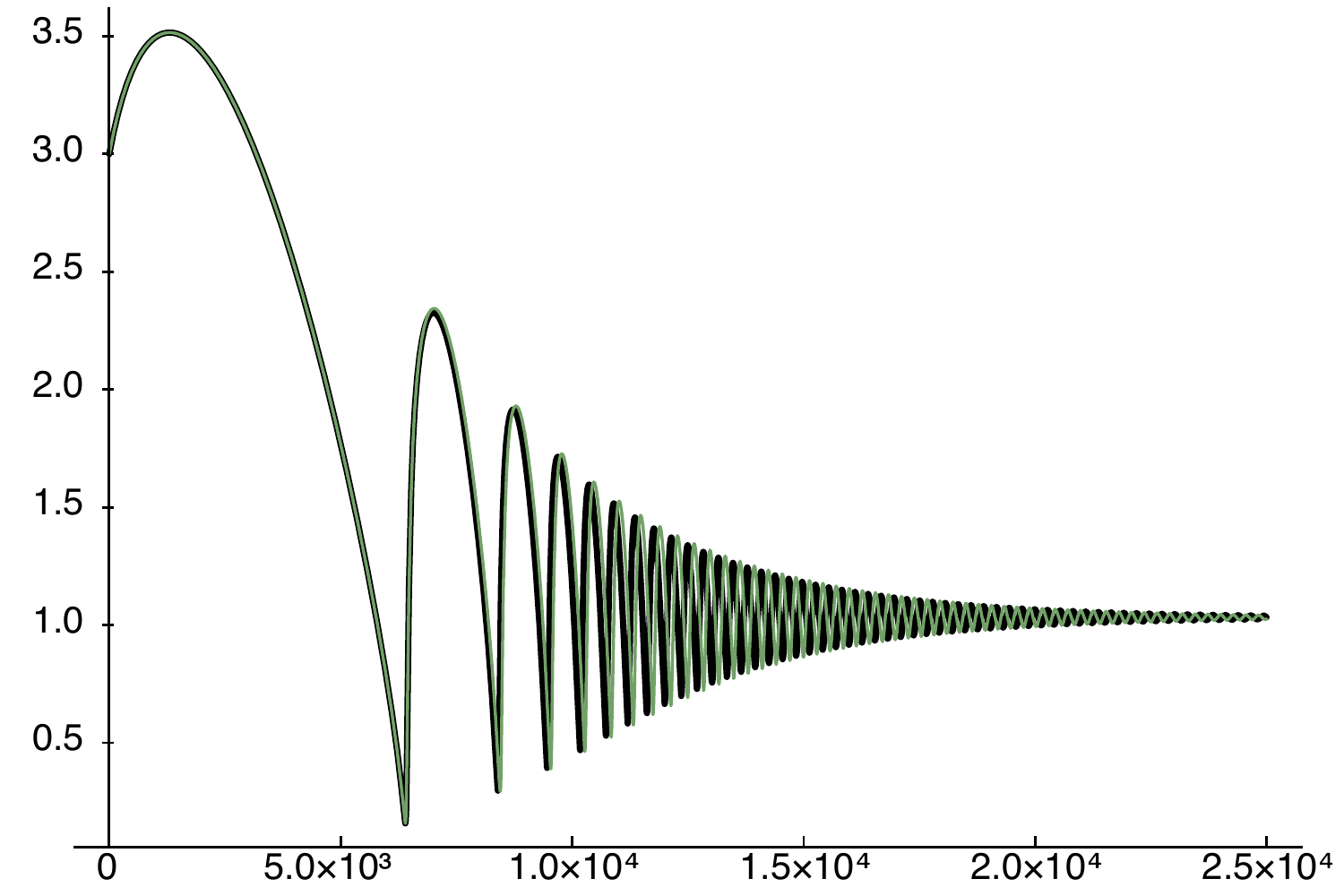}
    \put(48,-6){$t/T$}
    \put(-10,30){$y/R$}
    \put(6,70){(f)}
    \end{overpic}
    \end{minipage}
        \par\vspace{10mm}
	\caption{Time-resolved inertial particle trajectories from the Maxey--Riley equation (black) and the asymptotic expansion in small Stokes number (green) for particles initially located at $\plabel = (2,2)$ (left column) and $\plabel = (1,3)$ (right column). The top row (a,d) shows the trajectories over the first $20$ oscillation periods, and the middle (b,e) and lower (c,f) rows depict the time histories of the $x$ and $y$ components, respectively, sampled once per period.}
    \label{fig:MR_Asymptotic}
\end{figure}

In this section, we assess the accuracy of the asymptotic expansion of the inertial particle velocity field in small Stokes number, developed in equation~(\ref{inertialv}). For evaluation purposes, we compare the trajectories of inertial particles transported by this velocity field, with the trajectories of the same particles with velocity obtained from the full Maxey--Riley equation~(\ref{maxeyriley}). It should be noted that we are not yet assessing the mean transport algorithm in this section, so our comparison is made of the full time-resolved trajectory, computed from (\ref{fulltransport}) for both velocities. As discussed earlier in the paper, the Basset term is neglected in both forms of velocity. For the numerical integration of these trajectories, we use a fifth order Adams--Bashforth method with time step $\Omega \Delta t/(2\pi)  = 0.004$.

For the case of a single oscillating cylinder, we simulate two trajectories: one for a particle starting from $\plabel = (2,2)$ and tracked for $10000$ periods of oscillation, and another for a particle released from $\plabel = (1,3)$ and tracked for $25000$ periods. Both particles are released after the flow has reached its stationary periodic state. The comparisons of these trajectories are shown in Figure~\ref{fig:MR_Asymptotic}. The small portions of the full trajectories shown in the top row exhibit the characteristic fluctuations of these trajectories about a mean. To reveal this mean behavior more clearly, we sample these trajectories only once per cycle in the middle and lower rows, with the history of each component depicted in a separate plot. These plots exhibit the trapping behavior: both particles converge toward a fixed point inside the streaming cell along the 45-degree ray.

The plots in Figure~\ref{fig:MR_Asymptotic} show that the asymptotic expansion in small Stokes number has very accurately preserved the behavior of the Maxey--Riley equation. Though small errors accumulate over time, the trajectories apparently agree well even after $25000$ periods. Table~\ref{tab:1cylinder_asymptotic} reports a quantitative measure of this comparison, with error defined as the difference of the asymptotically-approximated trajectory components from those of the Maxey--Riley trajectory at the same instant, normalized by the current radial distance from. The error remains small throughout, and the final trapping location is predicted with less than one percent error.

\begin{table}[tb]
\centering
 \begin{tabular}{|c | c | c | c | } 
 \hline 
 
$\plabel$ & $ \mbox{error after } 1 \mbox{ period} (\%) $ & $t_f/T$ &  $ \mbox{error at } t_f(\%) $ \\\hline
$(2, 2)$ & $ (-1.19 \times 10^{-5}, 8.77 \times 10^{-5} ) $  & $10000$ & $ ( -6.91, -1.11 ) $\\
$ (1, 3)$  & $ (1.29 \times 10^{-4}, 2.04 \times10^{-5})$  &  $25000$ & $  (-0.693, -0.736) $ \\
\hline
\end{tabular}
\caption{Relative errors on particle position, after $1$ period and at the final time $t_f$, in two different inertial particles trajectories predicted by the small Stokes number expansion.}
\label{tab:1cylinder_asymptotic}
\end{table} 

\begin{figure}[t]
    \centering
        \begin{overpic}[width = 10cm]{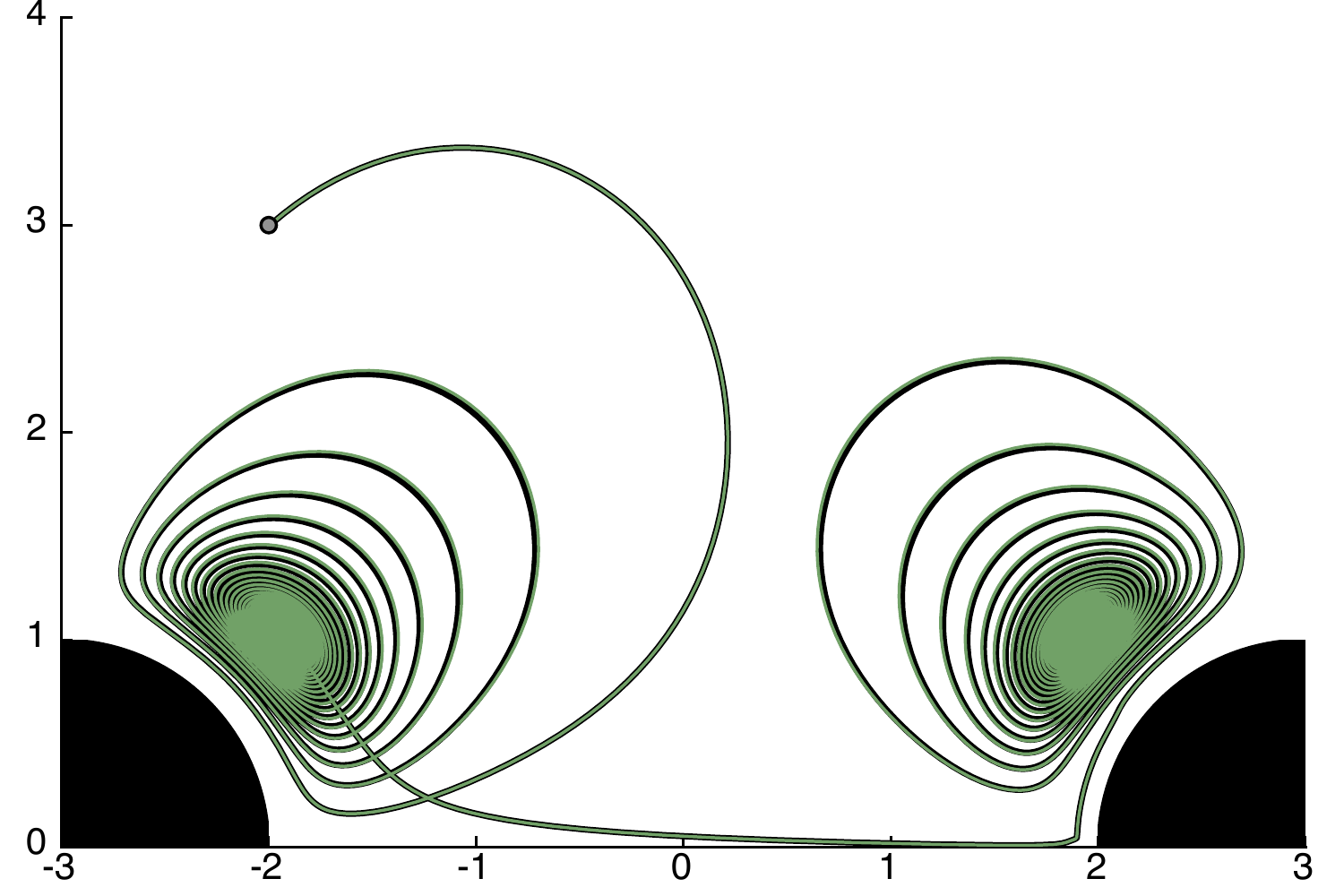}
        \put(48,-2){$x/R$}
        \put(-8,33){$y/R$}
        \end{overpic}
    \par\vspace{6mm}
\caption{Trajectory of an inertial particle initially located at $\plabel = (-2,3)$ over $65000$ periods of oscillation. Trajectories from the Maxey--Riley equation (black) and small Stokes number expansion (green) are both sampled once per period.}
\label{fig:MR_Asymptotic_2cylinders}
\end{figure}

Now, let us validate our small Stokes number expansion on the transport of inertial particles in the two-cylinder case. Here, we release particles just after the initiation of motion of the left cylinder. This case is potentially more challenging due to the transient behavior of the flow after each oscillator's motion is initiated. The results in Figure~\ref{fig:MR_Asymptotic_2cylinders}, which depicts the full trajectory sampled once per period for a particle released from $(-2,3)$, show that the particle is first trapped near the center of a streaming cell near the left cylinder at $(x,y) = (-1.98,1.03)$; and after the right cylinder starts its own motion, the particle is eventually drawn to a new trapping location at $(x,y) = (1.98,1.03)$. It should be noted that this problem requires the no-penetration constraint described in \eqn~(\ref{nopen}) when the inertial particle approaches the right oscillator. As observed in Figure~\ref{fig:MR_Asymptotic_2cylinders}, the particle is drawn toward this oscillator along the axis of symmetry. Without the explicit enforcement of this constraint the particle would spuriously pass across the oscillator surface. Instead, the particle remains offset from the oscillator by a small distance set by the smoothing parameter $\delta$ in this constraint and is quickly drawn into an orbit that spirals toward the trapping point. Throughout this sequence, the asymptotically-approximated trajectory agrees well with the Maxey--Riley trajectory, with error less than $0.01$ percent after the first transient phase. The final percentage error in the trapping location, after 65000 cycles, is $(1.76\times 10^{-2}, 1.19 \times 10^{-2})$.


\subsection{Mean fluid particle trajectories} 
\label{validation_xi_fluid}

 \begin{figure}[t]
 \centering
  \begin{overpic}[width = 0.55\textwidth]{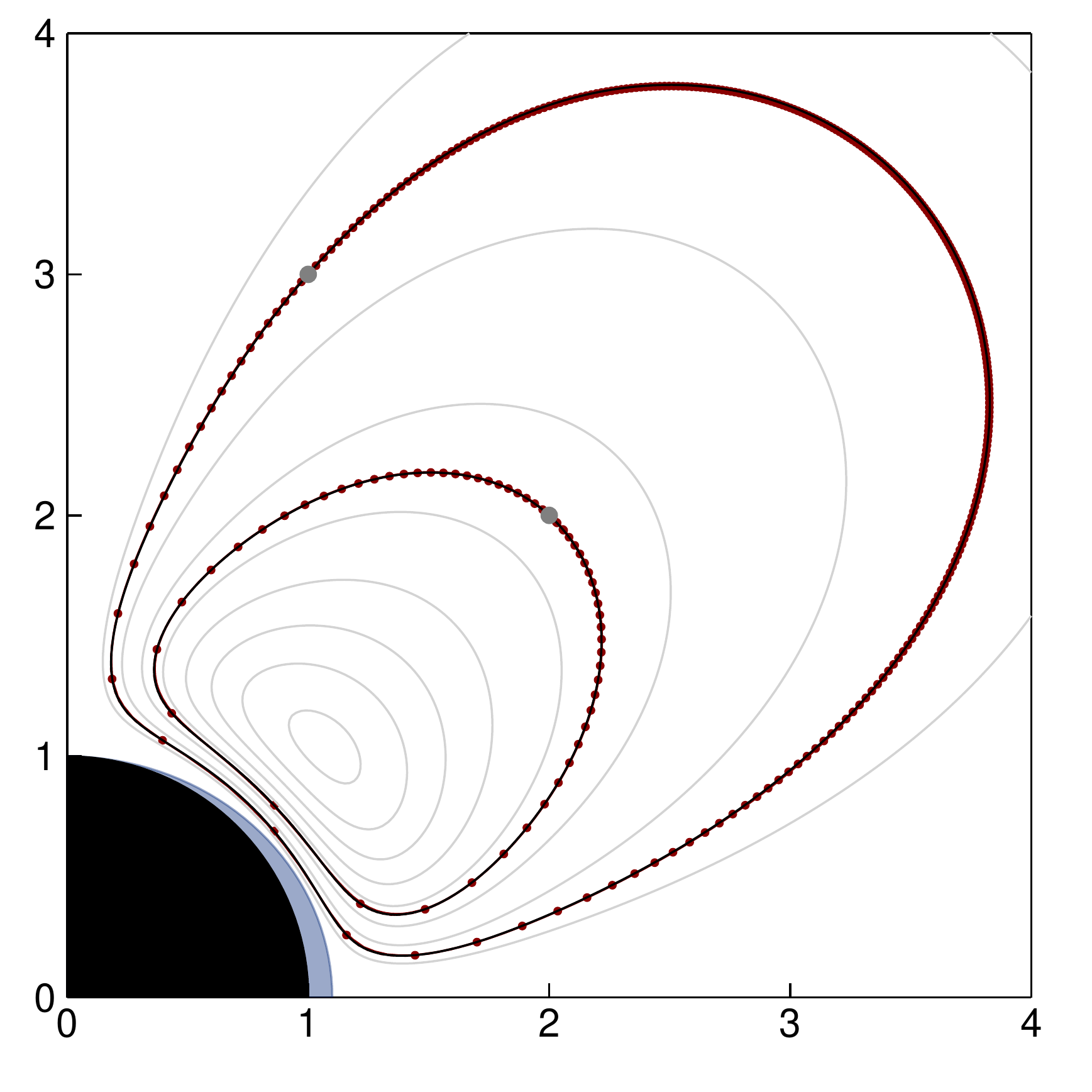}
            \put(48,0){$x/R$}
            \put(-10,48){$y/R$}
            \end{overpic}
     \caption{Lagrangian mean trajectories (magenta circles, spaced by 30 periods) generated for fluid particles started at $\plabel = (2,2)$ and $\plabel = (1,3)$ (denoted by the larger grey circle), compared with the Lagrangian streamlines, depicted as black lines. (Other Lagrangian streamlines are shown in light grey.) The black region shows the mean position of the cylinder, and the lighter shaded region in the vicinity of the cylinder shows the range of displacement of the cylinder over one oscillation cycle.}
\label{fig:Fluid_avg}
 \end{figure}

In this section, we verify our algorithm for computing mean particle trajectories by applying it to fluid particles. In the first case, we compute trajectories from the fluid velocity field generated by the single oscillating cylinder after it has achieved periodic behavior. As we discussed in Section~\ref{asympmeanv}, the mean trajectories of fluid particles are equivalently derived from the contours of a Lagrangian streamfunction field, $\lagrange{\stream}$, in \eqn~(\ref{streamL}). This alternative approach provides a natural target for verification. Examples of this comparison are shown in Figure~\ref{fig:Fluid_avg} for two different particles, both of whose mean trajectories have been integrated with a time step size of $10$ periods. The agreement is very good, and importantly, the trajectories generated by the algorithm are closed after each orbit to within small numerical error.

\begin{figure}[tp]
 \centering
    \begin{minipage}[h!]{0.45\textwidth}
        \centering
            \begin{overpic}[width = \textwidth]{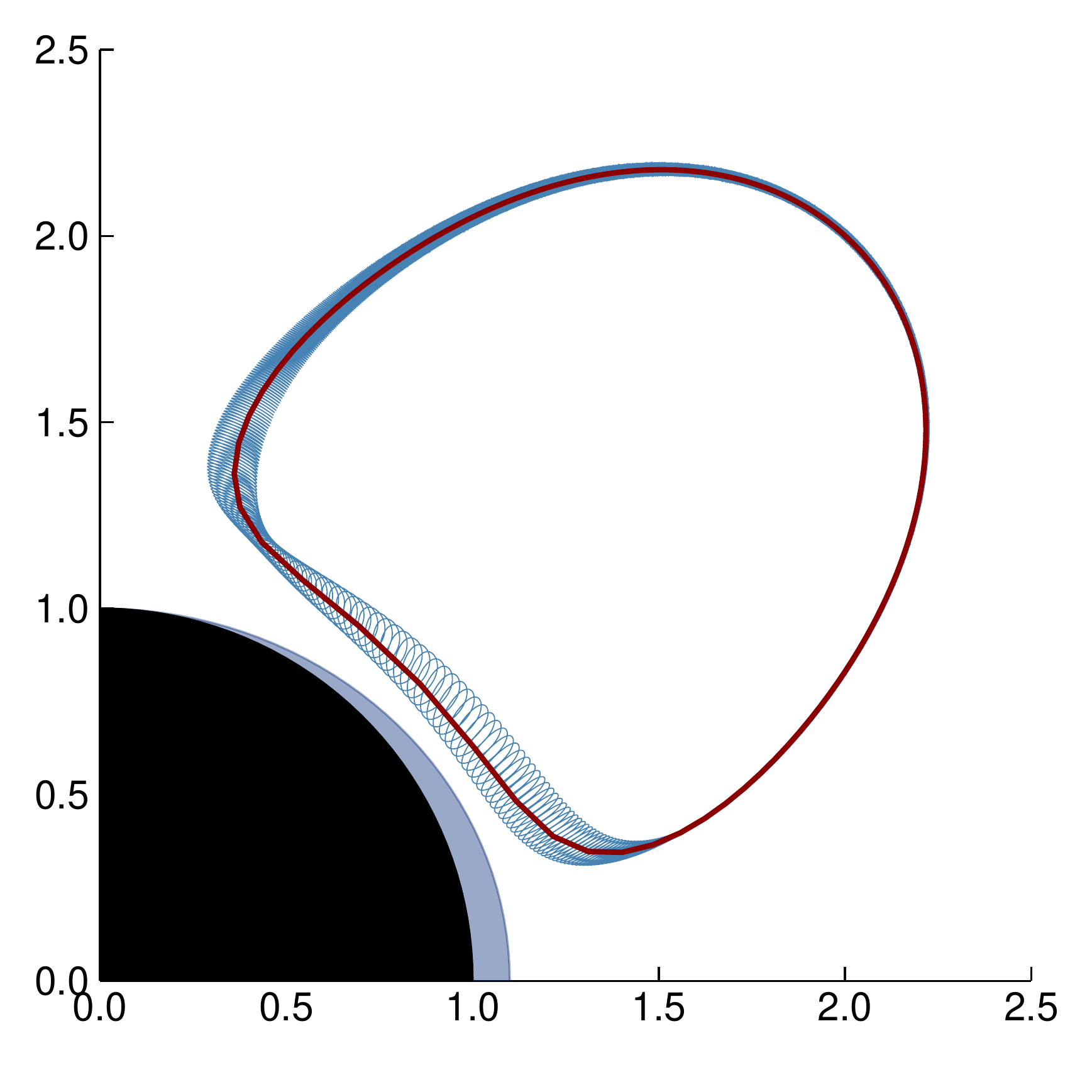}
            \put(48,0){$x/R$}
            \put(-10,48){$y/R$}
            \put(15,90){(a)}
            \end{overpic}
    \end{minipage}%
      \hspace{0.05\textwidth}
     \begin{minipage}[h!]{0.34\textwidth}
 \begin{overpic}[width = \textwidth]{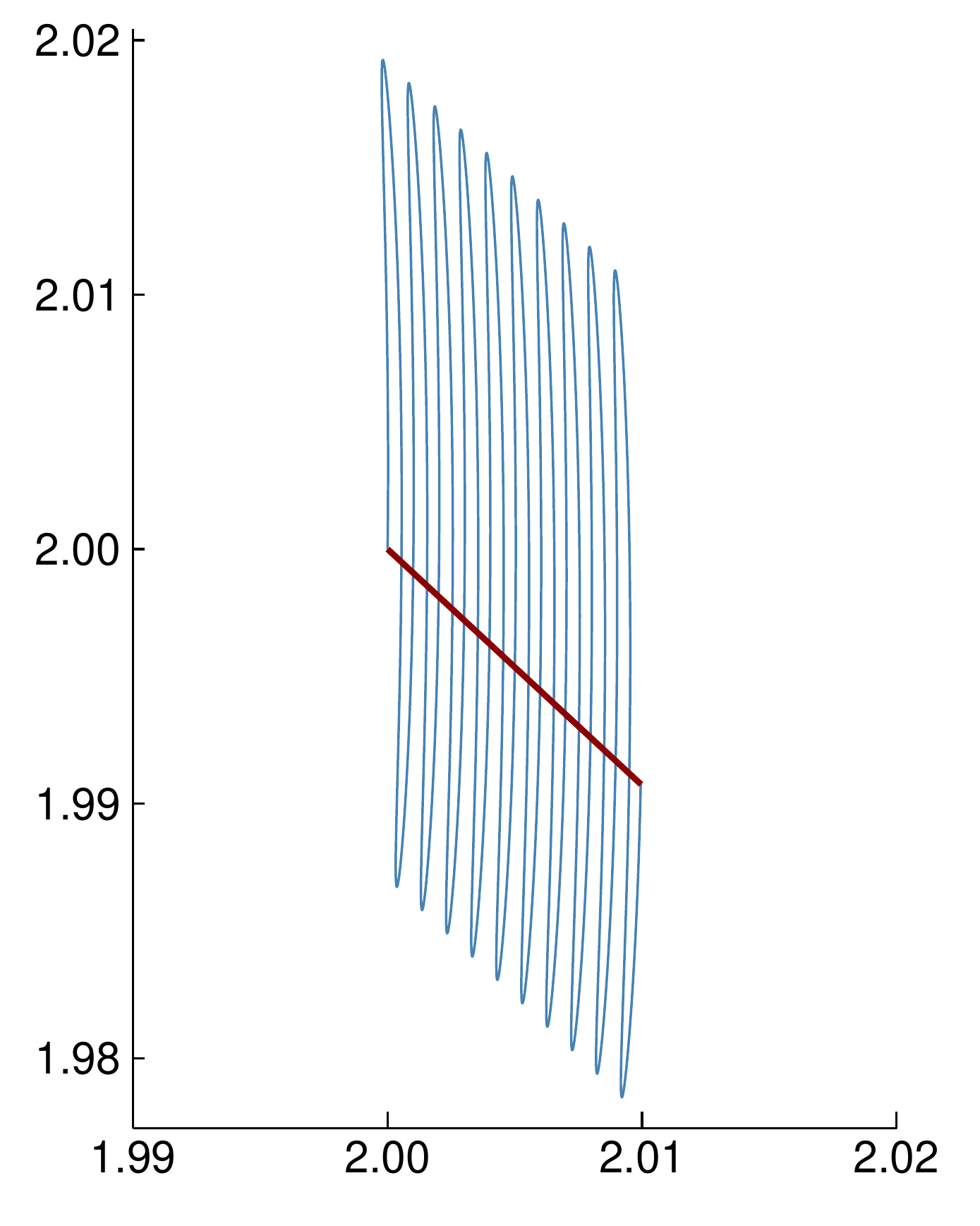}
            \put(38,-2){$x/R$}
            \put(-10,48){$y/R$}
            \put(15,90){(b)}
            \end{overpic}
  \end{minipage}%
  \par\vspace{0.1cm}
   \begin{minipage}[h!]{0.45\textwidth}
        \centering
            \begin{overpic}[width = \textwidth]{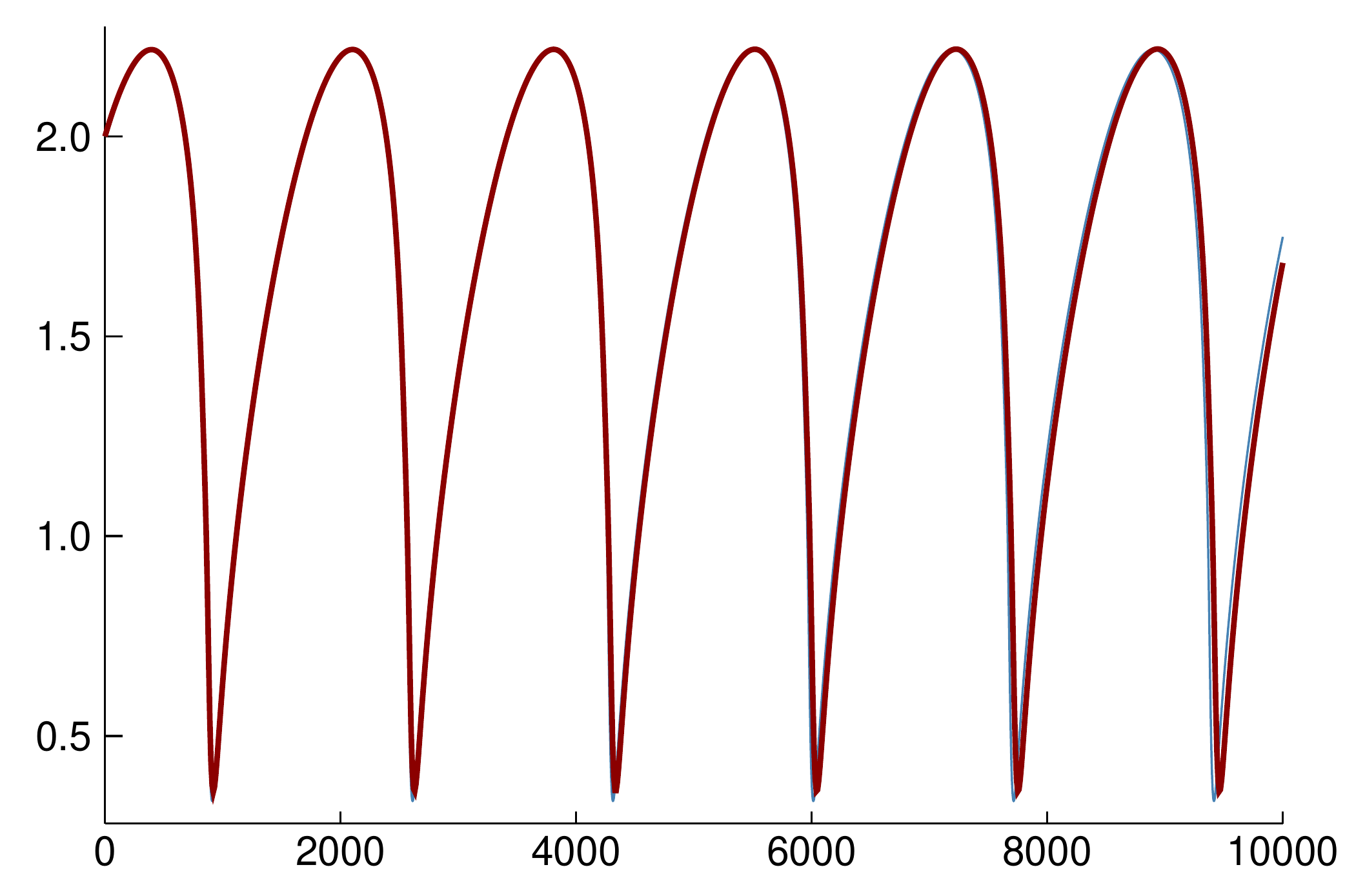}
            \put(48,-2){$t/T$}
            \put(-8,35){$x/R$}
            \put(10,65){(c)}
            \end{overpic}
    \end{minipage}%
\hspace{0.05\textwidth}
        \begin{minipage}[h!]{0.45\textwidth}
    	\centering
        \begin{overpic}[width= \textwidth]{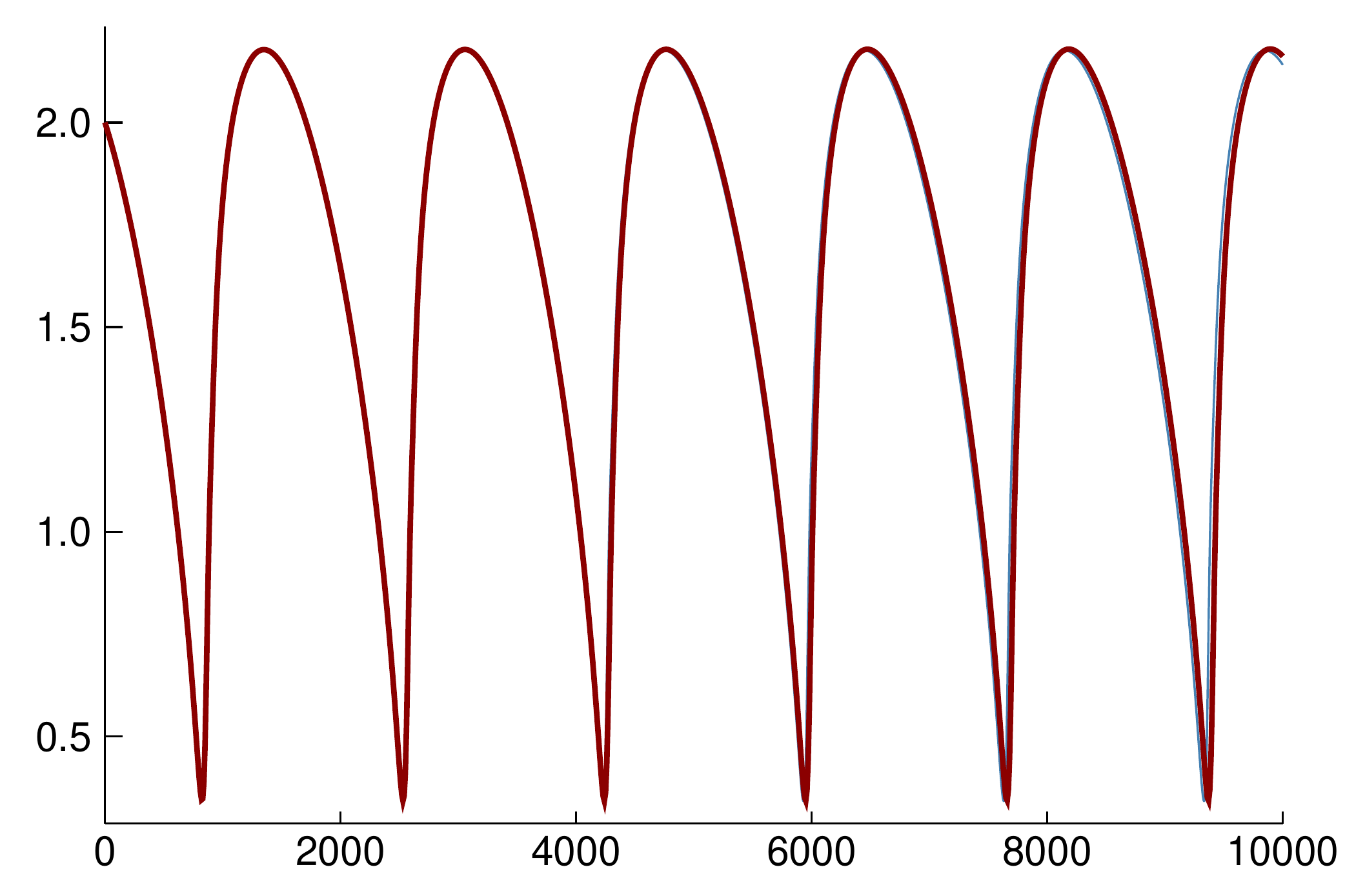}
            \put(48,-2){$t/T$}
            \put(-8,35){$y/R$}
            \put(10,65){(d)}
        \end{overpic}
    \end{minipage}%
    \caption{(a) Lagrangian mean trajectory (magenta) for fluid particle started at $\plabel = (2,2)$, compared with the full time-resolved trajectory (blue) for the same particle. (b) Magnified view of fluid particle trajectory. (c,d) Comparison of $x$ and $y$ components, respectively, of the Lagrangian mean trajectory and the full time-resolved trajectory (sampled once per cycle).}
\label{fig:Fluid_avg_transient}
 \end{figure}

In Figure~\ref{fig:Fluid_avg_transient} we compare the mean trajectory of the particle $\plabel = (2,2)$ with the full time-resolved trajectory. This latter trajectory is obtained from the same (numerically-computed) fluid velocity field, but by integrating the unsteady fluid velocity with 250 time steps per period and cubic spline interpolation of the Cartesian grid values. The full trajectory reveals the oscillations incurred by the particle as it orbits about the streaming cell. The mean trajectory from the algorithm displays the expected behavior, passing through the first point in each cycle, as shown in the small section of trajectory in Figure~\ref{fig:Fluid_avg_transient}(b). Indeed, when the full trajectory is sampled once per cycle (starting with its initial position), as shown in Figure~\ref{fig:Fluid_avg_transient}(c,d), the mean trajectory agrees well with it even after 10000 oscillation periods, corresponding to nearly 6 orbits.

\begin{figure}[tbp]
    \centering
    \begin{minipage}[h!]{0.49\textwidth}
    	\centering
        \begin{overpic}[width = 7cm]{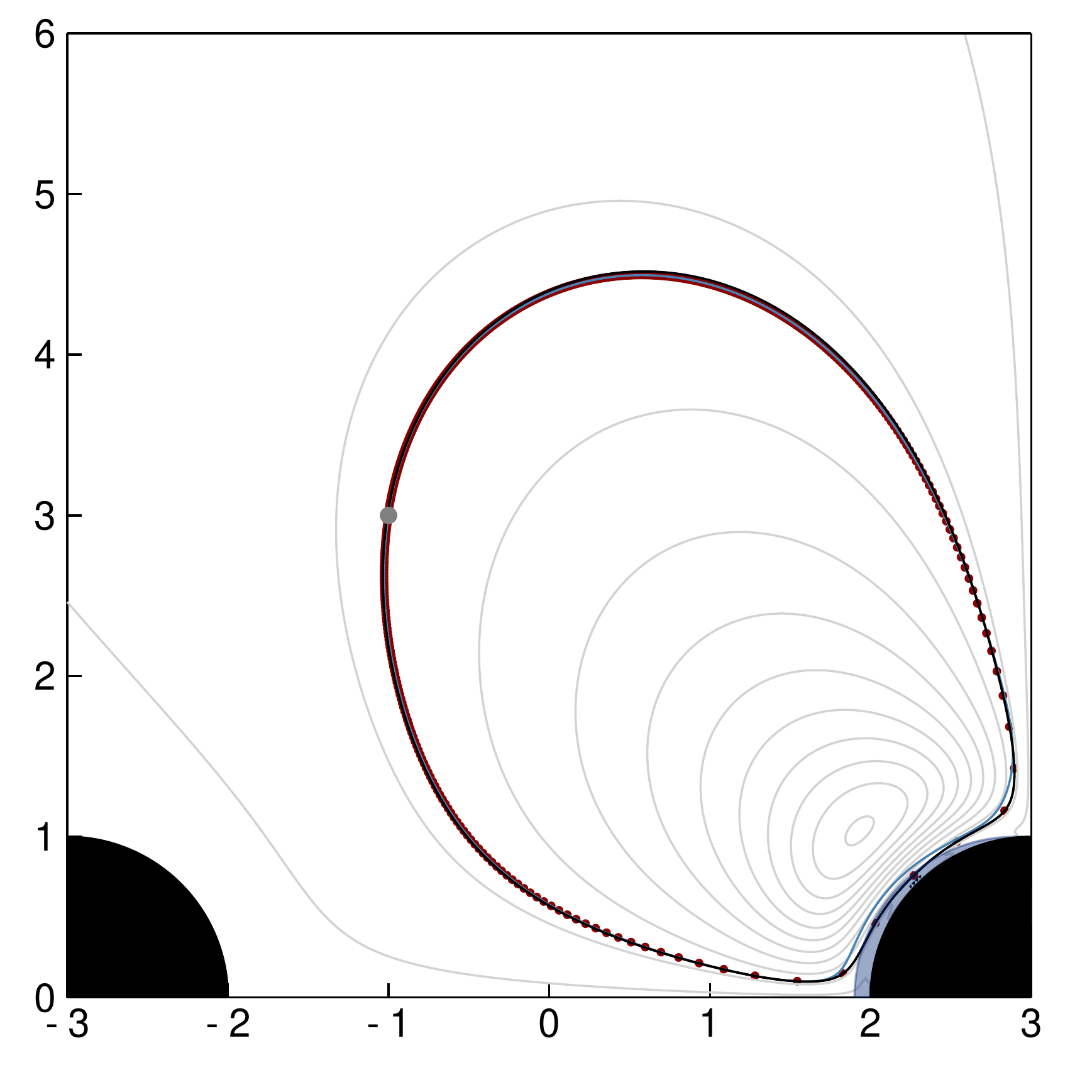}
        \put(43,-2){$x/R$}
        \put(-10,50){$y/R$}
        \end{overpic}
            \end{minipage}%
            ~
            \begin{minipage}[h!]{0.49\textwidth}
        \centering
            \begin{overpic}[width = 7cm]{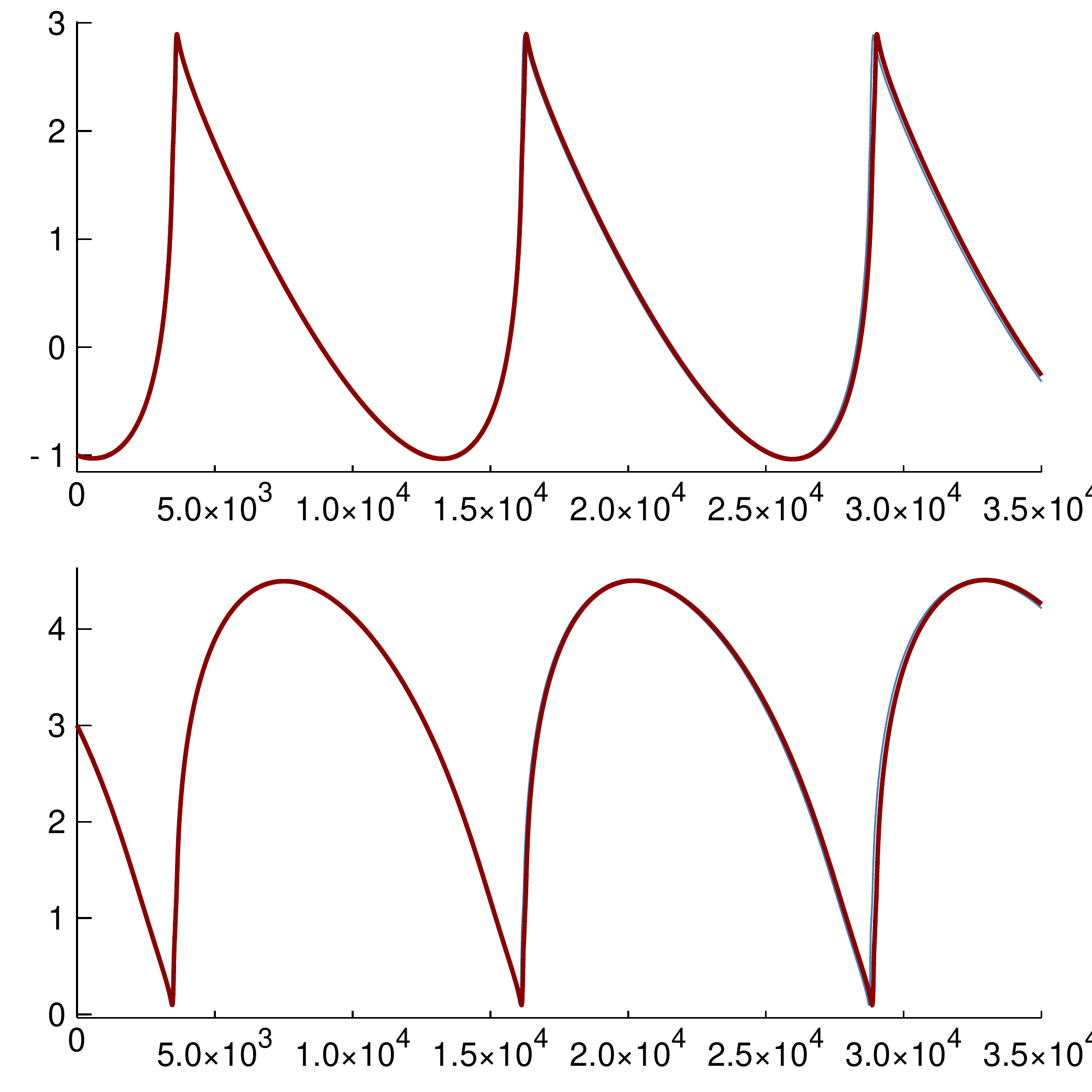}
            \put(48,-6){$t/T$}
            \put(-10,25){$y/R$}
            \put(-10,76){$x/R$}
            \end{overpic}
    \end{minipage}%
    \vspace{6.0mm} 
\caption{Left: Time resolved fluid particle trajectory, sampled once per cycle (blue), and Lagrangian mean trajectory (magenta circles, spaced by 30 periods), and Lagrangian mean streamlines (in light gray) for $\plabel = (-1, 3)$, with mean cylinder configurations depicted in black. Right: Time histories of the $x$ and $y$ components of both trajectories shown in the left panel.}
\label{fig:2cylinder_fluid_hybrid}
\end{figure}

In the second case, we use the configuration of two cylinders. As a target of comparison, we release a particle from $\plabel = (-1,3)$, near the left cylinder, after statistically stationary behavior has been achieved from the right cylinder's motion. The comparison with the full time-resolved trajectory is shown in Figure~\ref{fig:2cylinder_fluid_hybrid} and exhibits very good agreement. The particle is initially drawn toward the right cylinder and achieves a closed orbit about the streaming cell; each orbit requires approximately 13000 periods of oscillation.

%


\subsection{Mean inertial particle trajectories}
\label{validation_xi_inertial}

The previous section demonstrated that our proposed algorithm can successfully predict the mean trajectories of fluid particles. In this section, we apply the algorithm to inertial particle transport. Each trajectory computed from this algorithm is compared with the full time-resolved trajectory obtained from the same inertial particle velocity field, $\inertv$, derived from the fluid velocity field after it has reached a periodic state. Thus, the differences between these trajectories is due entirely to errors in truncating the asymptotic expansion \newstuff{in $\epsilon$} in the construction of the Lagrangian mean velocity in (\ref{flucteqnsimple})--(\ref{meanv1}).

\begin{figure}[t]
 \centering
 \begin{minipage}[h!]{0.45\textwidth}
        \centering
  \begin{overpic}[width = \textwidth]{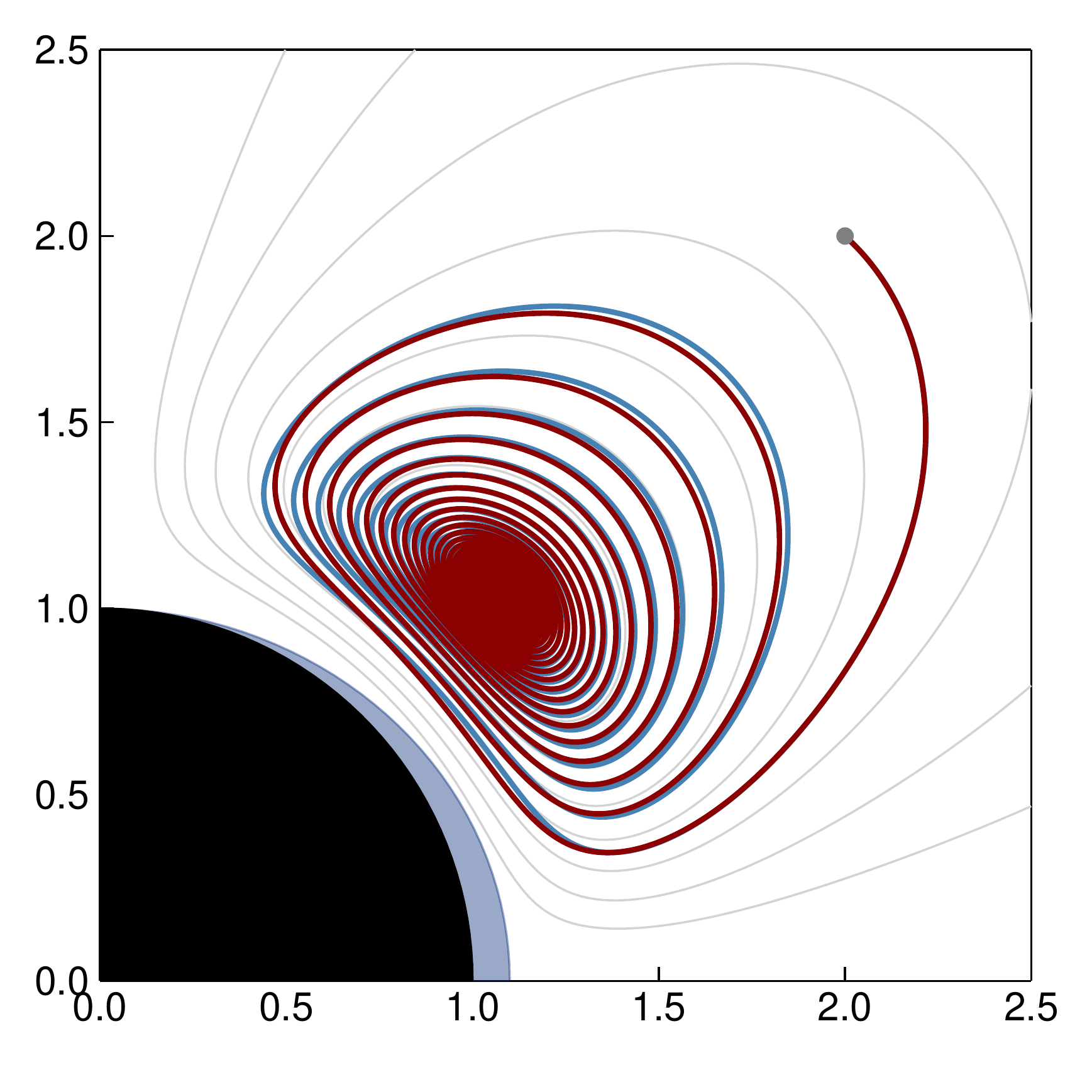}
            \put(48,0){$x/R$}
            \put(-10,48){$y/R$}
            \put(15,90){(a)}
            \end{overpic}
            \end{minipage}
            \hspace{0.05\textwidth}
            \begin{minipage}[h!]{0.45\textwidth}
        \centering
  \begin{overpic}[width = \textwidth]{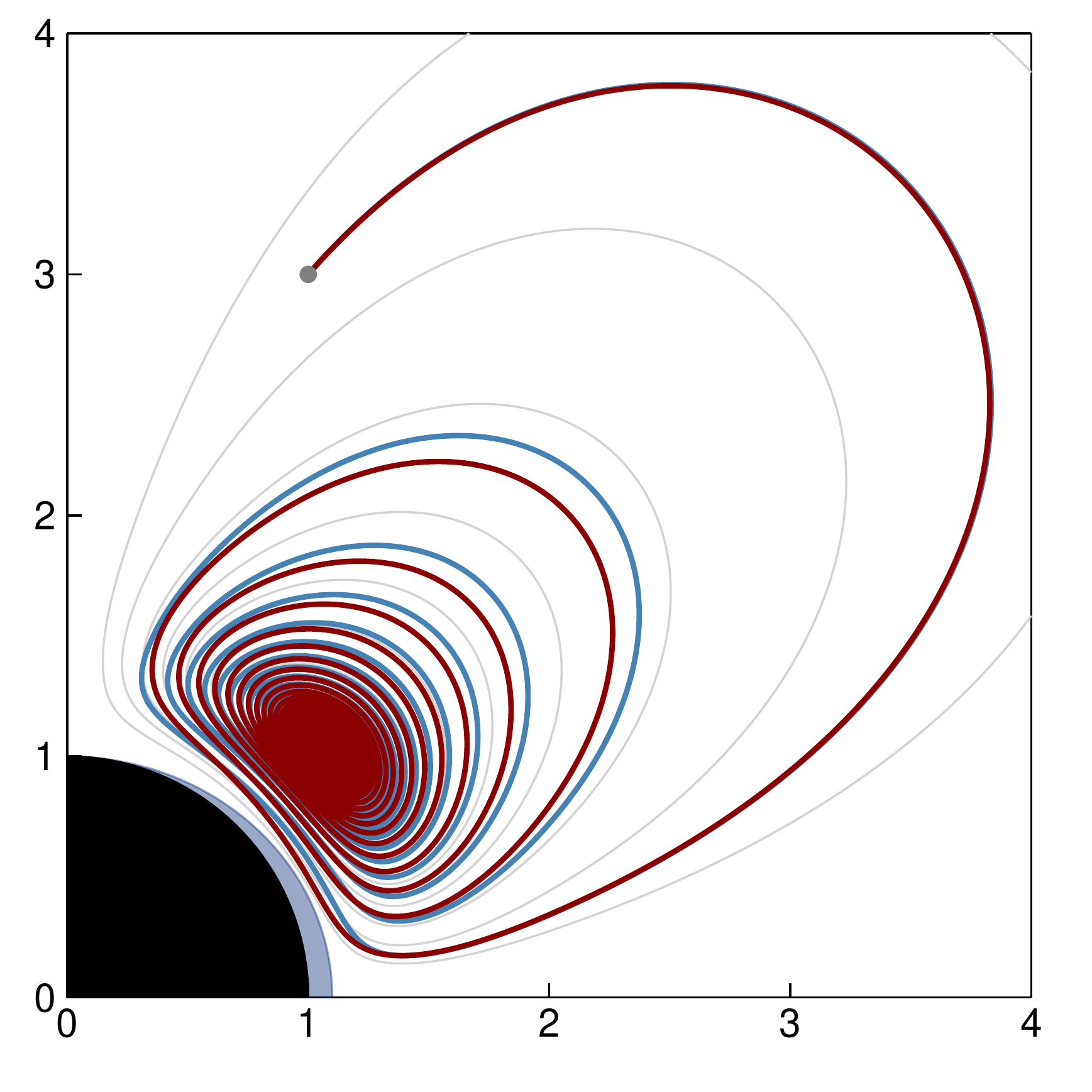}
            \put(48,0){$x/R$}
            \put(-10,48){$y/R$}
            \put(10,90){(b)}
            \end{overpic}
            \end{minipage}
     \caption{Lagrangian mean trajectories (magenta) generated for inertial particles started at (a) $\plabel = (2,2)$ and (b) $\plabel = (1,3)$ (denoted by the small circle), compared with the full time-resolved trajectories (blue) of the same particles, sampled once per cycle. The Lagrangian streamlines of the fluid are shown for reference as light grey lines. The black region shows the mean position of the cylinder, and the lighter shaded region in the vicinity of the cylinder shows the range of displacement of the cylinder over one oscillation cycle.}
\label{fig:Inertial_avg}
 \end{figure}

\begin{figure}[tp]
 \centering
   \begin{minipage}[h!]{0.45\textwidth}
        \centering
            \begin{overpic}[width = \textwidth]{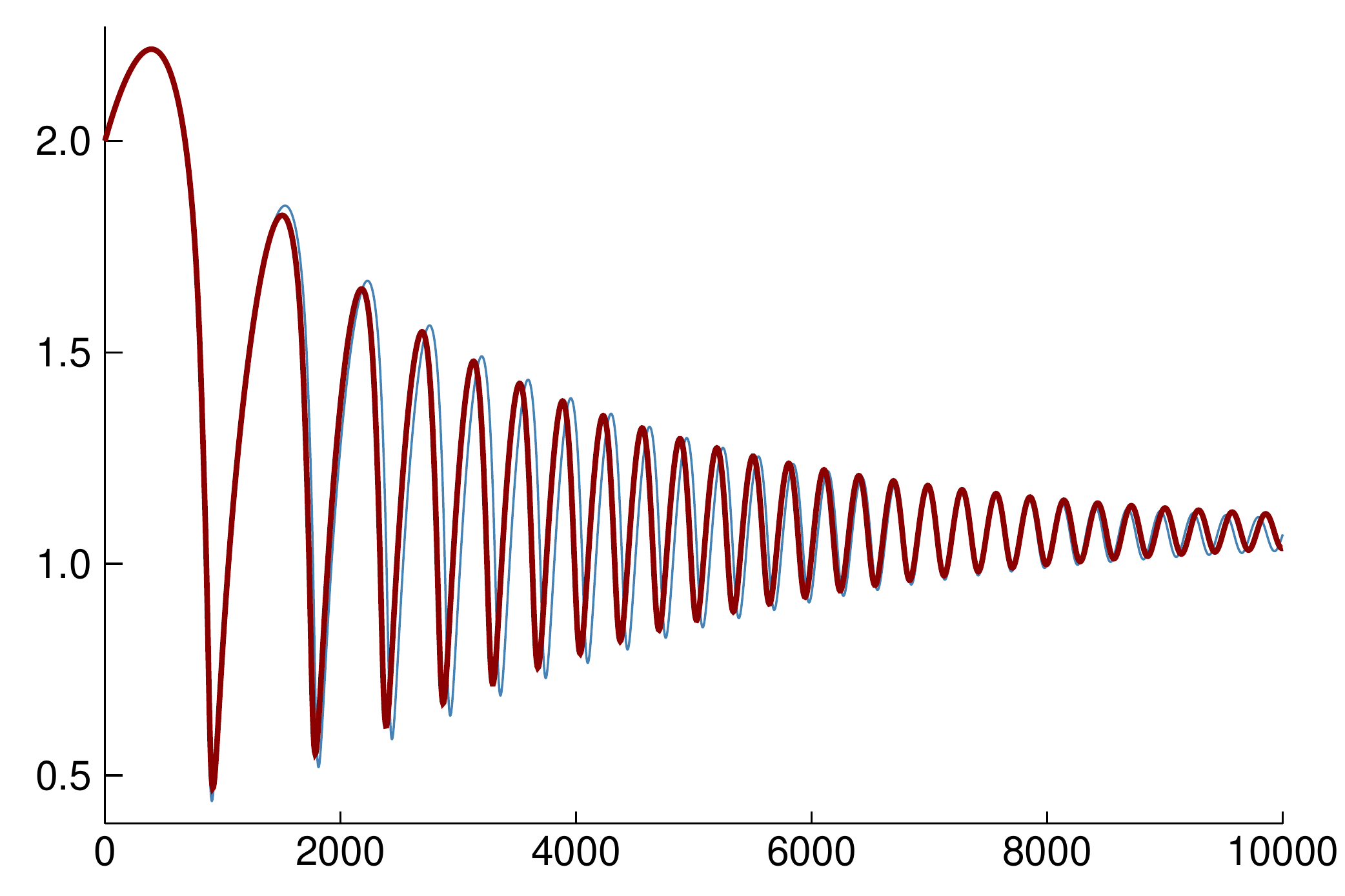}
            \put(48,-4){$t/T$}
            \put(-8,35){$x/R$}
            \put(10,65){(a)}
            \end{overpic}
    \end{minipage}%
\hspace{0.05\textwidth}
   \begin{minipage}[h!]{0.45\textwidth}
        \centering
            \begin{overpic}[width = \textwidth]{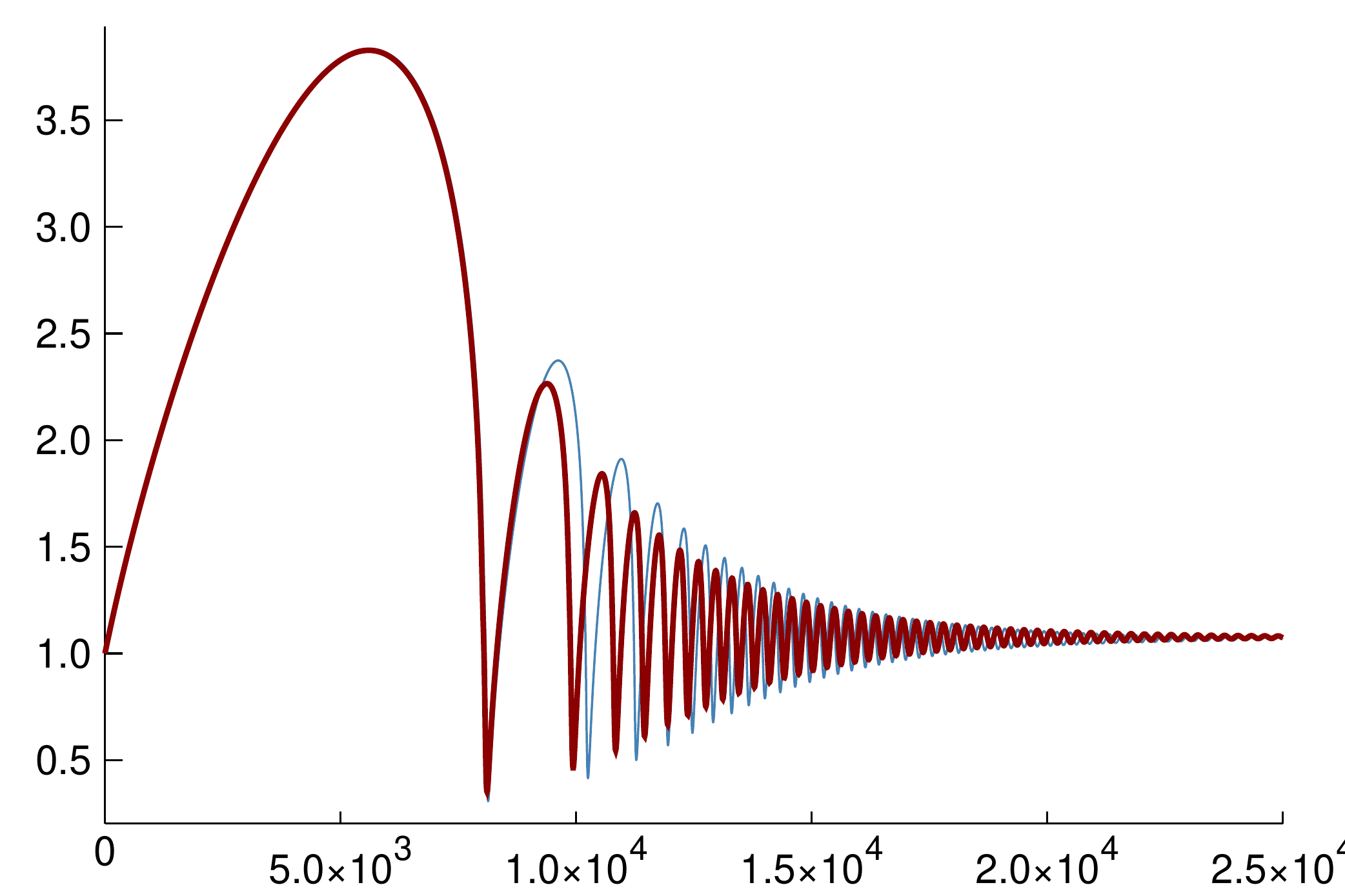}
            \put(48,-4){$t/T$}
            \put(-8,35){$x/R$}
            \put(10,65){(d)}
            \end{overpic}
    \end{minipage}%
    \par\vspace{0.4cm}
 \begin{minipage}[h!]{0.45\textwidth}
    	\centering
        \begin{overpic}[width= \textwidth]{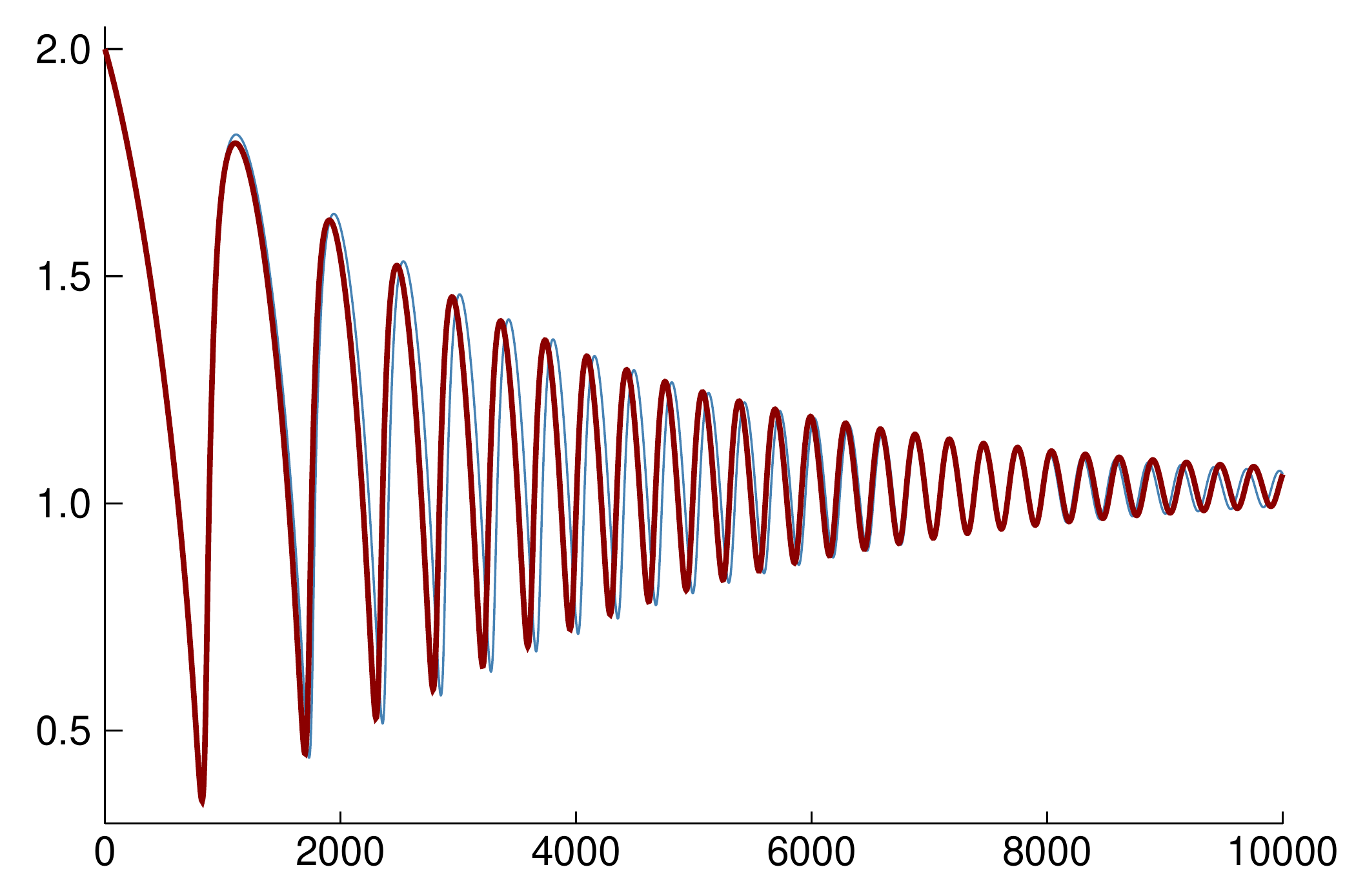}
            \put(48,-4){$t/T$}
            \put(-8,35){$y/R$}
            \put(10,65){(b)}
        \end{overpic}
    \end{minipage}%
\hspace{0.05\textwidth}
        \begin{minipage}[h!]{0.45\textwidth}
    	\centering
        \begin{overpic}[width= \textwidth]{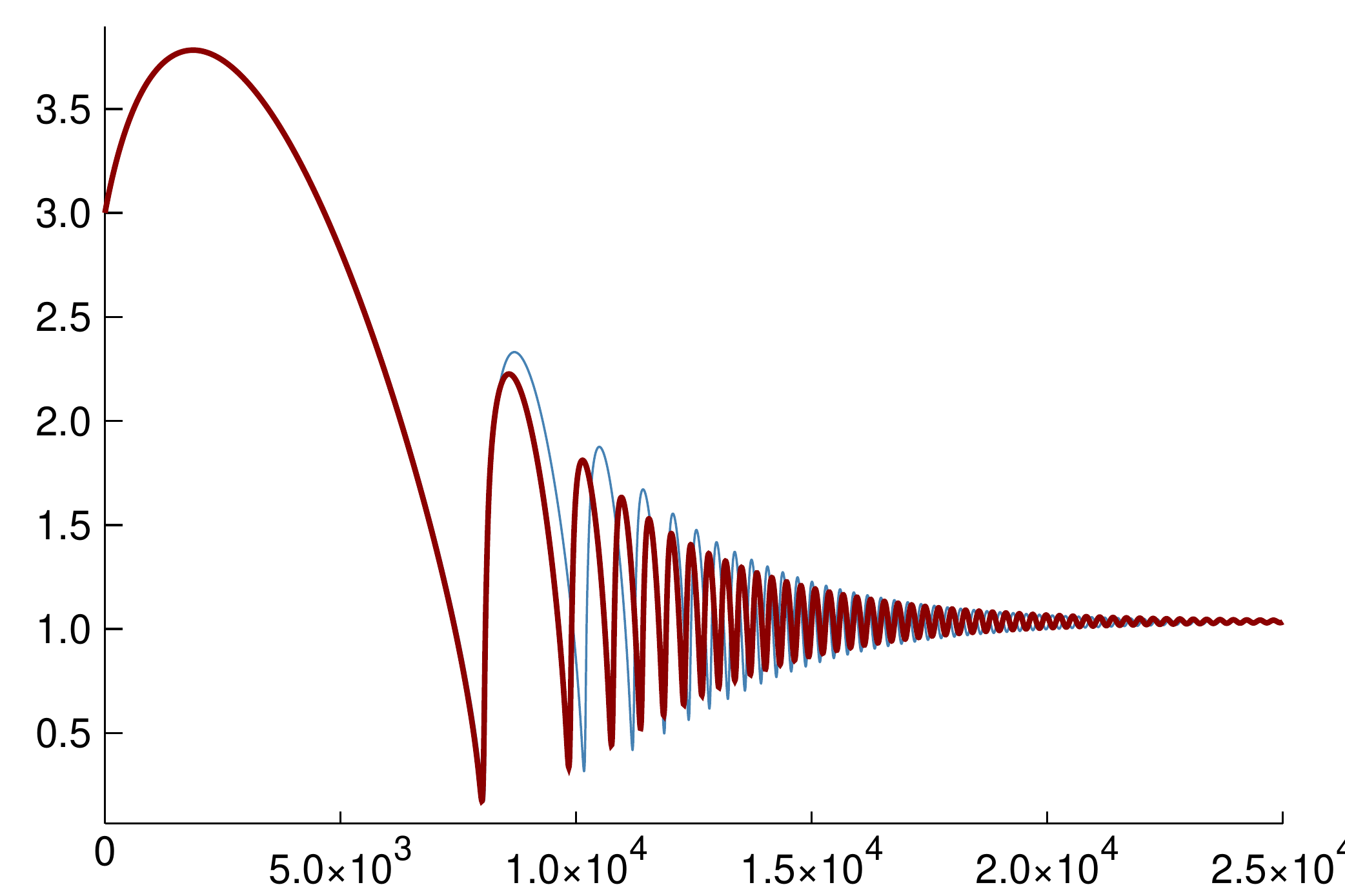}
            \put(48,-4){$t/T$}
            \put(-8,35){$y/R$}
            \put(10,65){(e)}
        \end{overpic}
    \end{minipage}%
      \par\vspace{0.3cm}
    \begin{minipage}[h!]{0.45\textwidth}
        \centering
            \begin{overpic}[width = \textwidth]{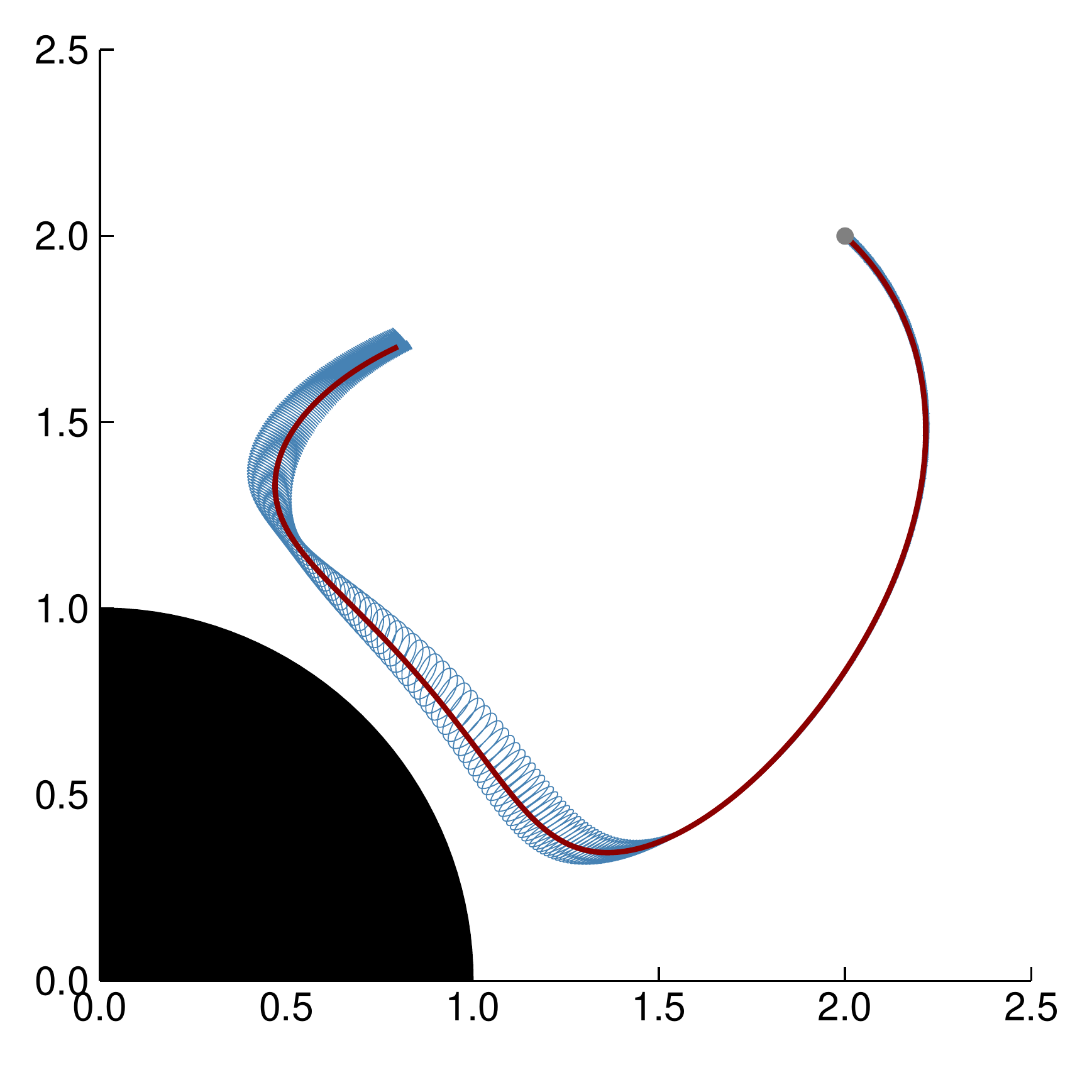}
            \put(48,0){$x/R$}
            \put(-10,48){$y/R$}
            \put(12,90){(c)}
            \end{overpic}
    \end{minipage}%
    \hspace{0.05\textwidth}
    \begin{minipage}[h!]{0.45\textwidth}
        \centering
            \begin{overpic}[width = \textwidth]{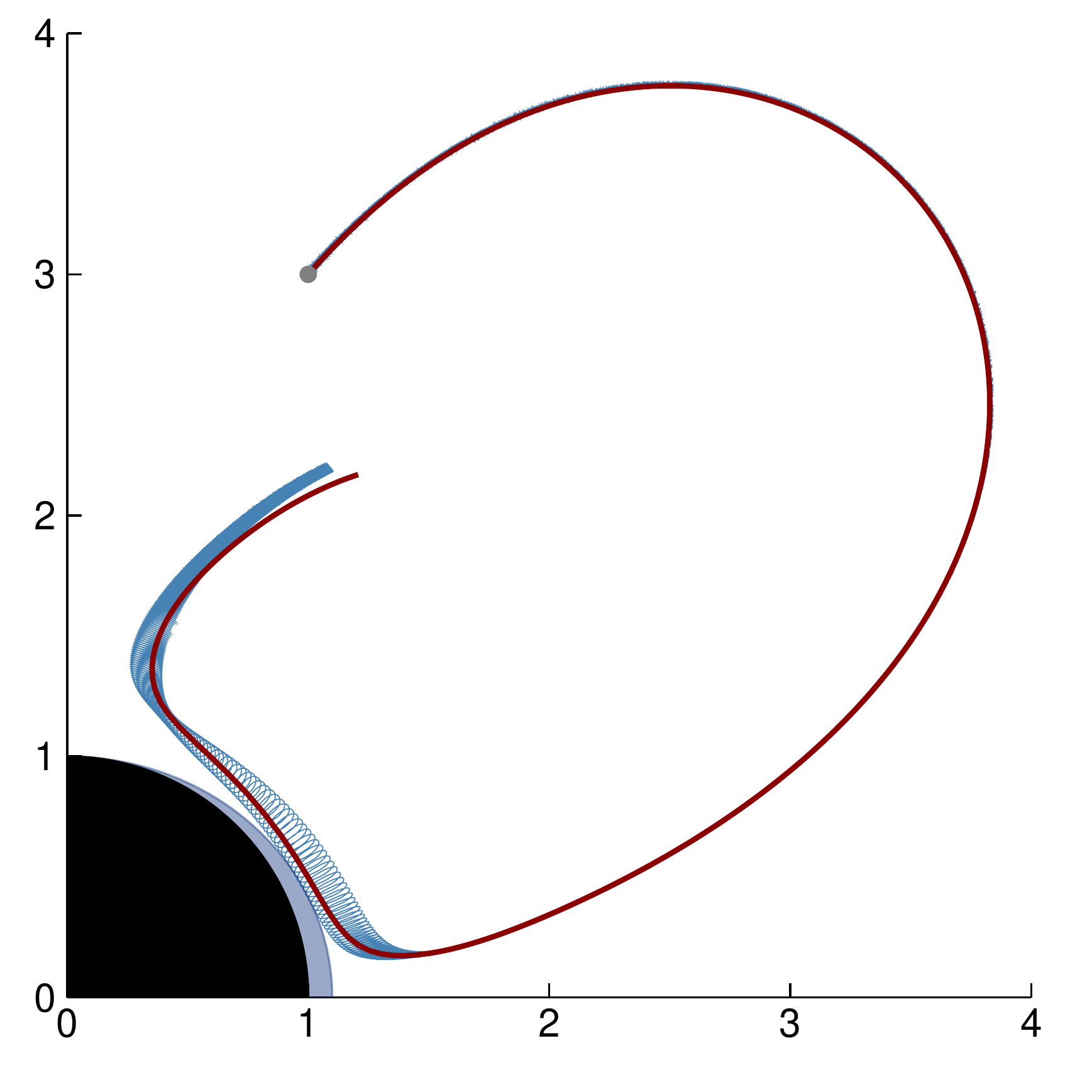}
            \put(48,0){$x/R$}
            \put(-10,48){$y/R$}
            \put(10,90){(f)}
            \end{overpic}
    \end{minipage}%
    \caption{Left column: (a,b) Comparison of $x$ and $y$ components, respectively, of the Lagrangian mean trajectory and the full time-resolved trajectory (sampled once per cycle) for particle $\plabel = (2,2)$. (c) Lagrangian mean trajectory (magenta) for inertial particle $\plabel = (2,2)$, compared with the full time-resolved trajectory (blue) for the same particle.  Right column: (d,e) Comparison of $x$ and $y$ components, respectively, of the Lagrangian mean trajectory and the full time-resolved trajectory (sampled once per cycle) for particle $\plabel = (1,3)$. (f) Lagrangian mean trajectory (magenta) for inertial particle started at $\plabel = (1,3)$, compared with the full time-resolved trajectory (blue) for the same particle.}
\label{fig:Inertial_avg_transient}
 \end{figure}

The panels in Figure~\ref{fig:Inertial_avg} depict the predicted mean trajectories of inertial particles initially released from $(2,2)$ and $(1,3)$, respectively, in the single cylinder configuration. Both exhibit good agreement with the full trajectory, though small errors incurred during the nearest approach to the cylinder (due to the aforementioned truncation of the expansion in $\epsilon$) tend to push the particle onto a slightly different orbit in each encirclement of the streaming cell. However, these errors are largely irrelevant, as both the mean trajectory and the full time-resolved trajectory converge on the same trapping point. This is more clearly revealed in panels Figure~\ref{fig:Inertial_avg_transient}(a,b) and (d,e), which show the respective comparisons of each coordinate's history along the trajectory (plotted once per cycle in the case of the full trajectory). The orbits of the streaming cell predicted by the mean transport algorithm are slightly faster than those of the full trajectories due to the slightly larger push toward the center experienced by the particle in the mean algorithm as it passes closest to the cylinder. To better illustrate the relationship between these two approaches to computing the trajectories, in Figure~\ref{fig:Inertial_avg_transient}(c) and (f) we plot every point along a portion of the full time-resolved trajectory and overlay the same portion of trajectory predicted by the mean transport algorithm. The algorithm visually tracks the center of oscillations, but skews slightly inward as it moves away from the cylinder. It can be observed that, for $\plabel = (1,3)$, the oscillations along the full trajectory overlap with the right-most configuration of the cylinder; however, those portions of the trajectory that overlap correspond to the phase in the cycle when the cylinder is in its left-most configuration.

\begin{figure}
    \centering
    \begin{overpic}[width = 0.45\textwidth]{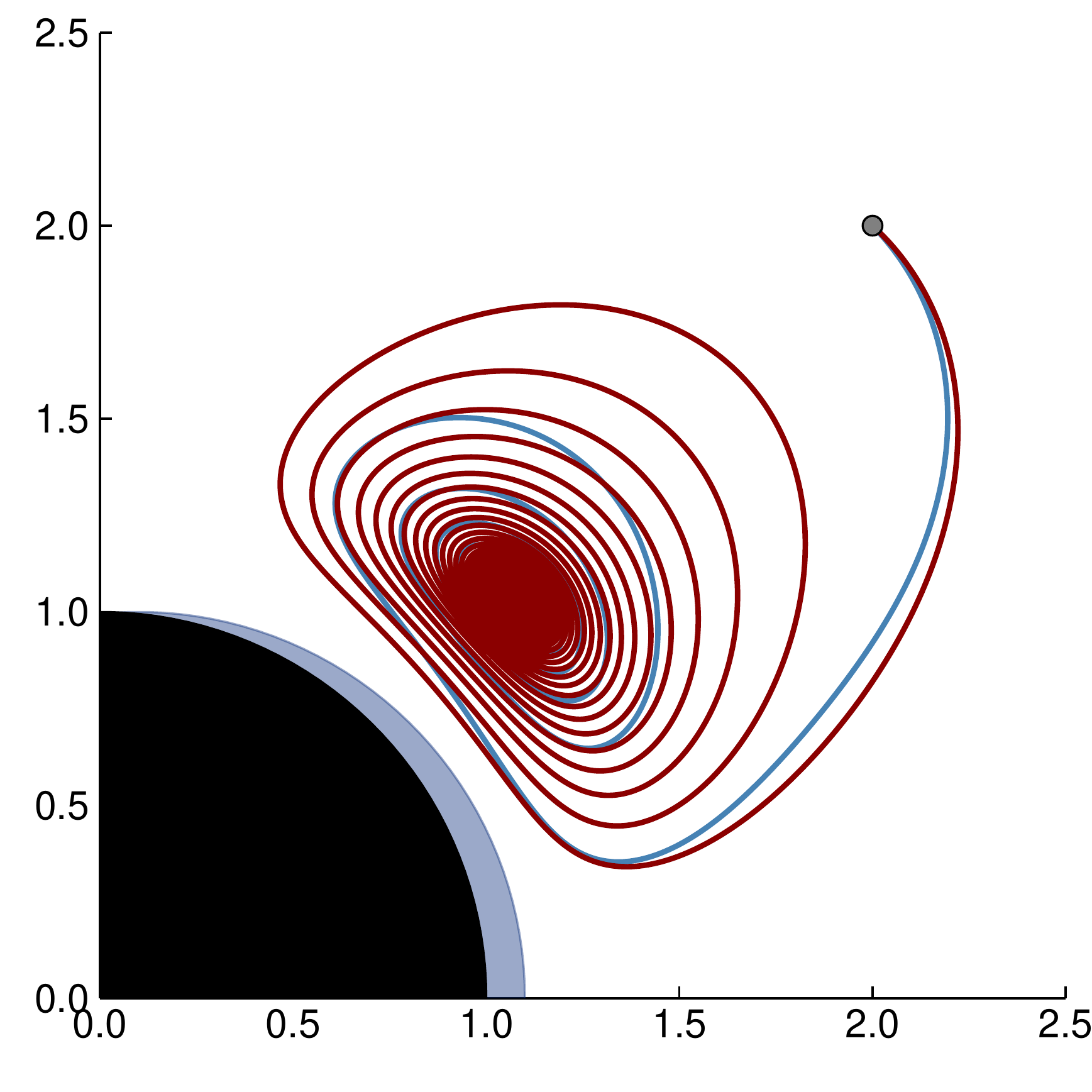}
            \put(48,0){$x/R$}
            \put(-10,48){$y/R$}
    \end{overpic}
    \caption{Lagrangian mean trajectories generated for inertial particles with $\rho_p/\rho_f = 0.95$ (magenta)  and $0.05$ (blue) started at $\plabel = (2,2)$}
    \label{fig:Inertial_avg_beta}
\end{figure}

Most of our results in this paper are focused on a single type of particle, nearly neutrally buoyant with $\rho_p/\rho_f = 0.95$ (or $\beta = 1.034$). To demonstrate the effect of density ratio, we compare this particle's mean trajectory in Figure~\ref{fig:Inertial_avg_beta} with that of a very light particle, with density ratio $\rho_p/\rho_f = 0.05$ (so that $\beta = 2.73$). The light particle is affected much more by the buoyancy term---the first term in (\ref{accforcenew})---which applies a centrifugal motion directed toward the center of the streaming cell. Interestingly, this contribution is most active during the intervals when the Saffman lift is not, on the outermost parts of the orbit when the particle's trajectory is most curved.

\begin{figure}
    \centering
    \begin{minipage}[h!]{0.7\textwidth}
    	\centering
                \begin{overpic}[width = \textwidth]{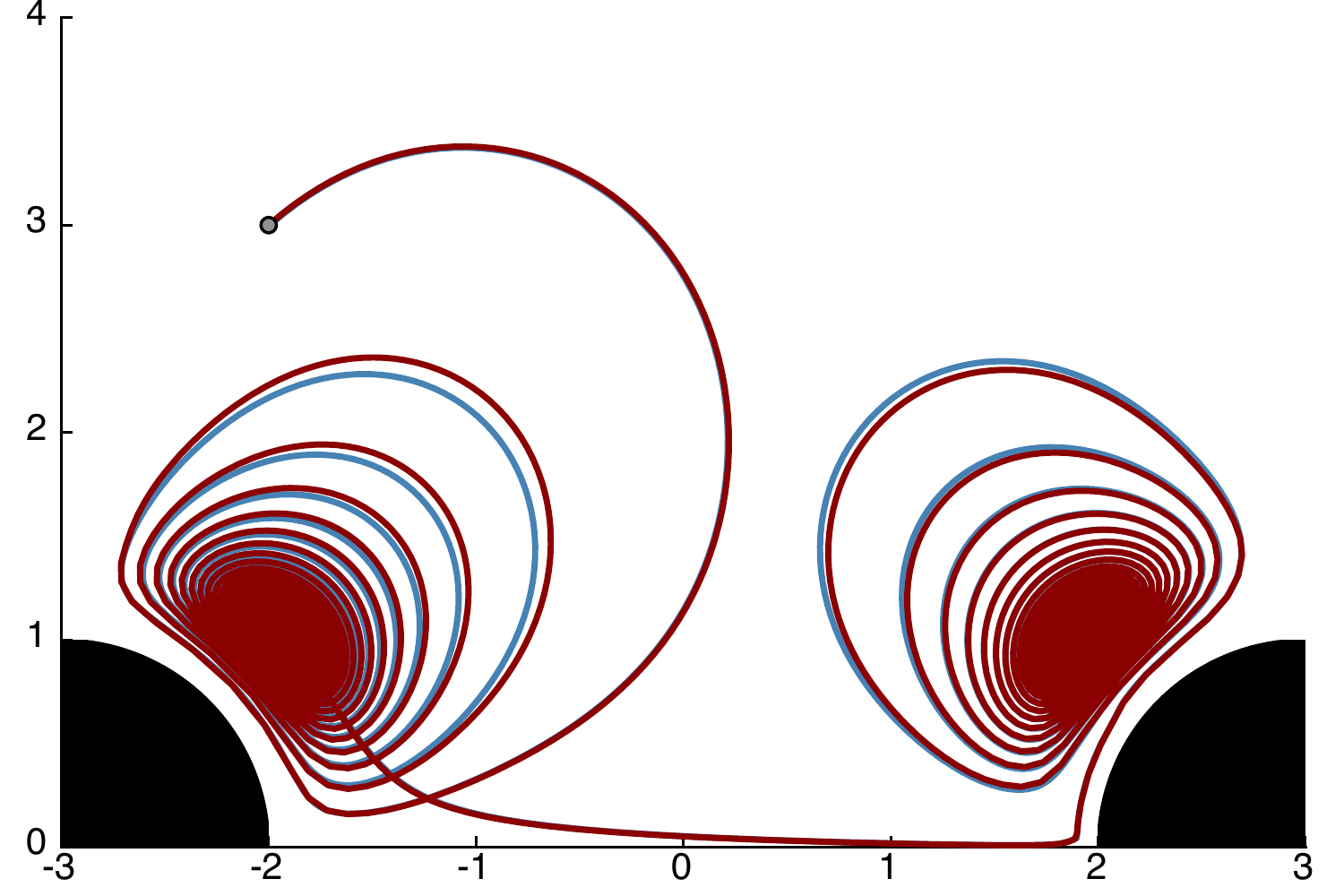}
        \put(48,-3){$x/R$}
        \put(-6,33){$y/R$}
        \put(10,60){(a)}
        \end{overpic}
    \end{minipage}%
    \par\vspace{15mm} 
    \begin{minipage}[h!]{0.45\textwidth}
    	\centering
        \begin{overpic}[width = 8cm]{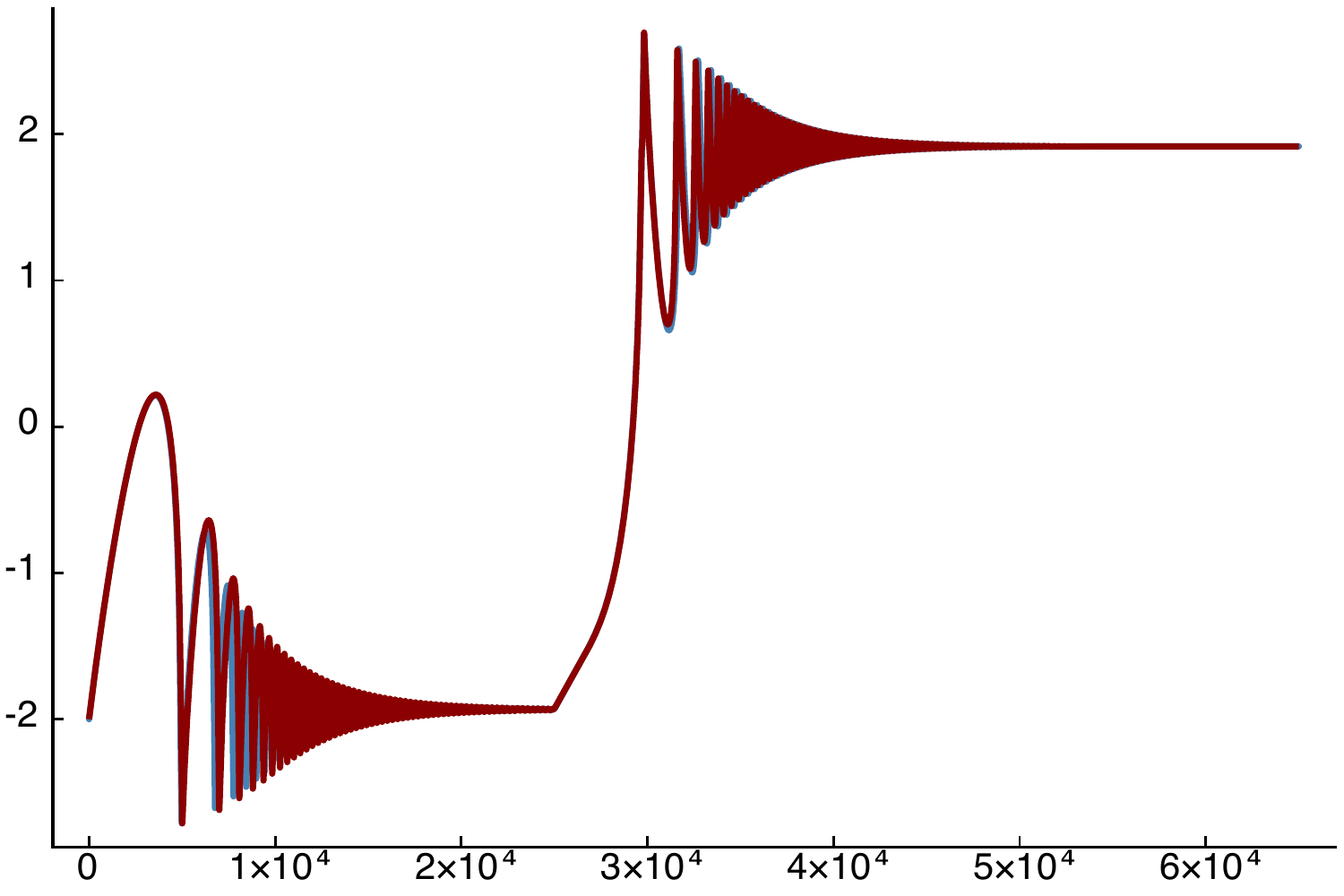}
            \put(48,-8){$t/T$}
            \put(-12,33){$x/R$}
            \put(10,60){(b)}
        \end{overpic}

    \end{minipage}%
    ~
    \hspace{0.05\textwidth}
    \begin{minipage}[h!]{0.45\textwidth}
        \centering
            \begin{overpic}[width = 8cm]{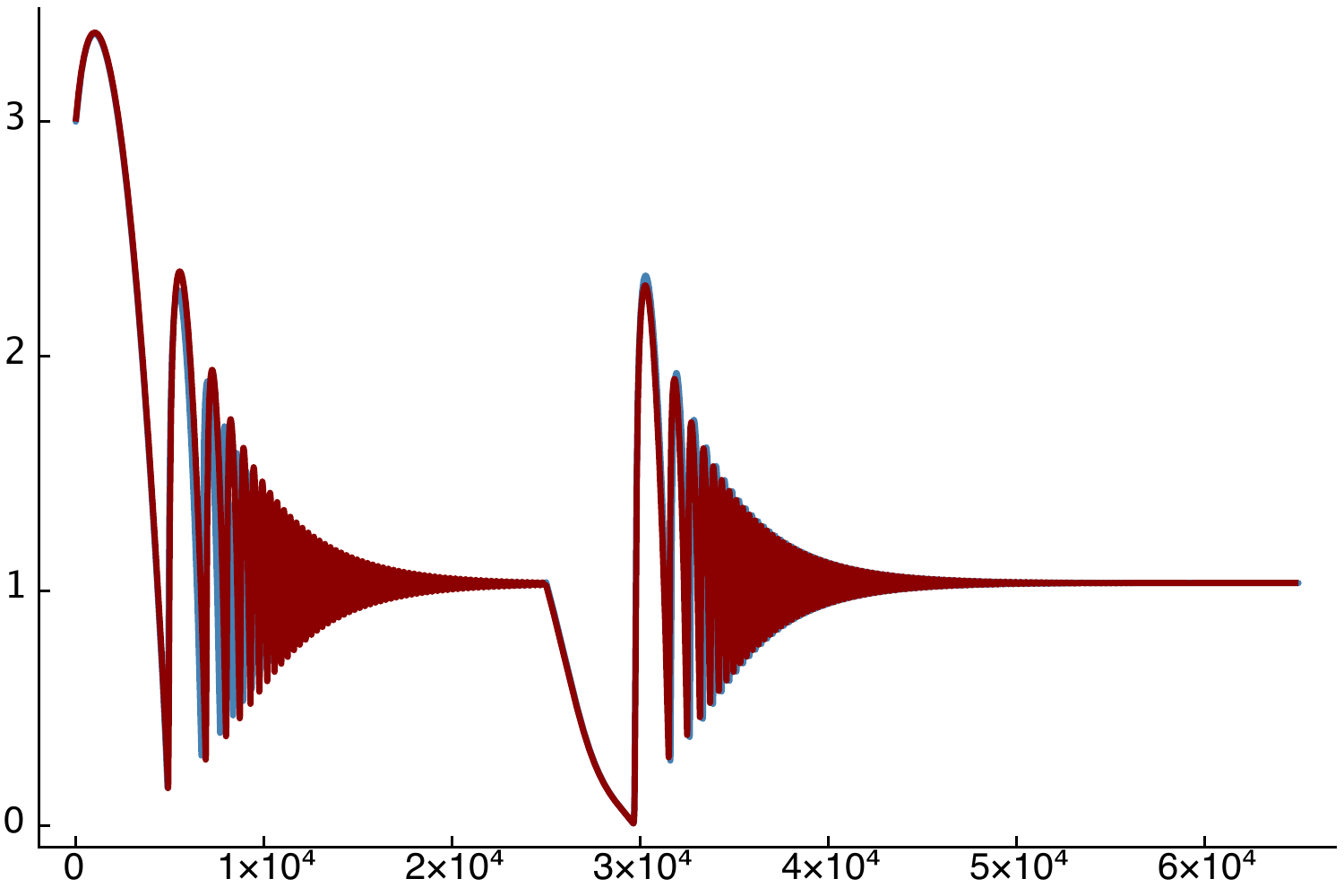}
            \put(48,-8){$t/T$}
            \put(-12,33){$y/R$}
            \put(10,60){(c)}
            \end{overpic}
    \end{minipage}%
        \vspace{6.0mm} 
\caption{(a) Full time resolved inertial particle trajectory, sampled once per cycle (blue), and Lagrangian mean trajectory (magenta) for a particle released from $\plabel = (-2, 3)$ in the sequential oscillator configuration.  (b,c) Time histories of the $x$ and $y$ trajectory components, respectively.}
\label{fig:2cylinder_inertial_hybrid}
\end{figure}
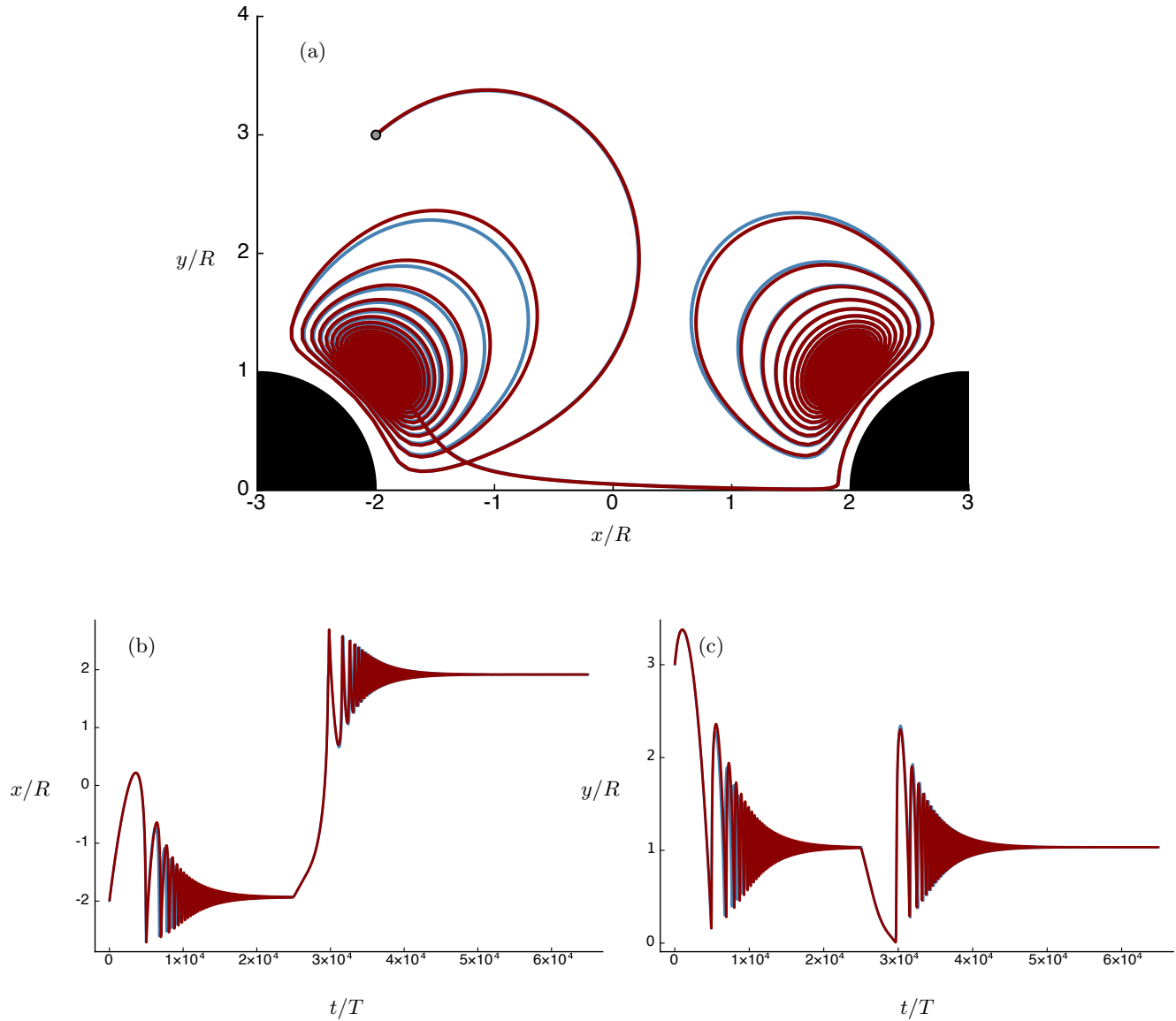

Figure~\ref{fig:2cylinder_inertial_hybrid} shows the results for a particle released from $(-2,3)$ in the two-cylinder configuration. Over the first 25000 periods, the particle is drawn toward the streaming cell of the left cylinder during that cylinder's motion. Then, when the left cylinder stops its motion and the right cylinder starts to oscillate, the particle is entrained into the cell nearest to the right cylinder over the ensuing 40000 periods. Both trajectories are predicted well by the mean transport algorithm. It should be noted that the no-penetration constraint is active for both the mean and the full trajectory predictions during the interval in which the particle reaches the right-most cylinder and is drawn along its boundary (at around 30000 periods).

\subsection{The effect of transient behavior}

 \begin{figure}[t]
    \centering
    \begin{minipage}[h!]{0.45\textwidth}
    	\centering
        \begin{overpic}[width = 7cm]{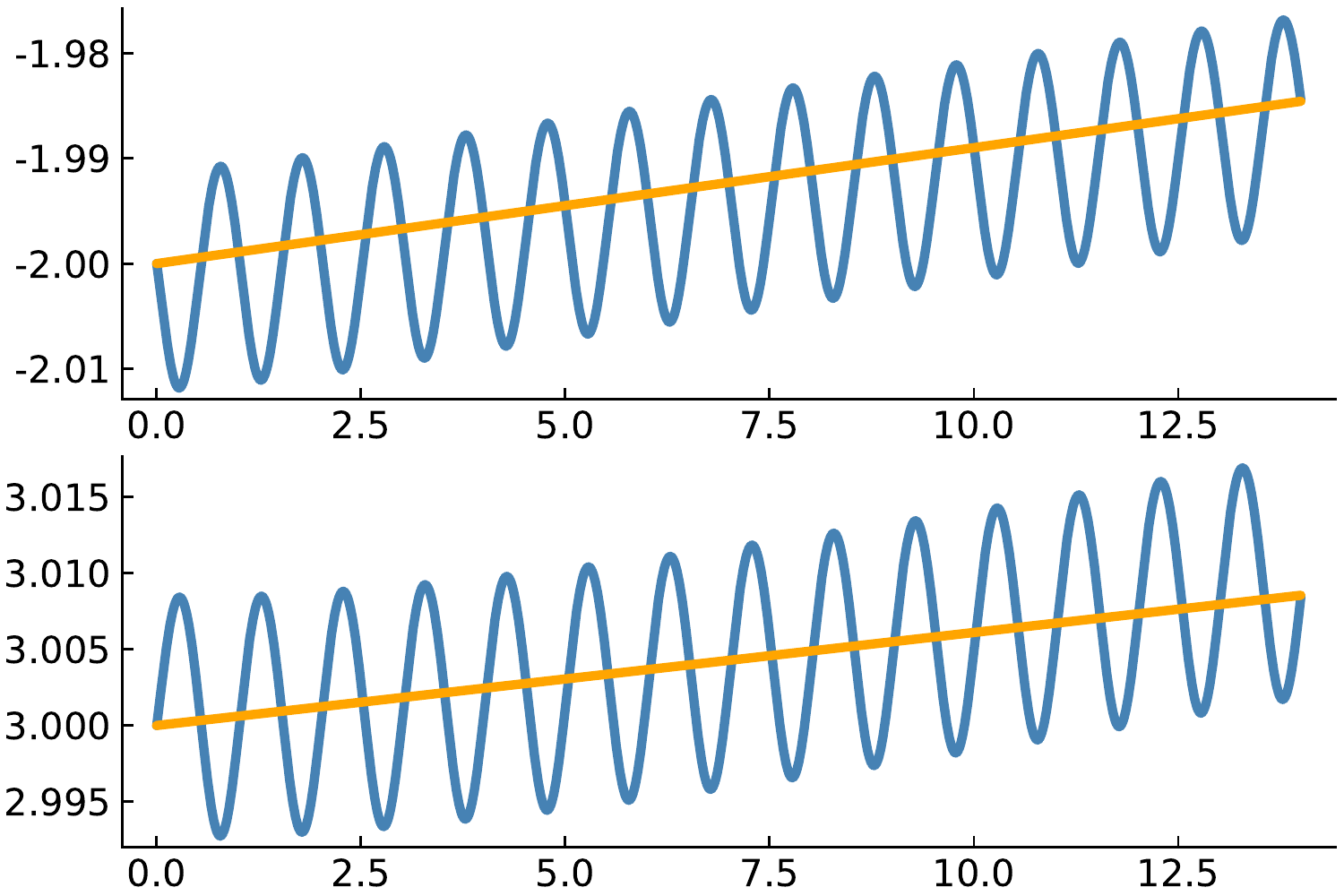}
        \put(48,-6){$t/T$}
        \put(-12,15){$y/R$}
        \put(-12,50){$x/R$}
        \end{overpic}

    \end{minipage}%
    ~
    \hspace{3mm}
    \begin{minipage}[h!]{0.45\textwidth}
        \centering
            \begin{overpic}[width = 7cm]{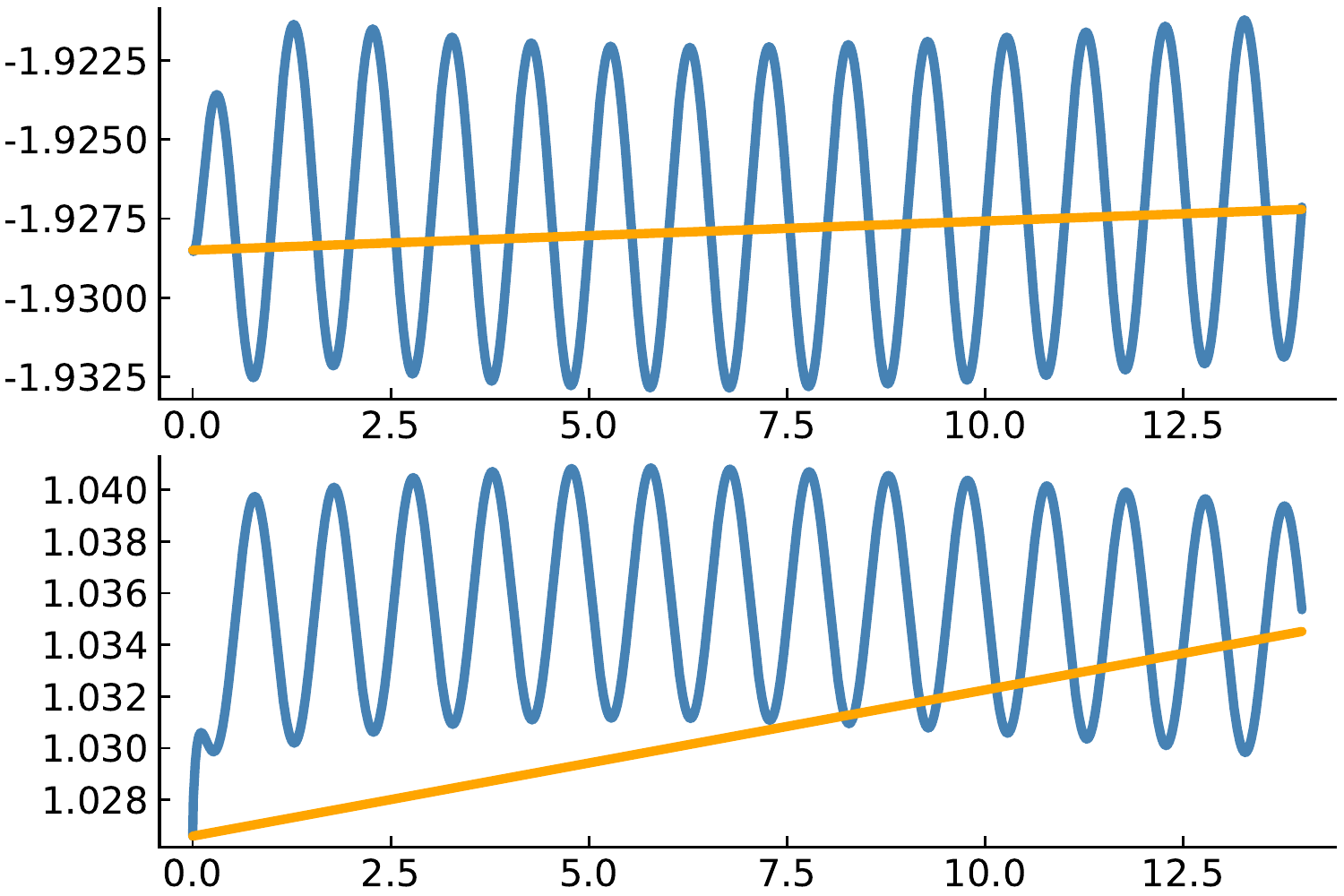}
            \put(48,-6){$t/T$}
            \put(-12,15){$y/R$}
            \put(-12,50){$x/R$}
            \end{overpic}
    \end{minipage}%
    \vspace{6mm}
	\caption{Time resolved inertial particles trajectories (blue) and mean transport algorithm (yellow) for $\plabel = (-2.0, 3.0)$ over the transient regime of the left cylinder (left) and $\plabel = (-1.9285,1.0266)$ over the transient regime of the right cylinder (right).}
    \label{fig:MR_Avg_transient_2cylinders}
    \vspace{6mm}
    \end{figure}

An important question that overlies the transport of particles in viscous streaming is the effect of transient behavior in the fluid during changes of oscillator motion. In the examples we have profiled in this paper, these changes in motion occur suddenly: each cylinder stops or starts instantaneously.  Such sudden changes provide an unambiguous context in which to assess the transient effects since the flow is necessarily approaching a well-defined stationary state when such transient effects will vanish. Some of the sudden change of motion is communicated everywhere instantaneously through pressure. The transient effects are due to viscous diffusion, which plays a particularly important role along directions transverse to the motion. As mentioned earlier, at $\Rey = 40$, this diffusion requires (empirically) around 20 oscillation cycles to spread the information about motion changes across the entire region of interest and thus establish stationary periodic behavior.

In Figure~\ref{fig:MR_Avg_transient_2cylinders} we examine the effects of transient behavior for two inertial particles over the first 14 oscillation periods. Each case depicts the inertial particle's trajectory predicted by the mean transport algorithm over one time step (in this example only, taken to be 14 periods) during a transient interval of the two-cylinder array---in the first case during the initial motion of the left cylinder, and in the second case during the newly-initiated motion of the right cylinder (after the left cylinder has stopped). The full time-resolved trajectory is depicted for reference in both cases. It is important to observe first that, as a result of the pressure-driven part of the flow, each particle's trajectory achieves approximately periodic behavior very quickly; the viscous adjustment takes longer. However, the mean transport algorithm predicts the behavior very well during such intervals: the particle's final location at the end of the step agrees well with where it is expected to be along the full trajectory. Furthermore, the plots show that the influence of transient behavior in the flow is likely negligible. In 14 periods, neither particle has moved more than $0.02R$ from its initial location. Even if we were to ignore the mean transport during this transient interval, our error would be equivalent to assuming that the particle had started at a negligibly different location and then been subject to a truly periodic flow.

\section{Conclusions}
\label{conclusion}

In this paper, we have developed simplified equations governing the mean transport of inertial particles in viscous streaming flows. In flows generated by weakly oscillating rigid surfaces, the motions of fluid and inertial particles exhibit two distinct time scales: a fast scale associated with the particle's oscillatory motion and a slow scale associated with its mean translation. Previous work by Chong et al.~\citep{chong2013inertial} has shown that the mean motion of small inertial particles in streaming flows is well described by the Maxey--Riley equation augmented with Saffman lift and with Fax\'en corrections retained, but with the Basset memory term neglected. Collectively, the Fax\'en correction and the Saffman lift effect the trapping of inertial particles in streaming cells generated near the oscillating object.

In this work, we have analyzed such transport with the help of three key tools. First, we conceived an Eulerian field for inertial particle velocity by asymptotically expanding the Maxey--Riley equation in the small Stokes number associated with small particles in moderate Reynolds number flows. This approach follows the earlier works of Maxey~\citep{maxey1987gravitational} and Ferry and Balachandar~\citep{ferry2001fast}, but importantly here, retains the essential Fax\'en term. This expansion has confirmed the observations made by Chong et al.~\citep{chong2013inertial}: A small neutrally-buoyant particle moves at leading order like a fluid particle, but in regions of shear near the oscillating body, the Fax\'en correction alters the particle's velocity from that of the fluid and the Saffman lift then causes it to move transversely to the shear, ultimately causing it to spiral toward a trapping point in the center of a streaming cell.

The second tool has been the Generalized Lagrangian Mean theory of Andrews and McIntyre~\citep{andrews1978exact}. This theory's exact distinction between the mean and fluctuating parts of a trajectory has allowed us to construct the Lagrangian mean velocity field, which is ultimately responsible for a particle's mean transport. This mean field receives an essential contribution from Stokes drift, based on an Eulerian disturbed displacement field that accompanies the time-varying velocity field.

The third important tool has been an expansion in the small oscillation amplitude. This expansion's effect on the fluid velocity field was already known from early work on streaming (e.g., Holtsmark et al.~\citep{holtsmark1954boundary}). However, with the availability now of the inertial particle velocity field and its subsequent decomposition into mean and fluctuating parts, we have been able to identify the dominant effects of small amplitude oscillation on mean inertial particle transport.  For fluid particles, the particle trajectories are directly obtained from the contours of a mean Lagrangian streamfunction field.

By applying the resulting algorithm to two basic oscillator flows, we have demonstrated that the approximations we have made by truncating these expansions have generally preserved the accuracy of the original treatment. Furthermore, the application of these tools has made a tremendous impact on the efficiency of computing mean particle trajectories. The previous approach to predicting such trajectories involved two straightforward steps: first, to compute the fluid velocity field until it reached stationary periodic behavior; and second, to advance the particle in this oscillatory flow field with the Maxey--Riley equation, with time steps that sufficiently resolve the fast scales. Each such time step requires the evaluation of the instantaneous forces on the particle, generally obtained by interpolating the velocity field and its derivatives from the computational grid. A full trajectory generally requires $O(10^{6})$ such time steps and is quite slow to compute. The new approach presented here, which constructs the aforementioned Eulerian fields to develop the Lagrangian mean velocity field, allows time steps that are $O(1000)$ times larger than the previous approach.

It is also important to stress that, in spite of our examples, no aspect of our treatment of this problem is limited to two-dimensional oscillators. Indeed, the cost reduction would be proportionally greater in three-dimensional problems, where the calculations of forces on full trajectories require more taxing interrogations and calculations of the flowfield data.

We have also shown in this work that the viscous transients in the fluid that arise after changes of oscillator motion have insignificant effect on mean particle transport. This observation depends on the distinction of time scales in this regime of Reynolds number and oscillator amplitude: a particle moves very little in the time it takes for viscous diffusion to communicate the oscillator's change of motion, so it is safe to assume that the fluid has already reached a stationary periodic state. We have not exploited this feature in this paper, but in a paper currently in development we will demonstrate that, by treating the underlying flow field as strictly periodic, we can solve for this flow field and then construct the Lagrangian mean velocity of either type of particle entirely in the frequency domain. This leads to a further substantial gain in computational efficiency.

\section{Acknowledgements}
Support by the National Science Foundation under grant 1538824 is gratefully acknowledged. The authors also would like to thank Chris Rackauckas for fruitful discussions on the manifold projection method for constrained ordinary differential equations. We would also like to note that all of the computational tools developed for this paper are available at \url{https://github.com/jdeldre/ViscousStreaming.jl}.

\appendix
\section{Appendices}


\subsection{Asymptotic expansion of the inertial particle velocity for small Stokes number}
\label{appendix_inertialasymptotics}

Here, we present the detailed asymptotic expansion of the inertial particle velocity field, $\inertv$, in small Stokes number $\tau$, by substituting the form (\ref{vform}) and seeking an expression for $\vdiff$ in terms of the fluid velocity field and its derivatives. With the form (\ref{vform}), we can rewrite equation (\ref{maxeyrileyeulerian}) in terms of $\vdiff$. First, note that the two time derivatives are related to each other by
\begin{equation}
    \ddt {} = \DDt {} + \left(\vfax + \tau^{1/2}\vdiff \right)\cdot\grad.
\end{equation}
It can be easily verified that
\begin{equation}
    \ddt \inertv = \DDt \fluidv + \left(\vfax + \tau^{1/2}\vdiff \right)\cdot\grad(\fluidv + \vfax) + \DDt \vfax + \tau^{1/2} \ddt \vdiff.
\end{equation}
Thus, by substituting this and shifting all terms involving $\vdiff$ to the left-hand side, multiplying by $\tau^{1/2}$, and manipulating the terms slightly, \eqn~(\ref{maxeyrileyeulerian}) can be rewritten as
\begin{align}
    \left[ 1 + \tau^{1/2} \beta^{1/2} \Lop + \tau \DDtcal{} + \tau^{3/2} \vdiff\cdot\grad  \right] \vdiff = -\beta^{1/2} \Saffman\left[ \vfax\right] + \tau^{1/2} \accforce,
    \label{maxeyrileyq}
\end{align}
where $\Lop = \Basset + \Saffman$, we have defined a differential operator,
\begin{equation}
\label{ddtcal}
    \DDtcal {\boldsymbol{f}} = \frac{\partial  \boldsymbol{f}}{\partial t}+ (\fluidv+\vfax)\cdot\grad \boldsymbol{f} + \boldsymbol{f} \cdot \grad\left(\fluidv+(1-\beta/5)\vfax \right),
\end{equation}
for some field vector $\boldsymbol{f}$, and $\accforce$ represents a fluid acceleration force,
\begin{equation}
\label{accforce}
    \accforce = (\beta-1) \DDt \fluidv + \frac{\beta}{5} \DDt \vfax - \DDtcal{\vfax}
\end{equation}

 To solve for $\vdiff$, we write it as an asymptotic sequence in powers of $\tau^{1/2}$:
\begin{equation}
\label{qexpand}
    \vdiff = \vdiff_{0} + \tau^{1/2}\vdiff_{1/2} + \tau\vdiff_{1} + + \tau^{3/2}\vdiff_{3/2} + O(\tau^2).
\end{equation}
Similarly, the operator on the left-hand side of (\ref{maxeyrileyq}) can be formally inverted by expanding it in powers of $\tau$ (with the help of (\ref{qexpand}) to replace $\vdiff$):
\begin{equation}
\begin{split}
    \left[ 1 + \tau^{1/2} \beta^{1/2} \Lop + \tau \DDtcal{} + \tau^{3/2} \vdiff\cdot\grad  \right]^{-1} &= 1 - \tau^{1/2}\beta^{1/2} \Lop - \tau \left(\DDtcal{} - \beta \Lop^2\right) \\ &+ \tau^{3/2} \left[ -\vdiff_0 \cdot \grad + \beta^{1/2} \left( \Lop \DDtcal{} + \DDtcal{} \Lop\right)\right] + O(\tau^{2}),
\end{split}
\end{equation}
where $\mathcal{L}^2$ denotes the composition of $\mathcal{L}$ with itself, i.e., $\mathcal{L}^2 =\mathcal{L} \circ \mathcal{L}$.

Applying this expanded inverse operator to the right-hand side of (\ref{maxeyrileyq}), substituting the expansion (\ref{qexpand}), and grouping terms with like powers of $\tau$, we obtain
\begin{align}
    \vdiff_0 &= -\beta^{1/2} \Saffman \left[ \vfax\right], \\
    \vdiff_{1/2} &= \accforce - \beta^{1/2} \Lop \left[ \vdiff_0\right], \\
    \vdiff_{1} &= -\beta^{1/2} \Lop \left[ \accforce\right] - \left(\DDtcal{} - \beta \Lop^2\right) \vdiff_0, \\
    \vdiff_{3/2} &= -\vdiff_0\cdot \grad \vdiff_0 + \beta^{1/2} \left( \Lop \DDtcal {\vdiff_0} + \DDtcal {} \Lop \vdiff_0 \right) - \left(\DDtcal{} - \beta \Lop^2\right) \accforce.
\end{align}
From these, the inertial particle velocity field can be reassembled:
\begin{equation}
\label{inertialvgen}
    \begin{split}
        \inertv &= \fluidv + \vfax - (\tau\beta)^{1/2} \Saffman \left[ \vfax\right] + \tau \left(\accforce + \beta \Lop \Saffman \left[ \vfax\right] \right) \\
        & - \tau^{3/2} \left[ \beta^{1/2} \Lop \left[ \accforce\right] -\beta^{1/2} \left(\DDtcal{} - \beta \Lop^2\right) \Saffman \left[ \vfax\right] \right] \\
        & - \tau^{2} \left[ \left(\DDtcal{} - \beta \Lop^2\right) \accforce +\beta \Saffman \left[ \vfax\right] \cdot \grad \Saffman \left[ \vfax\right] + \beta \left( \Lop \DDtcal {}\Saffman \left[ \vfax\right] + \DDtcal {} \Lop \Saffman \left[ \vfax\right] \right)  \right]+ O(\tau^{5/2}).
    \end{split}
\end{equation}

The equation above is valid in general circumstances of small $\tau$. However, we return now to the scaling of the Fax\'en correction velocity, $\vfax$. As we discussed earlier, when the particle is immersed in a laminar flow created by an oscillating cylinder, we expect $\vfax \sim \tau \fluidv$ due to the shear in the Stokes layer, whose thickness goes like $(\nu/\Omega)^{1/2}$. Formalizing this idea, we can define a rescaled Laplacian, $\lapnew$, such that
\begin{equation}
\label{laprescale}
    \lap = \Rey \lapnew,
\end{equation}
from which it follows that
\begin{equation}
    \vfax = \frac{1}{2} \tau \beta \lapnew \fluidv. 
\end{equation}
Then, with this explicit scaling of $\vfax$, and ignoring the Basset memory term $\Basset$ as justified by the analysis of Chong et al.~\citep{chong2013inertial}, we find that the particle velocity is given by
\begin{equation}
\begin{split}
\label{inertialv_app}
    \inertv = \fluidv & + \tau  \accforcenew - \tau^{3/2} \beta^{1/2} \Saffman \left[ \accforcenew\right] - \tau^2 \left[ \left(\DDtcal {} - \beta \Saffman^2 \right) \accforcenew - \frac{1}{10} \beta^2 \DDt{} \lapnew \fluidv \right] + O(\tau^{5/2}),
\end{split}
\end{equation}
where we have defined a simplified acceleration force
\begin{equation}
\label{accforcenew_app}
    \accforcenew \equiv (\beta-1) \DDt \fluidv  + \frac{1}{2} \beta \lapnew \fluidv,
\end{equation}
in which we remind the reader that the rescaled Laplacian $\lapnew$ is defined in (\ref{laprescale}). We have also simplified our earlier definition (\ref{ddtcal}) of the operator $\mathcal{D}/\mathcal{D}t$ in the appropriate manner to
\begin{equation}
    \DDtcal {\boldsymbol{f}} = \DDt {\boldsymbol{f}} + \boldsymbol{f}\cdot\grad\fluidv.
\end{equation}

\subsection{A review of Generalized Lagrangian Mean theory}
\label{appendix_glm}

In this section, we present the basic outline of the Generalized Lagrangian Mean theory of Andrews and McIntyre~\citep{andrews1978exact}. We present details only to the extent necessary for our results in the main body of the paper.

Consider a flow map, $\flowmap{\plabel}$, illustrated in Figure~\ref{fig:flowmap}, from material coordinates (i.e., Lagrangian label) $\plabel$ in a reference space $\refspace$ to a location in a space-time configuration space $\configspace \times [0,\infty)$. In this latter space, each slice $\configspace$ represents an evolved form of $\refspace$ at some later time $t$; we assume that $\configspace = \refspace$ at $t = 0$. We denote the velocity field associated with this map as $\genv$, and will refer to the infinitesimal bit of material associated with $\plabel$ as a `\particle'. We will use the notation $\material{\genv}(\plabel,t)$ to denote the velocity $\genv$ in its Lagrangian form, i.e.,
\begin{equation}
    \material{\genv}(\plabel,t) \equiv \ddp {\flowmap{\plabel}}{t}.
\end{equation}
As is typical, we regard the flow map as invertible, so that we can uniquely associate a Lagrangian label with any fixed location. To distinguish from later terminology, we will refer to the location $\flowmap{\plabel}$ as the {\em actual location} of the \particle~$\plabel$ at time $t$, and the velocity $\material{\genv}(\plabel,t)$ as the \particle's {\em actual velocity}.

We can always think of $\flowmap{\plabel}$ as a composition of two maps,
\begin{equation}
\label{composition}
    \flowmap{\plabel} = \secondmap{\meanflowmap{\plabel}},
\end{equation}
as illustrated in Figure~\ref{fig:flowmap}. The first, $\meanflowmap{\plabel}$, maps $\plabel$ in $\refspace$ to some location in another configuration space $\meanconfigspace \times [0,\infty)$ at time $t$; and the second,
\begin{equation}
    \secondmap{\x} \equiv \x + \fluctmap{\x},
    \label{secondmap}
\end{equation}
maps from that location in the slice $\meanconfigspace$ at $t$ to the actual location in the corresponding slice $\configspace$ at time $t$.

We will require that $\meanconfigspace$ coincides with $\refspace$ at $t = 0$, just as $\configspace$ does. In other words, the two spaces are identical at the initial instant. At all times, we assume that the mapping $\meanflowmap{\cdot}$---like $\flowmap{\cdot}$---is invertible. Also, just as $\flowmap{\cdot}$ is associated with the velocity field $\genv$, the flow map $\meanflowmap{\cdot}$ is generated by its own velocity field, $\genvmean$, i.e.,
\begin{equation}
\label{meanvel}
    \material{\genvmean}(\plabel,t) \equiv \ddp {\meanflowmap{\plabel}}{t}.
\end{equation}

From the definition (\ref{secondmap}) and its use in the composition (\ref{composition}), it is clear that $\fluctmap{\x}$ provides an additive correction from the location provided by mapping $\meanflowmap{\plabel}$ to the actual location of the \particle, provided by $\flowmap{\plabel}$:
\begin{equation}
    \label{XY}
    \fluctmap{\meanflowmap{\plabel}} = \flowmap{\plabel} - \meanflowmap{\plabel}.
\end{equation}
If we differentiate this with respect to time (keeping the Lagrangian label fixed), then by the chain rule we obtain
\begin{equation}
    \ddp {}{t}\fluctmap{\meanflowmap{\plabel}}  + \material{\genvmean}(\plabel,t) \cdot \grad\fluctmap{\meanflowmap{\plabel}}   = \material{\genv}(\plabel,t) - \material{\genvmean}(\plabel,t),
    \label{wW}
\end{equation}
which relates the velocity fields, $\genv$ and $\genvmean$, associated with each flow map. 

For GLM theory and our later applications, it will be very useful to regard $\fluct$ as an Eulerian vector field, $\fluctmap{\x}$. To make sense of this interpretation, let us make use of the inverse mapping $\plabel = \meanflowmapinv{\x}$, uniquely associating any fixed location $\x$ in $\meanconfigspace$ to the \particle~currently residing there at time $t$ via the mapping $\meanflowmap{\cdot}$. Then relation (\ref{XY}) can be written as
\begin{equation}
    \label{XYeuler}
    \fluctmap{\x} = \flowmap{\meanflowmapinv{\x}} - \x,
\end{equation}
showing that, at each point $\x$, $\fluctmap{\x}$ provides the actual location of the \particle~relative to $\x$ itself. In fact, it is clear from (\ref{composition}) that $\secondmap{\x} = \flowmap{\meanflowmapinv{\x}}$: the two sides of the equality just represent two different routes to the same map, as can be observed in Figure~\ref{fig:flowmap}.  Indeed, through this map formalism, any field quantity $\phi$ can be viewed from one of three perspectives: as an Eulerian field quantity in the configuration space $\meanconfigspace$,  $\push{\phi}(\cdot,t)$; as the `actual' Eulerian field in $\configspace$, $\phi(\cdot,t)$; or as a Lagrangian (i.e, \particle-centered) field, $\material{\phi}(\cdot,t)$, associating $\phi$ to \particles~in $\refspace$. They are related by
\begin{equation}
\label{phirelate}
    \push{\phi}(\x,t) \equiv \phi(\secondmap{\x},t)  \equiv \material{\phi}(\meanflowmapinv{\x},t).
\end{equation}

Similarly, we can rewrite the velocity relation (\ref{wW}) as
\begin{equation}
    \label{wWeuler}
    \ddp {}{t}\fluctmap{\x}  + \genvmean(\x,t) \cdot \grad\fluctmap{\x}   = \push{\genv}(\x,t) - \genvmean(\x,t),
\end{equation}
where we have defined the Eulerian velocity field, $\genvmean(\x,t) \equiv \material{\genvmean}(\meanflowmapinv{\x},t)$, associated with the flow map $\meanflowmap{\cdot}$. By $\push{\genv}(\x,t)$, we denote the actual velocity of the \particle~currently mapped to $\x$ by $\meanflowmap{\cdot}$:
\begin{equation}
    \push{\genv}(\x,t) \equiv \material{\genv}(\meanflowmapinv{\x},t).
\end{equation}
\Eqn~(\ref{wWeuler}) shows that this velocity differs from $\genvmean(\x,t)$ by a correction described by the rate of change of the $\fluct$ field measured along the $\meanflowmap{\cdot}$ trajectory passing through $\x$ at time $t$.

With this formalism in place, following Andrews and McIntyre \citep{andrews1978exact}, we can define the Lagrangian mean $\lagrange{(\cdot)}$ of any field $\push{\phi}(\x,t)$ as equal to the Eulerian mean along the trajectory followed by $\x$ in the configuration space $\configspace$ under the map $\secondmap{\x} \equiv \x + \fluctmap{\x}$:
\begin{equation}
\label{lagrangedef}
    \lagrange{\phi(\x,t)}  \equiv \euler{\phi(\x + \fluctmap{\x},t)} \equiv  \euler{\push{\phi}(\x,t)}.
\end{equation}

Thus far, we have not specified anything about the map $\meanflowmap{\cdot}$ and its associated field $\fluctmap{\x}$. GLM theory assigns $\fluctmap{\x}$ the role of a fluctuation field, and furthermore, asserts that it has zero Eulerian mean and that the velocity field $\genvmean(\x,t)$ is its own mean.
\begin{equation}
\label{axioms}
    \euler{\fluctmap{\x}} = 0,\quad \euler{\genvmean(\x,t)} = \genvmean(\x,t).
\end{equation}
We can immediately note that, by taking the mean of \eqn~(\ref{secondmap}) and applying the first of these axioms, we get
\begin{equation}
    \euler{\secondmap{\x}} = \x.
\end{equation}
That is, the location $\x$ in $\meanconfigspace \times[0,\infty)$ maps {\em on average} to the location $\x$ in $\configspace\times[0,\infty)$. Furthermore, by taking the Eulerian mean of the relationships (\ref{XYeuler}) and (\ref{wWeuler}), the axioms (\ref{axioms}) immediately imply other important relationships between the configuration spaces $\meanconfigspace\times[0,\infty)$ and $\configspace\times[0,\infty)$: The trajectory in $\meanconfigspace\times[0,\infty)$ described by the flow map $\meanflowmap{\plabel}$ is the mean of the actual trajectory $\flowmap{\plabel}$ in $\configspace\times[0,\infty)$, so that (\ref{XY}), rewritten trivially as
\begin{equation}
\label{lagrangereynolds}
    \flowmap{\plabel} = \meanflowmap{\plabel} + \fluctmap{\meanflowmap{\plabel}},
\end{equation}
represents a Reynolds decomposition of the trajectory, and $\fluct$ can be called the {\em disturbed displacement field}; and the velocity $\genvmean$ represents the Lagrangian mean of the actual velocity $\genv$,
\begin{equation}
\label{genmeanv}
    \genvmean(\x,t) = \lagrange{\genv(\x,t)},
\end{equation}
or simply, $\genvmean = \lagrange{\genv}$. From hereon, we will refer to $\meanconfigspace\times[0,\infty)$ as the {\em mean configuration space}.

If we define the Lagrangian mean material derivative as the rate of change while moving along a mean trajectory,
\begin{equation}
    \DDtL {} \equiv \ddp {}{t} + \lagrange{\genv}\cdot\grad,
\end{equation}
and the Lagrangian disturbance velocity as the difference between the actual velocity and the Lagrangian mean velocity,
\begin{equation}
\label{genvl}
    \genv^{l} \equiv \push{\genv} - \lagrange{\genv},
\end{equation}
then we can rewrite the velocity relationship (\ref{wWeuler}) as
\begin{equation}
\label{flucteqn}
    \DDtL{\fluct} = \genv^{l}.
\end{equation}
This equation provides the basis for generating the actual trajectory of the \particle~while following the \particle's mean trajectory. It is useful to note that the Lagrangian mean of this equation is identically zero. 

The analysis of other fields follows from the definitions thus far. For a general field $\phi$, if we differentiate the relationship (\ref{phirelate}) with respect to time and apply the chain rule, then, with the help of (\ref{wWeuler}), it can be shown that 
\begin{equation}
\DDtL{\push{\phi}}(\x,t) = \push{\left(\DDt \phi\right)}(\x,t) \equiv \left( \ddp {\phi}{t} + \genv\cdot\grad\phi \right)(\secondmap{\x},t).
\end{equation}
In other words, the rate of change of $\push{\phi}$ measured while moving along the mean trajectory is identical to the rate of change measured while moving along the actual trajectory. That is, no information has been lost while following a different trajectory. The Reynolds decomposition of this field follows naturally from the Lagrangian mean,
\begin{equation}
    \push{\phi}(\x,t) = \lagrange{\phi}(\x,t) + \phi^{l}(\x,t).
\end{equation}


\bibliographystyle{unsrt}
\bibliography{mybib}

\end{document}